\documentclass[twoside]{IEEEtran}
\usepackage{amssymb}
\usepackage{graphicx}

\begin{document}

\newtheorem{corollary}{Corollary}[section]
\newtheorem{definition}{Definition}[section]
\newtheorem{example}{Example}[section]
\newtheorem{lemma}{Lemma}[section]
\newtheorem{proposition}{Proposition}[section]
\newtheorem{statement}{Statement}[section]
\newtheorem{theorem}{Theorem}[section]
\newtheorem{property}{Property}
\newtheorem{fact}{Fact}[section]

\title{Partial Recovery of Quantum Entanglement\thanks{This work was partly supported by the Natural Science Foundation
of China (Grant Nos. 60503001, 60321002, and 60305005), and by
Tsinghua Basic Research Foundation (Grant No. 052220204). R. Duan
acknowledges the financial support of Tsinghua University (Grant
No. 052420003).}}
\author{Runyao Duan,\ \ \ Yuan Feng,\ \ \ and Mingsheng Ying\thanks{ The authors are with
the State Key Laboratory of Intelligent Technology and Systems,
Department of Computer Science and Technology, Tsinghua
University, Beijing, China 100084. E-mails:
dry02@mails.tsinghua.edu.cn (Runyao Duan), feng-y@tsinghua.edu.cn
(Yuan Feng), and yingmsh@tsinghua.edu.cn (Mingsheng Ying)}}
\date{\today}
\maketitle

\begin{abstract}
Suppose Alice and Bob try to transform an entangled state shared
between them into another one by local operations and classical
communications. Then in general a certain amount of entanglement
contained in the initial state will decrease in the process of
transformation. However, an interesting phenomenon called partial
entanglement recovery shows that it is possible to recover some
amount of entanglement by adding another entangled state and
transforming the two entangled states collectively.

In this paper we are mainly concerned with the feasibility of
partial entanglement recovery. The basic problem we address is
whether a given  state is useful in recovering entanglement lost
in a specified transformation. In the case where the source and
target states of the original transformation satisfy the strict
majorization relation, a necessary and sufficient condition for
partial entanglement recovery is obtained. For the general case we
give two sufficient conditions. We also give an efficient
algorithm for the feasibility of partial entanglement recovery in
polynomial time.

As applications, we establish some interesting connections between
partial entanglement recovery and the generation of maximally
entangled states, quantum catalysis, mutual catalysis, and
multiple-copy entanglement transformation.

\smallskip\
\end{abstract}

\section{Introduction}

\PARstart{Q}{uantum} entanglement is a valuable resource in
quantum information processing. It can implement some information
processing tasks that cannot be accomplished classically. As a
consequence, entanglement has been widely used in quantum
cryptography \cite{BB84}, quantum superdense coding \cite{BS92},
and quantum teleportation \cite{BBC+93}; see \cite{M00}, Chapter
12 for an excellent exposition. Due to the great importance of
quantum entanglement, a fruitful branch of quantum information
theory named quantum entanglement theory is currently being
developed.

Since quantum entanglement exists between different subsystems of
a composite system shared by spatially separated parties, a
natural constraint on the manipulation of entanglement is that the
separated parties are only allowed to perform local quantum
operations on their own subsystems and to communicate to each
other classically (LOCC). Using this restricted set of
transformations, the parties are often required to optimally
manipulate the entangled state. One of the central problems of
quantum entanglement theory is thus to find the conditions for
when an entangled state can be transformed into another one using
LOCC. This problem can be approached in two different, but
complementary, contexts: the finite regime and the asymptotic
regime. In the asymptotic regime Bennett and his collaborators
\cite{BBPS96} proposed a reversible protocol which shows that any
two bipartite entangled pure states with infinite copies can be
converted into each other without any loss of entropy of
entanglement. Since in practice one can only have finitely many
copies of an entangled state, it is of great interest to consider
the problem of entanglement transformation in a finite
(non-asymptotic) setting. Arguably, the most important step in the
finite regime was made by Nielsen in \cite{NI99}, where he
reported a necessary and sufficient condition for a bipartite
entangled pure state to be transformed into another pure one
deterministically using LOCC. Suppose two distantly located
parties, Alice and Bob, share an entangled state $|\psi\rangle$ in
$\mathcal{C}^{n}\otimes \mathcal{C}^n$ with Schmidt decomposition
$ |\psi\rangle=\sum_{i=1}^n \sqrt{\alpha_i}|i_A\rangle|i_B\rangle
$, where $\alpha_1 \geq \alpha_2 \geq \cdots \geq \alpha_n\geq 0$
are Schmidt coefficients and $\sum_{i=1}^n\alpha_i=1$.
$|i_A\rangle$ and $|i_B\rangle$ are orthonormal bases of Alice's
and Bob's systems, respectively. Suppose the two parties want to
transform $|\psi\rangle$ into another state $|\varphi\rangle$ with
Schmidt decomposition $|\varphi\rangle=\sum_{i=1}^n
\sqrt{\beta_i}|i'_A\rangle|i'_B\rangle $, where $\beta_1 \geq
\beta_2 \geq \cdots \geq \beta_n\geq 0$ and
$\sum_{i=1}^n\beta_i=1$. The orthonormal bases $|i_A\rangle$ and
$|i_A'\rangle$ (similarly, $|i_B\rangle$ and $|i_B'\rangle$) are
not necessarily the same. For the sake of convenience, we write
$\psi=(\alpha_1,\cdots,\alpha_n)$ and
$\varphi=(\beta_1,\cdots,\beta_n)$ for the respective ordered
Schmidt coefficient vectors of $|\psi\rangle$ and
$|\varphi\rangle$. Nielsen proved that Alice and Bob can achieve
this transformation of $|\psi\rangle$ to $|\varphi\rangle$ with
certainty using LOCC, written $|\psi\rangle\rightarrow
|\varphi\rangle$, if and only if $\psi\prec \varphi$. Here the
symbol `$\prec$' stands for `majorization relation', and
$\psi\prec \varphi$ holds if and only if
$$ \sum_{i=1}^l \alpha_i\leq \sum_{i=1}^l \beta_i {\rm  \ for\ all\ }1\leq l< n $$
and $\sum_{i=1}^n \alpha_i= \sum_{i=1}^n \beta_i$. If all
inequalities in the above equation hold strictly and $\sum_{i=1}^n
\alpha_i= \sum_{i=1}^n \beta_i$, then we say that $\psi$ is
strictly majorized by $\varphi$. Majorization is an interesting
and well-developed topic in linear algebra. For more details, we
refer the reader to \cite{MO79} and \cite{AU82}.

Nielsen's theorem establishes a connection between the theory of
majorization and entanglement transformation. It is of fundamental
importance in studying entanglement transformation and has many
interesting corollaries. For example, by taking limits the
asymptotic result of Bennett {\it et al} can be recovered from
Nielsen's theorem. Unlike the transformations in the asymptotic
regime, a direct implication of Nielsen's theorem is that the
amount of entanglement decreases during the deterministic
transformation. Let
$E(|\psi\rangle)=-\sum_{i=1}^n\alpha_i\log_2\alpha_i$ be the
entropy of entanglement of $|\psi\rangle$. Then by the properties
of majorization, it follows that $|\psi\rangle\rightarrow
|\varphi\rangle$ implies $E(|\varphi\rangle)\leq E(|\psi\rangle)$
\cite{NI99}. Indeed, these properties of majorization imply that
any well-behaved entanglement measures, such as Renyi's entropy,
or any other suitable concave functions, also decrease under LOCC.
Intuitively, this means that a certain amount of entanglement will
be lost in a LOCC transformation. It would be desirable to save
some entanglement lost and reduce the net loss of entanglement in
the transformation, since the saved entanglement can be used, for
example, to increase the classical capacity of a quantum channel
\cite{Shor02}.

The possibility of recovering lost entanglement was first observed
by Morikoshi \cite{FM00}. We outline Morikoshi's recovering scheme
as follows. Suppose Alice and Bob share an entangled state
$|\psi\rangle$ and they can transform it into $|\varphi\rangle$ by
LOCC. As we mentioned above, this process is generally a
dissipative one in the sense that the quantity of entanglement in
the target state is less than that in the source state. Suppose
now an auxiliary state $|\chi\rangle$ is supplied to Alice and
Bob. Instead of transforming $|\psi\rangle$ into $|\varphi\rangle$
directly, they perform collective operations on the joint state
$|\psi\rangle\otimes|\chi\rangle$, and transform it into another
joint state $|\varphi\rangle\otimes |\omega\rangle$. Of course, as
required by Nielsen's theorem, entropy of entanglement of the
whole system decreases too. But by choosing a suitable auxiliary
state $|\chi\rangle$, sometimes a state $|\omega\rangle$ with more
entropy of entanglement can be obtained. Intuitively, this process
enables part of entanglement lost in the original transformation
to be transferred to the auxiliary state, and it was termed {\it
partial entanglement recovery}. Morikoshi demonstrated that
partial entanglement recovery for a transformation between
$2\times 2$ states is always possible by using a $2\times 2$
auxiliary state.

Partial entanglement recovery for transformations between higher
dimensional states was considered by Bandyopadhyay \textit{et al.}
in \cite{SVF01}. To avoid trivial cases (a perfect recovery can
always be achieved by letting $|\chi\rangle=|\varphi\rangle$ and
$|\omega\rangle=|\psi\rangle$), a notion of \textit{genuine}
partial recovery was introduced. A partial recovery scheme is
genuine if the dimension of the auxiliary state is smaller than
that of the original source state. Then it was proven that for any
states $|\psi\rangle$ and $|\varphi\rangle$ such that $\psi$ is
strictly majorized  by $\varphi$ and $n>2$, a genuine partial
recovery is always possible by using only $2\times 2$-dimensional
auxiliary states. This extensively generalized the result in
\cite{FM00}. The possibility of genuine partial recovery for the
transformation of $|\psi\rangle$ to $|\varphi\rangle$ such that
$\psi$ is not strictly majorized by $\varphi$ was also examined
carefully in \cite{SVF01}. However, several fundamental problems
concerning partial entanglement recovery are still open. For
example, the existence of genuine partial recovery in the case of
$\alpha_n=\beta_n$ is unknown. Furthermore, the proof confirming
the existence of auxiliary states for partial entanglement
recovery presented in \cite{SVF01} is nonconstructive. In general,
this proof method does not provide a way to find these auxiliary
states efficiently.

In this paper, we study the feasibility of partial entanglement
recovery. We consider the problem of whether a given entangled
state can be used to recover some entanglement lost in a specified
transformation. Our motivations are twofold. The first one is more
theoretical. In some sense, the process of partial entanglement
recovery reveals a new kind of application of quantum
entanglement: it can be used to store some entanglement lost in
information processing tasks. So it is of great theoretical
interest to characterize the entanglement recovering ability of a
given entangled state, as it may lead to a better understanding of
some fundamental properties of quantum entanglement. In addition,
as will see later, the solution of the above problem leads us to a
rich mathematical structure and provides new insight into the
process of partial entanglement recovery. The second motivation is
more practical. Suppose we are required to perform a couple of
different entanglement transformations. In most applications the
available entangled states shared between two parties are
pre-specified and very limited. A solution to the above problem
will help us to determine whether partial entanglement recovery
for these transformations is possible with other pre-specified
entangled states. It is also worth noting that this problem is
more general than the ones discussed in \cite{FM00} and
\cite{SVF01}, and its solution automatically resolves many trivial
cases.

To state the above problem more formally, let us assume that
$|\psi\rangle$ and $|\varphi\rangle$ are the source state and the
target state of the specified transformation, respectively,  and
let $|\chi\rangle$ be the given auxiliary state. Furthermore,
suppose that $|\psi\rangle$ can be transformed into
$|\varphi\rangle$ with certainty using LOCC. Our goal is to
determine whether there exists another state $|\omega\rangle$
satisfying (i) the transformation of $|\psi\rangle\otimes
|\chi\rangle$ to $|\varphi\rangle\otimes |\omega\rangle$ can be
implemented with certainty using LOCC,  and (ii) $|\omega\rangle$
is more entangled than $|\chi\rangle$. Next, we clarify a subtle
point: what is meant by the statement that a state is more
entangled than another one? An exact mathematical definition is
needed. One way to do this is to use some measures of entanglement
such as entropy of entanglement mentioned above, as in \cite{FM00}
and \cite{SVF01}. Note that for deterministic transformations, a
single measure of entanglement is usually not enough to quantify
entanglement amount since there exist incomparable states
$|\psi\rangle$ and $|\varphi\rangle$, i.e., neither
$|\psi\rangle\rightarrow |\varphi\rangle$ nor
$|\varphi\rangle\rightarrow |\psi\rangle$ is possible \cite{NI99}.
So in the present paper, we adopt an alternative view-point: we
say that $|\omega\rangle$ is more entangled than $|\chi\rangle$ if
$|\omega\rangle\rightarrow |\chi\rangle$ and
$|\chi\rangle\nrightarrow |\omega\rangle$. By Nielsen's theorem,
this is equivalent to $\omega\prec \chi$ and $\chi\neq \omega$
(here both $\chi$ and $\omega$ are in non-increasing order). We
believe that this view-point is more reasonable than only
considering a single measure. Now the mathematical problem of the
feasibility of partial entanglement recovery can be clearly
formulated as follows:

\smallskip
\textit{Problem 1. Given a triple of states
$(|\psi\rangle,|\varphi\rangle, |\chi\rangle)$ such that
$\psi\prec \varphi$, determine whether there exists a state
$|\omega\rangle$ such that $\psi\otimes \chi\prec \varphi\otimes
\omega$, $\omega\prec \chi$, and $\chi\neq \omega$. }
\smallskip

If such a state $|\omega\rangle$ does exist, then we call it a
solution of Problem 1. In the above formulation we made no
additional assumptions on the dimension of $\chi$ except that it
is finite. So even in the case that the dimension of $\chi$ is
larger than or equal to that of $\psi$ (and $\varphi$), the above
problem still makes sense. This enables us to consider the process
of partial entanglement recovery in a general mathematical
framework. It is also worth pointing out that Problem 1 cannot be
directly solved by linear programming methods because the
majorization relation $\psi\otimes \chi\prec \varphi\otimes
\omega$ cannot be expressed by linear constraints, unless we know
how to order a tensor product of two probability vectors,
$\varphi\otimes \omega$. The main difficulty here is that the
order of $\varphi\otimes \omega$ is not related in any simple way
to the orders of $\varphi$ and $\omega$.

The principal aim of the present paper is to solve Problem 1
stated above. We first introduce three indices of uniformity for
bipartite entangled pure states. With the aid of these indices, we
prove that whether $|\chi\rangle$ can save some entanglement lost
for the transformation of $|\psi\rangle$ to $|\varphi\rangle$ only
depends on the target state $|\varphi\rangle$ and the presence of
the equalities in the majorization  relation $\psi\prec \varphi$.
To be concise, let $|\psi\rangle$ be a state with $m$ distinct
Schmidt coefficients, say, $\alpha_1'> \cdots> \alpha_m'\geq 0$.
If $m>1$, then the maximal local uniformity of $|\psi\rangle$,
denoted by $L_u(|\psi\rangle)$, is given by the maximal ratio of
$\alpha_{i+1}'$ and $\alpha_i'$ for all $1\leq i\leq m-1$. In
contrast, the global uniformity of $|\psi\rangle$, denoted by
$g_u(|\psi\rangle)$, is given by the ratio of $\alpha_m'$ and
$\alpha_1'$. In the special case of $m=1$, both indices are
defined to be $1$. These indices have many useful properties.
Indeed, they are key tools in studying partial entanglement
recovery. With these notions, Problem 1 is completely solved in
the case where $\psi$ is strictly majorized by $\varphi$ (Theorem
\ref{shannoncode}). We achieve this goal by considering two cases.
First, Problem 1 is examined carefully for a special case where
all nonzero Schmidt coefficients of $|\chi\rangle$ are identical,
i.e., $L_u(|\chi\rangle)=0$ or $L_u(|\chi\rangle)=1$. Second we
consider the general case that $0<L_u(|\chi\rangle)<1$ and prove:

(1) if $L_u(|\chi\rangle)>g_u(|\varphi\rangle)$, then
$|\chi\rangle$ can recover some entanglement lost for the
transformation of $|\psi\rangle$ to $|\varphi\rangle$;

(2) if $L_u(|\chi\rangle)=g_u(|\varphi\rangle)$, then there is
only a special form of $|\varphi\rangle$ for which $|\chi\rangle$
can save some entanglement lost in the transformation of
$|\psi\rangle$ to $|\varphi\rangle$; and

(3) if $L_u(|\chi\rangle)<g_u(|\varphi\rangle)$, then
$|\chi\rangle$ cannot recover entanglement lost in any
transformation with the target $|\varphi\rangle$.

It should be pointed out that the proof we present provides an
explicit construction of the resulting state $|\omega\rangle$. In
view of this, the above results are very useful in pursuing
practical applications of partial entanglement recovery. Some
interesting special cases of these results are also discussed.

For the case where $\psi$ is not strictly majorized by $\varphi$,
a complete solution of Problem 1 appears to be very difficult.
Nevertheless, two sufficient conditions for partial entanglement
recovery are presented (Theorems \ref{generaltheorem1} and
\ref{generaltheorem2}). Employing  these conditions as tools we
show that the genuine partial recovery is not always possible when
the dimension of the target state is larger than $2\times 2$. For
example, if $n=3$ and $\alpha_1=\beta_1$, $|\chi\rangle$ should be
at least a $3\times 3$ entangled state, which means any recovery
scheme cannot be genuine. (This result in fact has been obtained
implicitly in \cite{SVF01}). When $\alpha_1=\beta_1$ and
$\alpha_n=\beta_n$, we show that $4\times 4$-dimensional auxiliary
states are necessary and sufficient. In the case where $n=4$, a
genuine partial recovery is not possible. On the other hand, even
in these special cases, it still makes sense to consider whether
$|\chi\rangle$ is useful in recovering entanglement lost in the
transformation of $|\psi\rangle$ to $|\varphi\rangle$.

Besides the mathematical characterization of partial entanglement
recovery outlined above, we also present an algorithmic approach
to Problem 1. Let $n$ and $k$ be the dimensions of $\psi$ (as well
as $\varphi$) and $\chi$ (as well as $\omega$), respectively. Our
goal now is to design polynomial time algorithms in $n$ or/and $k$
to solve Problem 1. As mentioned above, the main difficulty in
solving Problem 1 lies in the fact that the order of the tensor
product $\varphi\otimes \omega$ cannot be determined by a simple
method even after we know the orders of $\varphi$ and $\omega$.
Thus one cannot apply standard linear programming techniques
directly. A naive enumeration of the possible orders of
$\varphi\otimes \omega$ yields about $(nk)!$ results, which is
intractable. A simple but powerful lemma is introduced to reduce
the number of orders of the tensor product. The basic idea behind
this lemma comes from the observation that for a fixed $\varphi$,
$\varphi\otimes \omega$ has at most $O((kn)^{2(k-1)})$ different
orders when $\omega$ varies. This number of the possible orders is
only a polynomial in $n$ when $k$ is treated as a constant. For
each possible order, we can employ linear programming methods to
solve the majorization inequality $\psi\otimes \chi\prec
\varphi\otimes \omega$. Consequently, an algorithm of time
complexity $O(n^{2k-1}\log_2n)$ is obtained (Theorem
\ref{algorithm1}). This algorithm is not efficient in the case
where $k$ can vary freely. Fortunately, by examining the
mathematical structure of partial entanglement recovery carefully,
we can further refine this algorithm into a new one with time
complexity $O(n^2k^4)$ (Theorem \ref{algorithm2}). Therefore we
can efficiently determine the feasibility of partial entanglement
recovery by using algorithmic methods.

To illustrate the utility of the above results, we show that
partial entanglement recovery also happens in situations such as
quantum catalysis, mutual catalysis, and multiple-copy
transformation. As an interesting application, we consider the
generation of maximally entangled states using the scheme of
partial entanglement recovery. We prove that any transformation
with the Schmidt coefficient vector of the source state being
strictly majorized by that of the target state can always
concentrate some partially entangled state into a maximally
entangled one. We also find a close connection between partial
entanglement recovery and quantum catalysis (see \cite{JP99, DF03,
DF04}). That is, if a transformation can be implemented with
certainty by using some quantum catalyst, then entanglement lost
in the transformation can be partially recovered by a suitable
auxiliary state. Moreover, we show that partial entanglement
recovery is directly connected to mutual catalysis \cite{XZZ02}.
As a consequence, a systematic construction of the instances with
mutual catalysis effect is sketched. When we consider the
possibility of partial entanglement recovery in multiple-copy
transformations (see \cite{SRS02}, \cite{DF03}, and \cite{DF04}),
we notice a very interesting phenomenon: although an auxiliary
state cannot be used to do partial entanglement recovery for a
single-copy transformation, it can recover some entanglement lost
in certain multiple-copy transformations.

The rest of the paper is organized as follows. Section II presents
some notations and  concepts, including the definitions of
uniformity indices. In Section III, we present a complete solution
to Problem 1 in the case that $\psi$ is strictly majorized by
$\varphi$. We consider general transformations in Section IV and
give two sufficient conditions for partial entanglement recovery.
Some special but interesting cases of these conditions are
investigated in detail. In Section V we discuss the feasibility of
partial entanglement recovery from an algorithmic viewpoint and
present two algorithms to solve Problem 1. To understand whether
partial entanglement recovery is possible in situations such as
quantum catalysis, mutual catalysis, and multiple-copy
transformation, we give more examples and discussions in Section
VI. In Section VII, we draw a brief conclusion. The proofs of some
lemmas and theorems are completed in Appendices.

\section{Preliminaries}
First, it is helpful to introduce some notations associated with
finite dimensional vectors. Let $x$ be an $n$-dimensional  vector.
The dimension of $x$ is denoted by ${\rm dim}(x)$, i.e., ${\rm
dim}(x)=n$. The notation $x^{\downarrow}$ is used to stand for the
vector that is obtained by rearranging the components of $x$ into
non-increasing order. Similarly, $x^{\uparrow}$ denotes the vector
that is obtained by rearranging the components of $x$ into
non-decreasing order. The notation $x^{\oplus k}$ denotes the
direct sum of $x$ with itself $k$ many times. In particular, for
constant $c$, $c^{\oplus k}$ is the $k$-dimensional vector $(c,c,
\ldots, c)$. If every component of $x$ is nonnegative, then we can
write
$$
x^\downarrow=({x_1'}^{\oplus k_1}, \ldots, {x_m'}^{\oplus k_m}),
$$
where $x_1'>\cdots >x_m'\geq 0$, $k_i\geq 1$ for each $i=1,\ldots,
m$, and $\sum_{i=1}^m k_i=n$. The above form of $x^{\downarrow}$
is usually called the compact form of $x$. It is obvious that the
compact form of a nonnegative vector is unique when the dimension
of the vector space under consideration is fixed.

The sum of the $m$ largest components of the vector $x$ is denoted
by $e_m(x)$. That is, $e_m(x)=\sum_{i=1}^m x^{\downarrow}_i$. It
is easily to verify that $e_m({x})$ is a continuous function of
$x$ for each $ m=1,\ldots, n$.

We say that $x$ is majorized by $y$, denoted by $x\prec y$, if
\begin{equation}\label{nielsentheorem}
e_m(x)\leq e_m(y), {\rm\ for\ all\ }1\leq m\leq n-1
\end{equation}
and $e_n(x)=e_n(y)$. If all inequalities in Eq.
(\ref{nielsentheorem}) are strict and $e_n(x)=e_n(y)$, then we
follow the terminology in \cite{SVF01} and say that $x$ is
strictly majorized by $y$, denoted by $x\lhd y$.

A vector $x$ is a segment of a vector $y$ if there exist $i\geq 1$
and $k\geq 0$ such that $x=(y_i, y_{i+1},\ldots, y_{i+k})$.

Now we apply the above terminology to bipartite entangled pure
states. Let $|\psi\rangle$ be an $n\times n$ entangled pure state
with ordered Schmidt coefficients $\alpha_1\geq \alpha_2\geq
\cdots\geq \alpha_n\geq 0$. As we have mentioned in the
introduction, the symbol $\psi$ is  used to denote the Schmidt
coefficient vector of $|\psi\rangle$, i.e.,
$\psi=(\alpha_1,\ldots, \alpha_n)$, which is just an
$n$-dimensional probability vector. We often identify  the compact
form of $\psi$ with the compact form of state $|\psi\rangle$. We
call $|\psi\rangle$ an $n\times n$ maximally entangled state if
the compact form of $|\psi\rangle$ reduces to
$(\frac{1}{n})^{\oplus n}$; otherwise we say that $|\psi\rangle$
is a partially entangled  state. If $\varphi'^{\downarrow}$ is a
segment of $\varphi^{\downarrow}$, then we call $|\varphi'\rangle$
an unnormalized state.

To apply Nielsen's theorem to unnormalized states, it can be
restated as: $|\psi\rangle\rightarrow |\varphi\rangle$ if and only
if $\psi\prec \varphi$.

We define $S(|\varphi\rangle)$ to be the set of all $n\times n$
entangled pure states $|\psi\rangle$ which can be directly
transformed into $|\varphi\rangle$ by LOCC. By Nielsen's theorem,
$S(|\varphi\rangle)=\{|\psi\rangle:{\psi}\prec {\varphi}\}$. We
 also define $S^{o}(|\varphi\rangle)$ to be the set of all
$n\times n$ states $|\psi\rangle$ such that $\psi$ is strictly
majorized by $\varphi$, i.e.,
$S^{o}(|\varphi\rangle)=\{|\psi\rangle:{\psi}\lhd {\varphi}\}$. It
should be noted that states $|\psi\rangle$ in $S(|\varphi\rangle)$
are required to have the same dimension as $|\varphi\rangle$. Such
a requirement forces us to distinguish $S(|\varphi\rangle)$ from
$S(|\varphi'\rangle)$ when $|\varphi\rangle$ and
$|\varphi'\rangle$ are essentially the same state but their
dimensions are different. For example, let
$\varphi=(0.5,0.25,0.25)$ and $\varphi'=(0.5,0.25,0.25,0)$. It is
obvious that the states $|\varphi\rangle$ and $|\varphi'\rangle$
are essentially the same. However, according to the above
definitions, $S(|\varphi\rangle)$ is completely different from
$S(|\varphi'\rangle)$. This design decision in defining
$S(|\varphi\rangle)$ enables us to considerably simplify the
presentation of our main results. The same remark also applies to
the definition of $S^o(|\varphi\rangle)$.

In this paper, the phrase `bipartite entangled pure state' is used
frequently. So, for convenience, sometimes we abbreviate it to
`state' or `quantum state'. This should not cause any confusion
from the context.

Now we introduce three notions which are key mathematical tools in
describing partial entanglement recovery.

\begin{definition}\label{uniformity}
Let $|\psi\rangle$ be an $n\times n$ partially entangled state
with compact form $(\alpha_1'^{\oplus k_1},\ldots,
\alpha_m'^{\oplus k_m})$, where $n=\sum_{i=1}^m k_i$ and $m>1$.
Then

(i) the minimal local uniformity of $|\psi\rangle$ is defined by
$$
 l_u(|\psi\rangle)={\rm min}\{\frac{\alpha_{i+1}'}{\alpha_i'}: 1\leq
 i<m\};
$$

(ii) the maximal local uniformity of $|\psi\rangle$ is defined by
$$
 L_u(|\psi\rangle)={\rm max}\{\frac{\alpha_{i+1}'}{\alpha_i'}: 1\leq
 i<m\};
$$

(iii) the global uniformity of $|\psi\rangle$ is defined by
$$
g_u(|\psi\rangle)=\frac{\alpha_m'}{ \alpha_1'}.
$$

\end{definition}

It is easy to see that the minimal local uniformity, the maximal
local uniformity, and the global uniformity of a  quantum state
$|\psi\rangle$ with $\psi^{\downarrow}=(\alpha_1,\ldots,
\alpha_n)$ may be rewritten in a slightly different way:
$$
 l_u(|\psi\rangle)={\rm min}\{\frac{\alpha_{i+1}}{\alpha_i}: 1\leq
 i<n\};
$$

$$
 L_u(|\psi\rangle)={\rm max}\{\frac{\alpha_{i+1}}{\alpha_i}: 1\leq
 i<n {\rm \ and\ } \alpha_i> \alpha_{i+1}\};
$$
$$
g_u(|\psi\rangle)=\frac{\alpha_n}{ \alpha_1}.
$$
The above rewriting will help us to simplify some proofs.

From the above rewriting of Definition \ref{uniformity}, it is
easy to see that both $l_u(|\psi\rangle)$ and $g_u(|\psi\rangle)$
are continuous with respect to $|\psi\rangle$. Thus it is
reasonable to define the minimal local uniformity and the global
uniformity of a maximally entangled state to be $1$. However, such
a continuous property does not hold for the maximal local
uniformity. To keep many properties of these indices  valid even
in the case that the quantum state under consideration is
maximally entangled, it is  convenient to define the maximal local
uniformity of a maximally entangled state  to be $1$. Also, for
the sake of convenience, when the dimension of the state under
consideration is one-dimensional, we define the uniform indices as
$1$.

In applying the above definitions of uniformity indices, it should
be noted that the dimension of $|\psi\rangle$ is somewhat
arbitrary, as one can append zeroes  to the vector $\psi$ and
thereby increase its dimension without changing the underlying
quantum state. Suppose that the number of nonzero components of
$\psi$ is $n$. If $|\psi\rangle$ is treated as an $n\times n$
state,  all the above three uniformity indices are positive.
However, if we append zeroes to $\psi$ and yield a state
$|\psi'\rangle$, then the uniformity indices of $|\psi'\rangle$
are changed rapidly. For example, let $\psi=(0.5,0.25,0.25)$ and
$\psi'=(0.5,0.25,0.25,0)$. It is obvious that  both the minimal
local uniformity and the global uniformity of $|\psi\rangle$ are
$0.5$. However, the minimal local uniformity and the global
uniformity of $|\psi'\rangle$ are changed into $0$. To avoid any
confusion that may be caused by the phenomenon that we just
mentioned in the above definition, the dimension of the states are
treated as fixed. In other words, if $|\psi'\rangle$ is obtained
from $|\psi\rangle$ by appending zeros in its Schmidt coefficient
vector, they may be thought of being two different states.
Therefore, it is reasonable to allow that sometimes
$l_u(|\psi\rangle)\neq l_u(|\psi'\rangle)$ (as well as
$L_u(|\psi\rangle)\neq L_u(|\psi'\rangle)$ and
$g_u(|\psi\rangle)\neq g_u(|\psi'\rangle)$).

Some simple but useful properties of the three indices defined
above are presented in the following:

\begin{lemma}\label{property1}\upshape
Let $|\psi\rangle$ and $|\varphi\rangle$ be two quantum states
with compact forms $\psi^{\downarrow}=(\alpha_1'^{\oplus
k_1},\ldots, \alpha_r'^{\oplus k_r})$ and
$\varphi^{\downarrow}=(\beta_1'^{\oplus l_1},\ldots,
\beta_s'^{\oplus l_s})$, respectively. Then

(1){\rm\ }  $0\leq l_u(|\psi\rangle), L_u(|\psi\rangle),
g_u(|\psi\rangle)\leq 1$.

(2){\rm\ } $l_u^{r-1}(|\psi\rangle)\leq g_u(|\psi\rangle)\leq
l_u(|\psi\rangle). $

(3){\rm\ } $g_u(|\psi\rangle)\leq L_u^{r-1}(|\psi\rangle)$.

(4){\rm\ }  $g_u(|\psi\rangle)\leq l_u(|\psi\rangle) \leq
L_u(|\psi\rangle)$.

(5){\rm\ } if $r=s$ and $\alpha_i'=\beta_i'$ for $i=1,\ldots, r$,
then $l_u(|\psi\rangle)=l_u(|\varphi\rangle)$. Similarly,
$L_u(|\psi\rangle)=L_u(|\varphi\rangle)$ and
$g_u(|\psi\rangle)=g_u(|\varphi\rangle)$.

(6){\rm\ } if  $|\psi\rangle\rightarrow |\varphi\rangle$, then
$g_u(|\psi\rangle)\geq g_u(|\varphi\rangle)$.

\end{lemma}

{\it Proof.} (1)--(5) follow immediately from Definition
\ref{uniformity}. (6) follows directly from Definition
\ref{uniformity} and the fact that if $|\psi\rangle\rightarrow
|\varphi\rangle$ then
$\alpha_1'\leq \beta_1'$ and $\alpha_r'\geq \beta_s'$. \hfill $\square$\\

We give some remarks on the above properties. (1) shows that the
three indices of minimal local uniformity, maximal local
uniformity and global uniformity are all between 0 and 1.
Moreover, they take value $1$ if the state is maximally entangled.
The minimal local uniformity and the global uniformity take the
value $0$ if the state in question has zero as one Schmidt
coefficient, while the maximal local uniformity takes value $0$ if
it is a maximally entangled state in a state space with lower
dimension, i.e., with a compact form $((\frac{1}{m})^{\oplus m},
0^{\oplus n-m})$ for some $m<n$. If $l_u(|\psi\rangle)=0$ or
$l_u(|\psi\rangle)=1$, i.e., $|\psi\rangle$ has zero as a Schmidt
coefficient or it is maximally entangled, then both the
inequalities in (2) hold with equalities. In the case that
$0<l_u(|\psi\rangle)<1$, the first equality in (2) holds if the
distinct Schmidt coefficients of $|\psi\rangle$ form a geometric
sequence; while the second equality holds if $|\psi\rangle$ has at
most two distinct Schmidt coefficients. The equality in (3) holds
if and only if the distinct Schmidt coefficients of $|\psi\rangle$
form a geometric sequence. (4) can be analyzed similarly. (5)
means that these indices only depend on distinct Schmidt
coefficients of the state. (6) indicates that global uniformity
is decreasing under LOCC.

In addition to these trivial properties displayed in Lemma
\ref{property1}, the following lemma presents three more
interesting properties of global uniformity and minimal local
uniformity:

\begin{lemma}\label{tensor} \upshape
Let $|\psi\rangle$ and $|\varphi\rangle$ be two quantum states.
Then

(1){\rm\ } $g_u(|\psi\rangle\otimes
|\varphi\rangle)=g_u(|\psi\rangle) g_u(|\varphi\rangle)$. In
particular, $g_u(|\psi\rangle^{\otimes m})=g_u^m(|\psi\rangle)$
for any $m\geq 1$.

(2){\rm\ } $l_u(|\psi\rangle\otimes |\varphi\rangle)\geq {\rm
min}\{l_u(|\psi\rangle), l_u(|\varphi\rangle)\}$.

(3){\rm\ } $l_u(|\psi\rangle)\leq l_u(|\psi\rangle^{\otimes
k})\leq {\rm
min}\{\frac{\alpha_2'}{\alpha_1'},\frac{\alpha_{r}'}{\alpha_{r-1}'}\}$,
where $\psi^{\downarrow}=(\alpha_1'^{\oplus k_1},\ldots,
\alpha_r'^{\oplus k_r}).$

\end{lemma}

\textit{Proof.} (1) follows immediately by Definition
\ref{uniformity}. (3) is a simple application of (2) and
Definition \ref{uniformity}. So it is enough to prove (2).

Let $\psi^{\downarrow}=(\alpha_1,\cdots, \alpha_m)$ and
$\varphi^{\downarrow}=(\beta_1,\cdots, \beta_n)$. Since the
minimal uniformity $l_u$ is a continuous functional,  we can
assume without loss of generality that all components of $\psi$
and $\varphi$ are positive. Let
$$a=\alpha_p\beta_q  {\rm \ \ and\ \ }b=\alpha_r\beta_s, {\rm\ \ } a\leq b,$$
be any two successive elements of the ordered probability vector
$(\psi\otimes \varphi)^{\downarrow}$. It is obvious that $r<m$ or
$s<n$. Suppose $r<m$ is the case,  let us try to prove
\begin{equation}\label{equation3}
      l_u(|\psi\rangle)\leq \frac{a}{b}.
\end{equation}
Indeed, from the definition of $l_u$ we have
\begin{equation}\label{equation4}
\begin{array}{rl}
      l_u(|\psi\rangle)&\leq \frac{\alpha_{r+1}}{\alpha_r}\\
      &=\frac{\alpha_{r+1}\beta_s}{\alpha_r\beta_s}\leq 1.
\end{array}
\end{equation}
Thus $\alpha_{r+1}\beta_s\leq b$. But, since $a$ and $b$ are
successive elements, $\alpha_{r+1}\beta_s$ cannot belong to the
interval $(a,b)$, that is,
\begin{equation}\label{equation5}
      \alpha_{r+1}\beta_s\leq a.
\end{equation}
From Eqs. (\ref{equation4}) and (\ref{equation5}) we get
immediately Eq. (\ref{equation3}).

If $r=m$ then we can be sure $s<n$. Thus we can apply analogous
arguments to prove that
$$l_u(|\varphi\rangle)\leq \frac{a}{b}.$$
In both cases one has
$${\rm min}\{l_u(|\psi\rangle),l_u(|\varphi\rangle)\}\leq \frac{a}{b}.$$
Since this is true for any successive $a\leq b$, we have proved
statement (2). \hfill $\square$\\

The above lemma deserves some more remarks.  Intuitively, (1)
shows that the global uniformity is multiplicative  under tensor
product. (2) means that the tensor product of two states is at
least as uniform as one of them. (3) provides an upper bound and a
lower bound respectively for the minimal local uniformity of any
state consisting of multiple copies of a given state. More
interestingly, it shows that the minimal local uniformity of a
$2\times 2$ or $3\times 3$ state remains invariant under tensor
products involving multiple copies.

One of the most interesting applications of the uniformity indices
introduced above is that they provide a characterization of when a
strict majorization relation holds.
\begin{lemma}\label{lginterior}\upshape
Let $|\varphi\rangle$ and $|\chi\rangle$ be  two quantum states,
and let $S^{o}(|\varphi\rangle)\otimes |\chi\rangle$ denote the
set of all states of the form $|\psi\rangle\otimes |\chi\rangle$
with $|\psi\rangle$ in $S^o(|\varphi\rangle)$, i.e.,
$S^{o}(|\varphi\rangle)\otimes |\chi\rangle=\{|\psi\rangle\otimes
|\chi\rangle:|\psi\rangle\in S^{o}(|\varphi\rangle) \}$. Then
$$
S^{o}(|\varphi\rangle)\otimes |\chi\rangle \subseteq
S^{o}(|\varphi\rangle\otimes |\chi\rangle) \Leftrightarrow
l_u(|\chi\rangle)>g_u(|\varphi\rangle).
$$
\end{lemma}

\textit{Proof.}  See Appendix A. \hfill $\square$\\

Roughly speaking, the above lemma shows that if the auxiliary
state $|\chi\rangle$ is  uniform enough, then the strict
majorization relation $\psi\otimes \chi\lhd \varphi\otimes \chi$
can be kept providing $\psi\lhd \varphi$, and vice versa. What we
would like to emphasize here is that the only constraint on the
source state $|\psi\rangle$ is $\psi\lhd \varphi$.

In the introduction we have frequently used the notion of partial
entanglement recovery. We present a rigorous definition as
follows.
\begin{definition}\upshape
Let $|\psi\rangle$ and $|\varphi\rangle$ be two $n\times n$
states, and let $|\chi\rangle$ be a $k\times k$ state. We say that
$|\chi\rangle$ can do partial entanglement recovery for the
transformation of $|\psi\rangle$ to $|\varphi\rangle$ if there
exists a $k\times k$ state $|\omega\rangle$ such that

(i){\rm\ } both the transformations of $|\psi\rangle\otimes
|\chi\rangle$ to $|\varphi\rangle\otimes |\omega\rangle$ and
$|\omega\rangle$ to $|\chi\rangle$ can be realized with certainty
under LOCC. That is, $|\psi\rangle\otimes
|\chi\rangle\rightarrow|\varphi\rangle\otimes |\omega\rangle$ and
$|\omega\rangle\rightarrow |\chi\rangle$;

(ii){\rm\ } the transformation of $|\chi\rangle$ to
$|\omega\rangle$ cannot be achieved with certainty under LOCC.
That is, $|\chi\rangle\nrightarrow |\omega\rangle$.
\end{definition}

Some remarks follow:
\begin{enumerate}

\item In the above definition, both the dimensions of the source
state $|\psi\rangle$ and the target state $|\varphi\rangle$ are
$n\times n$. Similarly, the dimensions of the auxiliary state
$|\chi\rangle$ and the resulting state $|\omega\rangle$ are both
$k\times k$. These constraints are reasonable since during the
transformation process the state space under consideration isn't
modified. Intuitively, $|\psi\rangle$ and $|\varphi\rangle$ are
two different states of the same two particles entangled between
Alice and Bob. The dimensions of these particles are assumed to be
finite and fixed. Similar arguments apply to the states
$|\chi\rangle$ and $|\omega\rangle$. This is in fact the reason
that we have to require that the dimension $|\varphi\rangle$ is
fixed in defining $S(|\varphi\rangle)$, $S^o(|\varphi\rangle)$,
$l_u(|\varphi\rangle)$, $L_u(|\varphi\rangle)$, and
$g_u(|\varphi\rangle)$, since all of them are introduced in this
paper to describe partial entanglement recovery.

\item  According to Nielsen's theorem, the above definition can be
rewritten as: a $k\times k$ auxiliary state $|\chi\rangle$ can do
partial entanglement recovery for a transformation of
$|\psi\rangle$ to $|\varphi\rangle$ if  there exists another
$k\times k$ state $|\omega\rangle$ such that all three relations
$\psi\otimes \chi\prec \varphi\otimes \omega$, $\omega\prec \chi$,
and $\chi^{\downarrow}\neq \omega^{\downarrow}$ hold
simultaneously.

\item  It is obvious that  $(x,0)\prec (y,0)$ if and only if
$x\prec y$. Without loss of generality, we can assume that the
number of nonzero Schmidt coefficients of $|\psi\rangle$ is $n$.
In other words, all Schmidt coefficients of the source state are
positive.
\end{enumerate}

For technical simplicity, we apply  the above discussions not only
to normalized but also to unnormalized states. Sometimes we shall
use a clause such as `$|\chi\rangle$ can save some entanglement
lost for the transformation of $|\psi\rangle$ to
$|\varphi\rangle$', and we shall say that $|\psi\rangle$  can
transfer some entanglement into the state $|\chi\rangle$ whenever
$|\chi\rangle$ can do partial entanglement recovery for the
transformation from $|\psi\rangle$ to some unspecified target
state $|\varphi\rangle$ to mean  that `$|\chi\rangle$ can do
partial entanglement recovery for the transformation of
$|\psi\rangle$ to $|\varphi\rangle$'.

\section{Partial entanglement recovery for a transformation between states with strict majorization}

In this section, we focus on  whether a given auxiliary state
$|\chi\rangle$ can do partial entanglement recovery for a
transformation of $|\psi\rangle$ to $|\varphi\rangle$ such that
$|\psi\rangle$ is in $S^o(|\varphi\rangle)$. A necessary and
sufficient condition for such a recovery is presented. Thus, a
complete characterization of such auxiliary states $|\chi\rangle$
is  obtained.

First, we define the distance between $|\psi\rangle$ and
$|\varphi\rangle$ to be the Euclidean distance between two
$n$-dimensional ordered probability vectors $\psi^{\downarrow}$
and $\varphi^{\downarrow}$, i.e,
$$\||\psi\rangle-|\varphi\rangle \|=\sqrt{\sum_{i=1}^n(\psi^{\downarrow}_i-\varphi^{\downarrow}_i)^2}.$$

Before presenting the main result of this section, we  prove a
useful theorem. Assuming that $|\psi\rangle$ is in
$S^{o}(|\varphi\rangle)$, we shall prove that if
$l_u(|\chi\rangle)>g_u(|\varphi\rangle)$ then a suitable
collective operation can transform the  joint state
$|\psi\rangle\otimes |\chi\rangle$ into another joint state
$|\varphi\rangle\otimes |\omega\rangle$ such that $|\omega\rangle$
is not `far from' $|\chi\rangle$. Surprisingly, this result does
not depend on which source state $|\psi\rangle$ we have chosen at
the beginning.

\begin{theorem}\label{superlginterior}\upshape
Let $|\psi\rangle$ and $|\varphi\rangle$ be  two  states with
$|\psi\rangle\in S^{o}(|\varphi\rangle)$. If $|\chi\rangle$ is an
auxiliary state such that
$l_u(|\chi\rangle)>g_u(|\varphi\rangle)$, then there exists a
positive number $\delta$ such that for any state $|\omega\rangle$
satisfying $\||\omega\rangle-|\chi\rangle\|<\delta$, it holds that
\begin{equation}\label{supercatalyst}
|\psi\rangle\otimes |\chi\rangle \rightarrow
|\varphi\rangle\otimes |\omega\rangle.
\end{equation}
\end{theorem}

\textit{Proof.} Since $l_u(|\chi\rangle)>g_u(|\varphi\rangle)$ and
$|\psi\rangle\in S^{o}(|\varphi\rangle)$, it follows from Lemma
\ref{lginterior} that
\begin{equation}\label{element}
{\psi\otimes \chi}\lhd {\varphi\otimes \chi}.
\end{equation}
Notice that a small enough perturbation on the right hand side of
Eq. (\ref{element}) will not change the relation `$\lhd$' since
every inequality in Eq. (\ref{nielsentheorem}) is strict. Thus it
is possible to take a sufficiently small  positive number $\delta$
such that for any state $|\omega\rangle$ satisfying
$\||\omega\rangle-|\chi\rangle\|<\delta$, the relation
$\psi\otimes \chi\prec \varphi\otimes \omega$ holds, which
confirms the validity of Eq. (\ref{supercatalyst}). With that we
complete the proof of Theorem \ref{superlginterior}.\hfill $\square$\\

The following simple corollary of Theorem \ref{superlginterior}
establishes a connection between uniformity indices  and partial
entanglement recovery.

\begin{corollary}\label{sbsupercatalysis}\upshape
If $g_u(|\varphi\rangle)<l_u(|\chi\rangle)<1$, then $|\chi\rangle$
can do partial entanglement recovery for any transformation of
$|\psi\rangle$ to $|\varphi\rangle$ with $|\psi\rangle\in
S^o(|\varphi\rangle)$.
\end{corollary}
Intuitively, if the minimal local uniformity of a partially
entangled pure state $|\chi\rangle$ is larger than the global
uniformity of $|\varphi\rangle$, then the transformation of
$|\psi\rangle$ to $|\varphi\rangle$ such that $\psi\lhd \varphi$
can always increase the entanglement degree of $|\chi\rangle$.

\begin{example}\label{example1}\upshape
Let  $|\psi\rangle$ and $|\varphi\rangle$ be two $2\times 2$
states with $\psi^{\downarrow}=(a,1-a)$ and
$\varphi^{\downarrow}=(b,1-b)$, where $\frac{1}{2}<a<b<1$. The
goal here is  to find a $2\times 2$ state that  can do partial
entanglement recovery for the transformation of $|\psi\rangle$ to
$|\varphi\rangle$.

Take an auxiliary state $|\chi(p)\rangle$ with
$\chi^{\downarrow}(p)=(p,1-p)$. It is easy to check that
$|\psi\rangle$ is in $S^o(|\varphi\rangle)$. By Corollary
\ref{sbsupercatalysis}, if $|\chi(p)\rangle$ satisfies
\begin{equation}\label{condition1}
g_u(|\varphi\rangle)<l_u(|\chi(p)\rangle)<1,
\end{equation}
then $|\chi(p)\rangle$ can be used to do partial entanglement
recovery for the transformation $|\psi\rangle\rightarrow
|\varphi\rangle$.  It is easy to see that Eq. (\ref{condition1})
is equivalent to
$$\frac{1-b}{b}<\frac{1-p}{p}<1,$$ or $\frac{1}{2}<p<b$. The desired
state $|\omega\rangle$ such that both $|\psi\rangle\otimes
|\chi\rangle\rightarrow |\varphi\rangle\otimes |\omega\rangle$ and
$|\omega\rangle\rightarrow |\chi\rangle$  hold can be taken as
$|\omega\rangle=|\chi(p-\epsilon)\rangle$ with a suitably small
positive number $\epsilon$. It is obvious that
$|\chi\rangle\nrightarrow |\omega\rangle$ whenever $\epsilon$ is
positive but small enough.
\hfill $\square$\\
\end{example}

In  Example \ref{example1}, the condition of $p<b$ means that
$|\chi(p)\rangle$ is more entangled than $|\varphi\rangle$. A
simple analysis shows that this condition is also necessary to
guarantee that $|\chi(p)\rangle$ does partial entanglement
recovery for the transformation of $|\psi\rangle$ to
$|\varphi\rangle$. So we rediscover the main result in
\cite{FM00}: for $2\times 2$-dimensional states, the auxiliary
state $|\chi\rangle$ can do nontrivial partial entanglement
recovery for a transformation with target state $|\varphi\rangle$
if and only if $|\chi\rangle$ is more entangled than
$|\varphi\rangle$.

\begin{example}\label{example2}\upshape
This is a generalization of Example \ref{example1}. Let
$|\psi\rangle$ and $|\varphi\rangle$ be two states such that
$|\psi\rangle$ is in $S^o(|\varphi\rangle)$. Our aim here is to
decide whether there exists some $2\times 2$ state that can do
partial entanglement recovery for the transformation of
$|\psi\rangle$ to $|\varphi\rangle$.

Take an auxiliary state $|\chi(p)\rangle$ with
$\chi^{\downarrow}(p)=(p,1-p)$. By Corollary
\ref{sbsupercatalysis}, if
\begin{equation}\label{condition2}
g_u(|\varphi\rangle)<l_u(|\chi(p)\rangle)<1
\end{equation}
then $|\chi(p)\rangle$ can do partial entanglement recovery for
the transformation of $|\psi\rangle$ to $|\varphi\rangle$.
Moreover, Eq. (\ref{condition2}) is equivalent to
\begin{equation}\label{condition3}
\frac{1}{2}<p<\frac{1}{1+g_u(|\varphi\rangle)}.
\end{equation}

Therefore, the entanglement lost in the transformation of
$|\psi\rangle$ to $|\varphi\rangle$  can always be partially
recovered by a $2\times 2$ state $|\chi(p)\rangle$ satisfying Eq.
(\ref{condition3}). Again, the desired state $|\omega\rangle$ such
that both $|\psi\rangle\otimes|\chi\rangle\rightarrow
|\varphi\rangle\otimes |\omega\rangle$ and
$|\omega\rangle\rightarrow |\chi\rangle$ can be taken as
$|\omega\rangle=|\chi(p-\epsilon)\rangle$ with a suitably small
positive number $\epsilon$.\hfill $\square$\\
\end{example}

In Example \ref{example2}, we show that the entanglement lost in a
transformation of $|\psi\rangle$ to $|\varphi\rangle$ such that
$|\psi\rangle\in S^o(|\varphi\rangle)$ can always be partially
recovered by a $2\times 2$ state $|\chi\rangle$, and the explicit
construction of such a state $|\chi\rangle$ is also presented.
This is a considerable refinement of Theorem 1 in \cite{SVF01}. We
also point out that in the proof of Theorem 1 in \cite{SVF01}, an
important constraint on $p$, i.e., $p\beta_n<(1-p)\beta_1$ or
$l_u(|\chi(p)\rangle)>g_u(|\varphi\rangle)$, is missing, therefore
the case (ii) in the proof of Theorem 1 in \cite{SVF01} is
possible if $x=s=n$ and $y=t=0$,  which makes  the proof there
invalid.

Corollary \ref{sbsupercatalysis} only provides us with a
sufficient condition for which $|\chi\rangle$ can be used to
receive some entanglement lost in a transformation of
$|\psi\rangle$ to $|\varphi\rangle$ with $|\psi\rangle \in
S^{o}(|\varphi\rangle)$. However, this condition is too strong to
be satisfied in many cases. Nevertheless, the following  theorem
gives a weaker condition, and indeed it provides a complete
characterization of states $|\chi\rangle$ that can be used to do
partial entanglement recovery for a transformation with target
state $|\varphi\rangle$ and source state $|\psi\rangle$ in
$S^{o}(|\varphi\rangle)$.

\begin{theorem}\label{shannoncode}\upshape
Let $|\psi\rangle$ and $|\varphi\rangle$ be two $n\times n$ states
such that $|\psi\rangle$ is in $S^o(|\varphi\rangle)$, and let
$|\chi\rangle$ be a $k\times k$ auxiliary state. Then
$|\chi\rangle$ can do partial entanglement recovery for the
transformation of $|\psi\rangle$ to $|\varphi\rangle$ if and only
if one of the following three cases holds:

(i){\rm\ \ } $L_u(|\chi\rangle)=0$ and $na\geq n'(a+1)$. Here $a$
and $n'$ are the numbers of nonzero Schmidt coefficients of
$|\chi\rangle$ and $|\varphi\rangle$, respectively;

(ii){\rm\ } $g_u(|\varphi\rangle)<L_u(|\chi\rangle)<1$;

(iii){\rm\ }$L_u(|\chi\rangle)=g_u(|\varphi\rangle)$ and
$\varphi^{\downarrow}=(\chi'^{\oplus m}/C)^{\downarrow}$. Here
$\chi'$ is a segment of $\chi^{\downarrow}$ with only two distinct
components, $C$ is a normalization factor, and $m\geq 1$.

Moreover, if  none of the above cases holds, then $|\chi\rangle$
cannot do partial entanglement recovery for any transformation of
$|\psi\rangle$ to $|\varphi\rangle$ such that $|\psi\rangle$ is in
$S(|\varphi\rangle)$.
\end{theorem}

\textit{Proof.}  See Appendix B. \hfill $\square$\\

To better understand the above theorem, we give the following
remarks:
\begin{enumerate}

\item The case that $L_u(|\chi\rangle)=1$ is not included in cases
(i), (ii), and (iii). Hence a simple corollary of Theorem
\ref{shannoncode} is that a maximally entangled state cannot be
used to do partial entanglement recovery. This is reasonable since
for such a state $|\chi\rangle$, there does not exist a $k\times
k$ state $|\omega\rangle$ which is more entangled than
$|\chi\rangle$.

\item The case that $L_u(|\chi\rangle)=0$ is slightly different
from the above case and is more interesting. Although
$|\chi\rangle$ is a maximally entangled state in a state space of
lower dimension $a\times a$ with $a<k$, it is only partially
entangled when it is considered as a $k\times k$ state. Hence it
is still possible to transform $|\psi\rangle\otimes |\chi\rangle$
into another state $|\varphi\rangle\otimes |\omega\rangle$, where
$|\omega\rangle$ is more entangled than $|\chi\rangle$. Case (i)
shows that the necessary and sufficient condition is that the
Schmidt numbers of $|\chi\rangle$ and $|\varphi\rangle$ satisfy a
simple inequality $na\geq n'(a+1)$. In some sense, the solution in
this case explains why the dimensions of the states need to be
fixed.

\item Case (ii) means that if $|\chi\rangle$ is partially
entangled and the maximal local uniformity of $|\chi\rangle$ is
larger than the global uniformity of $|\varphi\rangle$, then
$|\chi\rangle$ can be used to save some entanglement lost in the
transformation of $|\psi\rangle$ to $|\varphi\rangle$. This case
provides a feasible sufficient condition for partial entanglement
recovery.

\item Case (iii) is of special interest. It supplies the solution
at the critical point $L_u(|\chi\rangle)=g_u(|\varphi\rangle)$. As
we will see, the proof of this case is very complicated. We
include this case for the following two reasons. First, from the
aspect of the completeness of the solution. Including such a
special case enables us to completely solve the feasibility of
partial entanglement recovery for all $\psi$ and $\varphi$ with
$\psi\lhd \varphi$. Second, from the special form that
$|\varphi\rangle$ should satisfy. A careful observation shows that
$|\varphi\rangle$  has only two different Schmidt coefficients and
should be constructed by repeating a segment of
$\chi^{\downarrow}$ finitely many times. In our opinion this
provides new insight into the process of partial entanglement
recovery. In addition, in the proof of this case we have
extensively employed the techniques introduced in the present
paper and the properties of majorization. Hopefully, these proof
techniques will be useful in solving other problems in quantum
entanglement theory.
\end{enumerate}

In sum, Theorem \ref{shannoncode} provides a necessary and
sufficient condition under which $|\chi\rangle$ can do partial
entanglement recovery for some transformation with the target
state $|\varphi\rangle$. Therefore it can be treated as a basic
result about partial entanglement recovery. In view of Theorem
\ref{shannoncode}, it seems reasonable to use  maximal local
uniformity to describe the partial entanglement recovery power of
an auxiliary state.

It is worth noting that in the above theorem, the choice of
$|\psi\rangle$ has some free degree. That is, if $|\chi\rangle$
can be used to do partial entanglement recovery for a
transformation of $|\psi\rangle$ to $|\varphi\rangle$ such that
$|\psi\rangle$ is in $S^o(|\varphi\rangle)$, then for any
$|\psi'\rangle\in S^o(|\varphi\rangle)$, $|\chi\rangle$ can also
recover entanglement lost in the transformation of $|\psi'\rangle$
to $|\varphi\rangle$.

Theorem \ref{shannoncode} has many interesting corollaries. We
only consider the following  one where the auxiliary state
$|\chi\rangle$ is $2\times 2$-dimensional.
\begin{corollary}\label{qubitpair}\upshape
If $|\chi\rangle$ and $|\varphi\rangle$ are two partially
entangled states with $\chi^{\downarrow}=(p,1-p)$ and
$\varphi^{\downarrow}=(\beta_1,\ldots, \beta_n)$,   then
$|\chi\rangle$ can be used to do partial entanglement recovery for
the transformation of $|\psi\rangle$ to $|\varphi\rangle$ such
that $|\psi\rangle$ is in $S^o(|\varphi\rangle)$ if and only if
one of the following three cases holds:

(i){\rm \ \ } $p=1$ and $n\geq 2n'$, where $n'$ is the number of
nonzero components of $\varphi$;

(ii){\rm \ \ }$\frac{1}{2}<p<\frac{\beta_1}{\beta_1+\beta_n}$;

(iii){\rm\ }$p=\frac{\beta_1}{\beta_1+\beta_n}$ and
$|\varphi\rangle$ has a special form such that
$\varphi^{\downarrow}=((p,1-p)^{\oplus k}/k)^{\downarrow}$ for
some $k\geq
 1$.

Moreover, if none of  (i)--(iii) is satisfied, then $|\chi\rangle$
cannot do partial entanglement recovery for any transformation of
$|\psi\rangle$ to $|\varphi\rangle$ such that $|\psi\rangle$ is in
$S(|\varphi\rangle)$.
\end{corollary}

The most interesting part of the Corollary \ref{qubitpair} is case
(iii). The following example demonstrates this point.
\begin{example}\label{example3}\upshape
Let  $|\chi\rangle$, $|\varphi'\rangle$, $|\varphi''\rangle$, and
$|\varphi'''\rangle$ be  four states with
$\chi^{\downarrow}=(p,1-p),$
$\varphi'^{\downarrow}=(p,p,1-p,1-p)/2,$
$\varphi''^{\downarrow}=(p,p,1-p)/(1+p)$, and
$\varphi'''^{\downarrow}=(p,1-p,1-p)/(2-p)$, where $
\frac{1}{2}<p<1$. Obviously,
$$L_u(|\chi\rangle)=g_u(|\varphi'\rangle)=g_u(|\varphi''\rangle)=g_u(|\varphi'''\rangle)=\frac{1-p}{p}.$$

By Corollary \ref{qubitpair}, it is easy to see that
$|\chi\rangle$ can do partial entanglement recovery for any
transformation of $|\psi\rangle$ to $|\varphi'\rangle$ with
$|\psi\rangle\in S^o(|\varphi'\rangle)$ since
$\varphi'^{\downarrow}=((p,1-p)^{\oplus 2}/2)^{\downarrow}$.

However, again by the above corollary,  $|\chi\rangle$ cannot
recover anything for any transformations with the target states
$|\varphi''\rangle$ or $|\varphi'''\rangle$.\hfill $\square$\\
\end{example}

Until now we only deal with the transformations of $|\psi\rangle$
to  $|\varphi\rangle$ such that $\psi$ is strictly majorized by
$\varphi$. What about the other cases?  In next section, we will
prove two more general theorems about partial entanglement
recovery where $|\psi\rangle$ and $|\varphi\rangle$ only need to
satisfy the non-strict majorization relation $\psi\prec \varphi$.

\section{Partial entanglement recovery for a general transformation}

In this section we deal with partial entanglement recovery for a
class  of more general transformations. Before proceeding to the
main results, it will be helpful to introduce some notations. Let
$x$ and $y$ be two finite dimensional vectors. We write
$x\sqsubset y$ or $y\sqsupset x$ if
$x^{\downarrow}_1<y^{\downarrow}_1$ and $
x^{\uparrow}_1>y^{\uparrow}_1$. Roughly speaking, $x\sqsubset y$
means that the values of the extreme components of  $x$ are
strictly bounded by those of $y$. We use the formal expression
$\frac{x'}{x''}\sqsupset \frac{y'}{y''}$ as a convenient rewriting
of $x'\otimes y''\sqsupset x''\otimes y'$.

For simplicity, in this section we only deal with vectors that are
already in non-increasing order. That is, for a finite dimensional
vector $x$, we assume that $x=x^{\downarrow}.$

We now introduce the following concept.

\begin{definition}\label{decomposition}\upshape
A decomposition of  a vector $x$ is  a sequence of vectors
$x^1,\ldots, x^m$ satisfying

(i) each of these vectors has  dimension at least one, i.e., ${\rm
dim}(x^i)\geq 1$; and

(ii)  $x$ is the direct sum of these vectors, i.e.,
$x=(x^1,\ldots, x^m)$, or simply, $x=\oplus_{i=1}^m x^m$.
\end{definition}

The following simple lemma  provides a special  decomposition of
two vectors $x$ and $y$ such that $x\prec y$.
\begin{lemma}\label{majdecom}\upshape
If $x$ and $y$ are vectors satisfying $x\prec y$, then $x$ and $y$
can be uniquely decomposed as $x=(x^1,\ldots, x^m)$ and
$y=(y^1,\ldots, y^m)$ such that

(i) $x^i\lhd  y^i$ or $x^i=y^i$ for each $i=1,\ldots, m$; and

(ii) there does not exist an index $i$ such that $x^i=y^i$ and
$x^{i+1}=y^{i+1}$ hold simultaneously.

\end{lemma}
\textit{Proof.} The proof is simple, and the details are omitted. \hfill $\square$\\

The decompositions of $x$ and $y$ in Lemma \ref{majdecom} are
called  the {\it normal decompositions} of $x$ and $y$.

Motivated by Lemma \ref{majdecom}, we shall define two index sets
$I_{x,y}$ and $D_{x,y}$ for any vectors $x$ and $y$ satisfying
$x\prec y$. Suppose that $x$ and $y$ have normal decompositions as
in Lemma \ref{majdecom}. Then we define
$$I_{x,y}=\{i:x^i=y^i {\rm\ and\ } 1\leq i\leq m\}$$
and
$$D_{x,y}=\{i:x^i\lhd y^i {\rm\ and\ } 1\leq i\leq m\}.$$
It is obvious that
$$I_{x,y}\cap D_{x,y}=\emptyset {\rm\ and\ } I_{x,y}\cup D_{x,y}=\{1,\ldots, m\}.$$

One can  easily  check  that $x\lhd y$ is equivalent to
$I_{x,y}=\emptyset$ and $D_{x,y}=\{1\}$.

In what follows,  we only consider the auxiliary state with
positive Schmidt coefficients, as our major purpose here is to
find the states that can do partial entanglement recovery for a
given transformation. For simplicity, the maximally entangled
state is also not considered.

With these preliminaries, we present one of the main results in
this section, which gives a sufficient condition under which
$|\chi\rangle$ can do partial entanglement recovery for a
transformation of $|\psi\rangle$ to $|\varphi\rangle$.
\begin{theorem}\label{generaltheorem1}\upshape

Let $|\psi\rangle$ and $|\varphi\rangle$ be two states with normal
decompositions $\psi=(\psi^1,\ldots,\psi^m)$ and
$\varphi=(\varphi^1,\ldots, \varphi^m)$ such that $\psi$ is
majorized by $\varphi$ , and let $|\chi\rangle$ be an auxiliary
state  with a similar decomposition to $|\psi\rangle$ and
$|\varphi\rangle$, say, $\chi=(\chi^1,\ldots, \chi^m).$ If
\begin{equation}\label{embdedcond}
\frac{\chi^i}{\chi^j}\sqsupset\frac{\varphi^i}{\varphi^j}, {\rm\
for\ all\ }i\in I_{\psi,\varphi}{\rm\ and\ }j\in D_{\psi,\varphi}
 \end{equation}
and
\begin{equation}\label{moreuniformcon}
{\rm min}\{l_u(|\chi^i\rangle):1\leq i\leq m\}>{\rm
max}\{g_u(|\varphi^i\rangle):i\in D_{\psi,\varphi}\},
\end{equation}
then $|\chi\rangle$ can do partial entanglement recovery for the
transformation of $|\psi\rangle$ to $|\varphi\rangle$.

Moreover, if $|\chi\rangle$ satisfies Eqs. (\ref{embdedcond}) and
(\ref{moreuniformcon}), then there exists a positive number
$\delta$ such that for any state $|\omega\rangle$ with a
decomposition $\omega=(\omega^1,\ldots, \omega^m)$ satisfying

(i) {\rm\ }$\omega^i=\chi^i$ for any $i\in I_{\psi,\varphi}$; and

(ii){\rm\ }$\|\omega^i-\chi^i\|<\delta$ and
$\sum\omega^i=\sum\chi^i$ for any $i\in D_{\psi,\varphi},$ the
transformation of $|\psi\rangle\otimes |\chi\rangle$ to
$|\varphi\rangle\otimes |\omega\rangle$ can be realized with
certainty by LOCC, i.e., $|\psi\rangle\otimes
|\chi\rangle\rightarrow |\varphi\rangle\otimes |\omega\rangle.$
\end{theorem}

\textit{Proof.} See Appendix C.\hfill$\square$\\

Here we give some remarks:
\begin{enumerate}

\item  In the above theorem whether $|\chi\rangle$ can save
entanglement lost for the transformation of $|\psi\rangle$ to
$|\varphi\rangle$ does not directly depend on the choice of the
source state $|\psi\rangle$. It only depends on the decomposition
of the target state $|\varphi\rangle$ and the index sets
$I_{\psi,\varphi}$ and $D_{\psi,\varphi}$. For this reason, in
what follows, it is not necessary to specify the source state
$|\psi\rangle$ clearly. We only need to give a decomposition of
$\varphi$ and two index sets $I$ and $D$. Based on these
conditions, we can identify a class of auxiliary states
$|\chi\rangle$ that can do partial entanglement recovery for any
transformation of $|\psi\rangle$ to $|\varphi\rangle$ with source
state $|\psi\rangle$ satisfying $I_{\psi,\varphi}=I$ and
$D_{\psi,\varphi}=D$.

\item When $|\chi\rangle$ can be used to do partial entanglement
recovery for the transformation of $|\psi\rangle$ to
$|\varphi\rangle$, the more entangled state $|\omega\rangle$
generated from $|\chi\rangle$  is also explicitly given by the
above theorem. It should be noted that in general the resulting
state $|\omega\rangle$ is determined by the states $|\chi\rangle$,
$|\psi\rangle$ and $|\varphi\rangle$ together, although the choice
of $|\chi\rangle$ doesn't depend on the source state
$|\psi\rangle$. In other words, sometimes there may not exist a
universal state $|\omega\rangle$ in the sense that
$|\psi'\rangle\otimes |\chi\rangle\rightarrow
|\varphi\rangle\otimes |\omega\rangle$, $|\omega\rangle\rightarrow
|\chi\rangle$ and $|\chi\rangle\nrightarrow |\omega\rangle$ hold
for all states $|\psi'\rangle$ with
$D_{\psi',\varphi}=D_{\psi,\varphi}$ and $I_{\psi',
\varphi}=I_{\psi,\varphi}$.

\end{enumerate}

We now examine some special cases of Theorem
\ref{generaltheorem1}. The first special case is that both $D$ and
$I$ are singletons.

\begin{corollary}\label{boundary}\upshape
Let  $|\varphi\rangle$ and $|\chi\rangle$ be two states with
decompositions $\varphi=(\varphi^1,\varphi^2)$ and
$\chi=(\chi^1,\chi^2)$, and let $I=\{1\}$ and $D=\{2\}$. If
$|\varphi\rangle$ and $|\chi\rangle$ satisfy
\begin{equation}\label{embedcond3}
 \frac{\chi^1}{\chi^2}\sqsupset \frac{\varphi^1}{\varphi^2}
\end{equation}
and
\begin{equation}\label{condlgu3}
{\rm
min}\{l_u(|\chi^1\rangle),l_u(|\chi^2\rangle)\}>g_u(|\varphi^2\rangle),
\end{equation}
then $|\chi\rangle$ can do partial entanglement recovery for any
transformation of $|\psi\rangle$ to $|\varphi\rangle$ with source
state $|\psi\rangle$ such that  $I_{\psi,\varphi}=I$ and
$D_{\psi,\varphi}=D$.

A corresponding result for the dual of  case $I=\{2\}$ and
$D=\{1\}$ can be obtained by exchanging $\chi^1$ with $\chi^2$ and
$\varphi^1$ with $\varphi^2$ in Eq. (\ref{embedcond3}) and Eq.
(\ref{condlgu3}), respectively.
\end{corollary}

For the sake of convenience,  for two $n$-dimensional vectors $x$
and $y$ with $x\prec y$, we define $\Delta_{x,y}$ as the set of
all indices $m$ such that the inequality in Eq.
(\ref{nielsentheorem}) holds with an equality, i.e.,
$$
\Delta_{x,y}=\{m: e_m(x)=e_m(y) {\rm\ and\ } 1\leq m \leq n-1\}.
$$
Note that $1\in \Delta_{x,y}$ is equivalent to $x_1=y_1$ and
$n-1\in \Delta_{\psi,\varphi}$ is equivalent to $x_n=y_n$.

Now we present  two examples to illustrate the use of Corollary
\ref{boundary}.
\begin{example}\label{example4}\upshape
Let $|\psi\rangle$ and $|\varphi\rangle$ be two $n\times n$ states
such that $|\psi\rangle$ is in $S(|\varphi\rangle)$ $(n>2)$.
Assume that $\Delta_{\psi,\varphi}=\{1\}$. We hope to find an
auxiliary state $|\chi\rangle$ with the minimal dimension to do
partial entanglement recovery for the transformation of
$|\psi\rangle$ to $|\varphi\rangle$.

To be more specific, let $\varphi=(\beta_1,\ldots,\beta_n)$. Since
$\Delta_{\psi,\varphi}=\{1\}$, it is obvious that $\varphi$ has a
normal decomposition $\varphi=(\varphi^1,\varphi^2)$, where
$\varphi^1=(\beta_1)$ and $\varphi^2=(\beta_2,\ldots, \beta_n)$.
Moreover, $I_{\psi,\varphi}=\{1\}$ and $D_{\psi,\varphi}=\{2\}$.
Take an auxiliary state $|\chi\rangle$  with
$\chi=(\chi^1,\chi^2)$, where $\chi^1=(\gamma_1)$,
$\chi^2=(\gamma_2,\gamma_3)$,  and $\gamma_1>\gamma_2>\gamma_3>0$.

By Corollary \ref{boundary}, if $|\chi\rangle$ satisfies Eqs.
(\ref{embedcond3})--(\ref{condlgu3}), then $|\chi\rangle$ can do
partial entanglement recovery for the transformation of
$|\psi\rangle$ to $|\varphi\rangle$.  So we have
\begin{equation}\label{1condRu}
{\gamma_1}/{\gamma_2}>{\beta_1}/{\beta_2},
\end{equation}
\begin{equation}\label{1condru}
{\gamma_1}/{\gamma_3}<{\beta_1}/{\beta_n},
\end{equation}
and
\begin{equation}\label{1condlgu}
{\gamma_3}/{\gamma_2}>{\beta_n}/{\beta_2}.
\end{equation}

By Eqs.  (\ref{1condRu}) and (\ref{1condlgu}), we can take
positive numbers $\lambda$ and $\mu$ such that
\begin{equation}\label{gamma1}
\gamma_1=\gamma_2(1+\lambda)\beta_1/\beta_2
\end{equation}
and
\begin{equation}\label{gamma3}
\gamma_3=\gamma_2(1+\mu)\beta_n/\beta_2.
\end{equation}

Substituting Eqs.  (\ref{gamma1}) and (\ref{gamma3}) into Eq.
(\ref{1condru}) yields $0<\lambda<\mu$. Moreover, the constraint
$\gamma_2>\gamma_3$ and Eq. (\ref{gamma3}) yield
$\mu<(\beta_2-\beta_n)/\beta_n$. $\gamma_2$ is used to make the
following  normalization condition satisfied:
\begin{equation}\label{normalization}
\sum_{i=1}^3\gamma_i=1.
\end{equation}
Notice that $\beta_2>\beta_n$. One can easily check that such
$(\gamma_1,\gamma_2,\gamma_3)$ satisfying Eqs.
(\ref{gamma1})--(\ref{normalization})  is a solution of the system
of inequalities defined by Eqs. (\ref{1condRu})--(\ref{1condlgu}).
The parameters $\lambda$ and $\mu$ satisfy
$0<\lambda<\mu<(\beta_2-\beta_n)/\beta_n$.

Thus by Corollary \ref{boundary}, the $3\times 3$ auxiliary state
$|\chi\rangle$ can do partial entanglement recovery for the
transformation of $|\psi\rangle$ to $|\varphi\rangle$. Moreover,
the state $|\omega\rangle$ such that both $|\psi\rangle\otimes
|\chi\rangle\rightarrow |\varphi\rangle\otimes |\omega\rangle$ and
$|\omega\rangle\rightarrow|\chi\rangle$ hold can be chosen as
$\omega=(\gamma_1,\gamma_2-\epsilon, \gamma_3+\epsilon)$ with a
sufficiently small  positive number $\epsilon$.\hfill
$\square$\\
\end{example}

We point out that the existence of such an auxiliary state
$|\chi\rangle$ with $\chi=(\gamma_1,\gamma_2,\gamma_3)$ has been
proven in Theorem 3 in \cite{SVF01}, where $|\chi\rangle$ is of
the  form $\chi(p,q)=(p,q,1-p-q)$, $p\geq q\geq 1-p-q\geq 0$.
However, an important constraint on the $p$ and $q$ or the region
$R$, i.e., $(1-p-q)\beta_2>q\beta_n$,  is missing in \cite{SVF01}.
Thus an additional case which is not included in Case (i) and Case
(ii) in \cite{SVF01} is possible, which makes that proof invalid.

A dual case of Example \ref{example4} is as follows:
\begin{example}\label{example5}\upshape
Let $|\psi\rangle$ and $|\varphi\rangle$ be two $n\times n$ states
such that $|\psi\rangle$ is in $S(|\varphi\rangle)$ $(n>2)$.
Assume that $\Delta_{\psi,\varphi}=\{n-1\}$. We hope to find an
auxiliary state $|\chi\rangle$ with the minimal dimension to do
partial entanglement recovery for the transformation of
$|\psi\rangle$ to $|\varphi\rangle$.

To be more specific, let $\varphi=(\beta_1,\ldots,\beta_n)$. Since
$\Delta_{\psi,\varphi}=\{n-1\}$,  it is easy to check that
$\varphi$ has a normal decomposition
$\varphi=(\varphi^1,\varphi^2)$, where
${\varphi^1}=(\beta_1,\ldots,\beta_{n-1})$ and
${\varphi^2}=(\beta_n)$. Similarly, let an auxiliary sate
$|\chi\rangle$ have a decomposition $\chi=(\chi^1,\chi^2)$, where
${\chi^1}=(\gamma_1,\gamma_2)$, $\chi^2=(\gamma_3)$, and
$\gamma_1>\gamma_2>\gamma_3>0$. By Eqs. (\ref{embedcond3}) and
(\ref{condlgu3}) again, noticing that $I_{\psi,\varphi}=\{2\}$ and
$D_{\psi,\varphi}=\{1\}$, we have the following system of
inequalities:
\begin{equation}\label{n-1condRu}
{\gamma_3}/{\gamma_1}>{\beta_n}/{\beta_1},
\end{equation}
\begin{equation}\label{n-1condru}
{\gamma_3}/{\gamma_2}<{\beta_n}/{\beta_{n-1}},
\end{equation}
and
\begin{equation}\label{n-1condlgu}
{\gamma_2}/{\gamma_1}>{\beta_{n-1}}/{\beta_1}.
\end{equation}
By using a similar argument as in Example \ref{example4}, we can
take
\begin{equation}\label{n-1gamma2}
 \gamma_2=\gamma_1(1+\mu)\beta_{n-1}/\beta_1
\end{equation}
and
\begin{equation}\label{n-1gamma3}
\gamma_3=\gamma_1(1+\lambda)\beta_n/\beta_1,
\end{equation}
where $0<\lambda<\mu<(\beta_1-\beta_{n-1})/\beta_{n-1}$.
($\lambda<\mu$ is deduced by substituting Eqs.  (\ref{n-1gamma2})
and (\ref{n-1gamma3}) into Eq. (\ref{n-1condru}),
$\mu<(\beta_1-\beta_{n-1})/\beta_{n-1}$ comes from Eq.
(\ref{n-1gamma2}) and $\gamma_1>\gamma_2$). $\gamma_1$ is taken to
validate the following normalization condition
\begin{equation}
\sum_{i=1}^3\gamma_i=1.
\end{equation}
Since $\beta_1>\beta_{n-1}$, one can easily check that such a
state $|\chi\rangle$ with $\chi=(\gamma_1,\gamma_2,\gamma_3)$ is a
solution of the inequalities system defined by Eqs.
(\ref{n-1condRu})-(\ref{n-1condlgu}). Thus by Corollary
\ref{boundary}  $|\chi\rangle$ can do partial entanglement
recovery for the transformation of $|\psi\rangle$ to
$|\varphi\rangle$. Again, the desired state $|\omega\rangle$  such
that both $|\psi\rangle\otimes |\chi\rangle\rightarrow
|\varphi\rangle\otimes |\omega\rangle$ and
$|\omega\rangle\rightarrow|\chi\rangle$ hold can be chosen as
$\omega=(\gamma_1-\epsilon, \gamma_2+\epsilon, \gamma_3)$ with a
suitably small positive number  $\epsilon$.\hfill
$\square$\\
\end{example}

If one of the cases $\Delta_{\psi,\varphi}=\{1\}$ or
$\Delta_{\psi,\varphi}=\{n-1\}$ occurs, we can always use a
$3\times 3$ state $|\chi\rangle$ to partially recover entanglement
lost in the transformation of $|\psi\rangle$ to $|\varphi\rangle$.
The explicit construction of such $|\chi\rangle$ has also been
presented in the above examples.

The following corollary is another important special case of
Theorem \ref{generaltheorem1}:
\begin{corollary}\label{boundary2}\upshape
Let $|\varphi\rangle$ and $|\chi\rangle$ be two states with
$\varphi=(\varphi^1,\varphi^2,\varphi^3)$ and
$\chi=(\chi^1,\chi^2,\chi^3)$. $I=\{1,3\}$ and $D=\{2\}$. If
$|\chi\rangle$ and $|\varphi\rangle$ satisfy
\begin{equation}\label{embedcond121}
\frac{\chi^1}{\chi^2}\sqsupset \frac{\varphi^1}{\varphi^2},
\end{equation}
\begin{equation}\label{embedcond122}
\frac{\chi^3}{\chi^2}\sqsupset \frac{\varphi^3}{\varphi^2},
\end{equation}
and
\begin{equation}\label{condlgu12}
{\rm
min}\{l_u(|\chi^1\rangle),l_u(|\chi^2\rangle),l_u(|\chi^3\rangle)\}>g_u(|\varphi^2\rangle),
\end{equation}
then $|\chi\rangle$ can do partial entanglement recovery for any
transformation of $|\psi\rangle$ to $|\varphi\rangle$ with
$I_{\psi,\varphi}=I$ and $D_{\psi,\varphi}=D$.
\end{corollary}

A very interesting application of the above corollary is the
following:
\begin{example}\label{example6}\upshape
Let $|\psi\rangle$ and $|\varphi\rangle$ be two $n\times n$ states
such that  $|\psi\rangle$ is in $S(|\varphi\rangle)$ ($n>3$).
Assume $\Delta_{\psi,\varphi}=\{1, n-1\}$. Our purpose here is to
find an auxiliary state to do partial entanglement recovery for
the transformation of $|\psi\rangle$ to $|\varphi\rangle$.

Take a $4\times 4$ auxiliary  state $|\chi\rangle$ with
$\chi=(\gamma_1,\gamma_2,\gamma_3, \gamma_4)$. Let us decompose
$\varphi$ and $\chi$, respectively, into
$\varphi=(\varphi^1,\varphi^2,\varphi^3)$ and
$\chi=(\chi^1,\chi^2,\chi^3)$, where $\varphi^1=(\beta_1)$,
$\varphi^2=(\beta_2,\ldots, \beta_{n-1})$, $\varphi^3=(\beta_n)$,
$\chi^1=(\gamma_1)$, $\chi^2=(\gamma_2,\gamma_3)$,
$\chi^3=(\gamma_4)$, and $\gamma_1>\gamma_2>\gamma_3>\gamma_4>0$.
Since $\Delta_{\psi,\varphi}=\{1,n-1\}$, it is easy to check that
$I_{\psi,\varphi}=\{1,3\}$ and $D_{\psi,\varphi}=\{2\}$. Thus  by
Corollary \ref{boundary2}, $|\chi\rangle$ can do partial
entanglement recovery for the transformation of $|\psi\rangle$ to
$|\varphi\rangle$ if Eqs.  (\ref{embedcond121})--(\ref{condlgu12})
hold. A routine calculation leads to the following solution of
Eqs.  (\ref{embedcond121})--(\ref{condlgu12}):
$$\gamma_1=\gamma_2(1+\lambda)\beta_1/\beta_2,$$
$$\gamma_3=\gamma_2(1+\eta)(1+\mu)\beta_{n-1}/\beta_2,$$
$$\gamma_4=\gamma_2(1+\mu)\beta_n/\beta_2,$$ where
$1+\lambda<(1+\eta)(1+\mu)<\beta_2/\beta_{n-1}$, $\lambda,$ $\mu$,
and $\eta$ are all positive real numbers, and $\gamma_2$ is used
to validate the  normalization condition
$$\sum_{i=1}^4\gamma_i=1.$$  So such an auxiliary state
$|\chi\rangle$ for partial entanglement recovery  always exists.

Ultimately,  to partially recover entanglement lost in the
transformation of $|\psi\rangle$ to $|\varphi\rangle$, it is
sufficient to use an auxiliary state $|\chi\rangle$ with dimension
$4\times 4$. Again, the more entangled state $|\omega\rangle$
generated from $|\chi\rangle$ after the recovery process can be
chosen as $\omega=(\gamma_1,\gamma_2-\epsilon, \gamma_3+\epsilon,
\gamma_4)$, where $\epsilon$ is a sufficiently small positive
number.\hfill
$\square$\\
\end{example}

In \cite{SVF01}, it is  proven that any $3\times 3$ state cannot
be used to partially recover entanglement lost in the
transformation of $|\psi\rangle$ to $|\varphi\rangle$ with
$\psi\prec\varphi$ and $\Delta_{\psi,\varphi}=\{1,n-1\}$. By the
above example, we are able to show that $4\times 4$ auxiliary
states are necessary and sufficient to do partial entanglement
recovery for this special case.

In practice,  we hope that the dimension of the auxiliary state
$|\chi\rangle$ is as small as possible. In Theorem
\ref{generaltheorem1}, if there are two successive integers $i$
and $i+1$ both contained in $D_{\psi,\varphi}$ (in
$I_{\psi,\varphi}$ this case cannot happen), we in fact can
combine $\chi^i$ with $\chi^{i+1}$  to reduce the dimension of
$\chi$.  So a careful investigation of the structure of
$D_{\psi,\varphi}$ is necessary.

Let us  see a simple example. Suppose that for states
$|\psi\rangle$ and $|\varphi\rangle$,
$I_{\psi,\varphi}=\{1,4,7,12\}$ and
$D_{\psi,\varphi}=\{2,3,5,6,8,9,10,11\}$. By the construction in
Theorem \ref{generaltheorem1}, we should use an auxiliary state
$|\chi\rangle$ with $\chi=(\chi^1,\ldots, \chi^{12})$, where each
$\chi^i$ ($i\in D_{\psi,\varphi}$) has  dimension at least $2$.
Thus the vector $\chi$ has dimension at least
$|I_{\psi,\varphi}|+2|D_{\psi,\varphi}|=20$. If we combine the
successive integers in $D_{\psi,\varphi}$ together, we have
$D'_{\psi,\varphi}=\{\{2,3\}, \{5,6\}, \{8,9,10,11\}\}$, and the
dimension of $\chi$ is reduced to
$|I_{\psi,\varphi}|+2|D'_{\psi,\varphi}|=10$.

More formally, suppose that $|\psi\rangle$ and $|\varphi\rangle$
are two states with normal decompositions
$\psi=(\psi^1,\ldots,\psi^m)$ and
$\varphi=(\varphi^1,\ldots,\varphi^m)$. Let
$$I_{\psi,\varphi}=\{k_1,\ldots, k_p\},{\rm \ }1\leq p\leq
 m,$$ where
$$0=k_0<k_1<\cdots<k_p<k_{p+1}=m+1.$$  We define $$D'_{\psi,\varphi}=\{D_i:D_i\neq \emptyset {\rm\ and\ } 0\leq
i\leq p\},$$ where $$D_i=\{s:k_i+1\leq s\leq k_{i+1}-1\}.$$ The
constraint $D_i\neq \emptyset$ in the definition of
$D'_{\psi,\varphi}$ needs a careful explanation.  For any $1\leq
i\leq p$, we have that $k_i\in I_{\psi,\varphi}$ implies
$k_i+1\notin I_{\psi,\varphi}$. So $D_i\neq \emptyset$ in this
case. However, if $k_1=1$ or $k_p=m$ then we have $D_0=\emptyset$
or $D_p=\emptyset$, respectively. To avoid these two trivial
cases, the constraint $D_i\neq \emptyset$ is necessary. In
particular, if $I_{\psi,\varphi}=\emptyset$ then
$D'_{\psi,\varphi}=\{D_{\psi,\varphi}\}=\{\{1,\ldots, m\}\}$. For
the sake of convenience, we also define
$$I'_{\psi,\varphi}=\{\{i\}:i\in I_{\psi,\varphi}\}.$$ In the following discussions,
we shall use the elements of $I'_{\psi,\varphi}$ and
$D'_{\psi,\varphi}$ as indices. We  define the natural order of
the elements in $I'_{\psi,\varphi}\cup D'_{\psi,\varphi}$ as
$$D_0<\{k_1\}<D_1<\{k_2\}<\cdots<\{k_p\}<D_{p},$$
where we assume that any term which doesn't exist should be
omitted automatically without affecting the orders of other terms.

Suppose that $J$ is a finite set of integers. We use the notations
${\rm max\ }J$ and ${\rm min\ }J$ to denote the maximal and the
minimal elements of $J$, respectively. For any real function
$f(.)$ defined on $J$, the expression ${\rm\ arg\ min}_{k\in
J}f(k)$ denotes the index $i\in J$ such that $f(i)={\rm min}_{k\in
J}f(k)$ (here we assume that there is a unique $i$ of $J$ that can
attain the minimum).

Now we can present another condition for the existence of partial
entanglement recovery, which complements Theorem
\ref{generaltheorem1}.

\begin{theorem}\label{generaltheorem2}\upshape
Let $|\psi\rangle$ and $|\varphi\rangle$ be two states with normal
decompositions $\psi=(\psi^1,\ldots,\psi^m)$ and
$\varphi=(\varphi^1,\ldots, \varphi^m)$ such that $\psi$ is
majorized by  $\varphi$, and let  $|\chi\rangle$ be an auxiliary
state  with a decomposition
$\chi=(\chi^{D_0},\chi^{\{k_1\}},\chi^{D_1},\ldots,
\chi^{\{k_p\}},\chi^{D_{p}})=\oplus_{i\in I'_{\psi,\varphi}\cup
D'_{\psi,\varphi}}\chi^i.$ If
\begin{equation}\label{embdedcond1}
\frac{\chi^{\{i\}}}{\chi^J}\sqsupset\frac{\varphi^i}{\varphi^{J_i}}
\end{equation}
${\rm for\ all\ }i\in I_{\psi,\varphi}$,  $J\in
 D'_{\psi,\varphi}$ and $J_i={\rm \ arg\ min}_{k\in J} |i-k|,$
\begin{equation}\label{moreuniformcond1}
{\rm min}\{l_u(|\chi^J\rangle):J\in D'_{\psi,\varphi}\}>{\rm
max}\{g_u(|\varphi^i\rangle):i\in D_{\psi,\varphi}\},
\end{equation}
and
\begin{equation}\label{moreuniformcond2}
\begin{array}{rl}
{\rm min}\{l_u(|\chi^{\{i\}}\rangle):i\in I_{\psi,\varphi}\}>&{\rm
max}\displaystyle\bigcup_{J\in
D'_{\psi,\varphi}}\{g_u(|\varphi^{{\rm max\ } J}\rangle),\\
&g_u(|\varphi^{{\rm min\ } J}\rangle)\},
\end{array}
\end{equation}\\
then $|\chi\rangle$ can do partial entanglement recovery for the
transformation of $|\psi\rangle$ to $|\varphi\rangle$.

Moreover, if $|\chi\rangle$ satisfies Eqs.
(\ref{embdedcond1})-(\ref{moreuniformcond2}), then  there exists a
positive number $\delta$ such that for any state $|\omega\rangle$
with a decomposition $\omega=\oplus_{i\in I'_{\psi,\varphi}\cup
D'_{\psi,\varphi}}\omega^i$ satisfying

(i) {\rm\ }$\omega^i=\chi^i$ for any $i\in I'_{\psi,\varphi}$; and

(ii){\rm\ }$\|\omega^i-\chi^i\|<\delta$ and
$\sum\omega^i=\sum\chi^i$ for any $i\in D'_{\psi,\varphi},$ the
transformation of $|\psi\rangle\otimes |\chi\rangle$ to
$|\varphi\rangle\otimes |\omega\rangle$ can be realized with
certainty by LOCC, i.e., $ |\psi\rangle\otimes
|\chi\rangle\rightarrow |\varphi\rangle\otimes |\omega\rangle. $
\end{theorem}
\textit{Proof.} The proof is similar to Theorem \ref{generaltheorem1}, and we omit the details. \hfill $\square$\\

The key idea in the above theorem is to let all vectors between
$\varphi^{k_i+1}$ and $\varphi^{k_{i+1}-1}$ correspond to a single
$\chi^{D_i}$.  This reduces the dimension of $\chi$ efficiently.

An interesting special case of Theorem \ref{generaltheorem2} is
when the majorization $\psi\prec \varphi$ splits into $m$ strict
majorizations: $\psi^i\lhd \varphi^i$. We state this result in the
following:

\begin{corollary}\label{22}\upshape
Let $|\psi\rangle$ and $|\varphi\rangle$ be two states such that
$|\psi\rangle$ is in $S(|\varphi\rangle)$. Suppose that $\psi$ and
$\varphi$ have normal decompositions $\psi=(\psi^1,\ldots,\psi^m)$
and $\varphi=(\varphi^1,\ldots, \varphi^m)$, and let
$I_{\psi,\varphi}=\emptyset$.  If $|\chi\rangle$ is an auxiliary
state such that
$$l_u(|\chi\rangle)>{\rm max}\{g_u(|\varphi^i\rangle): 1\leq i\leq m\},$$ then $|\chi\rangle$ can do partial entanglement recovery
for the transformation of $|\psi\rangle$ to $|\varphi\rangle$.
\end{corollary}

\textit{Proof.} In fact, in this special case,
$D'_{\psi,\varphi}=\{D_{\psi,\varphi}\}=\{\{1,\ldots, m\}\}$,
$I'_{\psi,\varphi}=\emptyset$. Thus, by Theorem
\ref{generaltheorem2}, to do partial entanglement recovery, the
only non-trivial condition that $|\chi\rangle$ should satisfy is
Eq. (\ref{moreuniformcond1}), which is exactly the  assumption
of the present corollary. \hfill $\square$\\

It is easy to check that in the above corollary $|\chi\rangle$ can
be chosen as a $2\times 2$ state. However, by Theorem
\ref{generaltheorem1}, we can only find a state $|\chi\rangle$ of
dimension at least  $2m\times 2m$.

By summarizing  Theorems \ref{shannoncode}, \ref{generaltheorem1},
and  \ref{generaltheorem2}, we have the following:
\begin{theorem}\upshape
Suppose that $|\psi\rangle$ and $|\varphi\rangle$ are two $n\times
n$ states such that $\psi\prec \varphi$. We can always find an
auxiliary state $|\chi\rangle$ to do partial entanglement recovery
for the transformation of $|\psi\rangle$ to $|\varphi\rangle$,
where the dimension of $\chi$ is between $2\times 2$ and $n\times
n$. Moreover, such a state $|\chi\rangle$ can only depend on the
target state $|\varphi\rangle$ and the presence of equalities in
the majorization $\psi\prec \varphi$.
\end{theorem}

{\it Proof.} The proof follows immediately  from Theorems
\ref{shannoncode}, \ref{generaltheorem1}, and
\ref{generaltheorem2}. \hfill
$\square$\\

The upper bound $n\times n$ cannot always be reduced to
$(n-1)\times (n-1)$. We have seen in Example \ref{example6} that
when $n=4$, an auxiliary state $|\chi\rangle$ of dimension
$4\times 4$ is needed to do partial entanglement recovery for the
transformation of $|\psi\rangle$ to $|\varphi\rangle$ such that
$\Delta_{\psi,\varphi}=\{1,3\}$.

We conclude this section by  giving an  example to illustrate the
use of Theorem \ref{generaltheorem2}. This example is taken from
\cite{SVF01b}.

\begin{example}\label{example7}\upshape
Let $|\psi\rangle$ and $|\varphi\rangle$ be two $n\times n$ states
such that $|\psi\rangle$ is in $S(|\varphi\rangle)$ $(n>6)$.
Assume that $\Delta_{\psi,\varphi}=\{2,3,5\}$. The goal here is to
find an auxiliary state $|\chi\rangle$ to do partial entanglement
recovery for the transformation of $|\psi\rangle$ to
$|\varphi\rangle$.

To be specific, let $\varphi=(\beta_1,\ldots, \beta_n)$. It is
easy to check that $\varphi$ has a normal decomposition
$\varphi=(\varphi^1,\varphi^2,\varphi^3,\varphi^4)$, where
$\varphi^1=(\beta_1,\beta_2)$, $\varphi^2=(\beta_3)$,
$\varphi^3=(\beta_4,\beta_5)$, and $\varphi^4=(\beta_6,\ldots,
\beta_n)$. Also, $I_{\psi,\varphi}=\{2\}$ and
$D_{\psi,\varphi}=\{1,3,4\}$. So $I'_{\psi,\varphi}=\{\{2\}\}$ and
$D'_{\psi,\varphi}=\{\{1\},\{3,4\}\}$.

Take  a $5\times 5$ auxiliary state $|\chi\rangle$ with
$\chi=(\chi^{\{1\}},\chi^{\{2\}},\chi^{\{3,4\}})$, where
$\chi^{\{1\}}=(\gamma_1,\gamma_2)$, $\chi^{\{2\}}=(\gamma_3)$,
$\chi^{\{3,4\}}=(\gamma_4,\gamma_5)$, and
$\gamma_1>\cdots>\gamma_5>0$. By Theorem \ref{generaltheorem2},
Eq. (\ref{embdedcond1}) yields
 \begin{equation}\label{appcond1}
\frac{\chi^{\{2\}}}{\chi^{\{1\}}}\sqsupset
\frac{\varphi^2}{\varphi^1}
\end{equation}
and
\begin{equation}\label{appcond2}
\frac{\chi^{\{2\}}}{\chi^{\{3,4\}}}\sqsupset
\frac{\varphi^2}{\varphi^3}.
\end{equation}
Eq. (\ref{moreuniformcond1}) yields
\begin{equation}\label{appcond3}
\begin{array}{rl}
{\rm min}\{l_u(|\chi^{\{1\}}\rangle),l_u(|\chi^{\{3,4\}}\rangle)\}
>&{\rm max}\{g_u(|\varphi^1\rangle),\\
&g_u(|\varphi^3\rangle),g_u(|\varphi^4\rangle)\}.
\end{array}
\end{equation}
Eq. (\ref{moreuniformcond2}) yields
 \begin{equation}\label{appcond4}
l_u(|\chi^{\{2\}}\rangle)>{\rm
max}\{g_u(|\varphi^1\rangle),g_u(|\varphi^3\rangle),g_u(|\varphi^4\rangle)\},
\end{equation}
which is automatically satisfied since
$l_u(|\chi^{\{2\}}\rangle)=1$ while the right hand side of Eq.
(\ref{appcond4}) is strictly less than $1$.

More explicitly, we have

\begin{equation}\label{appcond5}
\frac{\gamma_3}{\gamma_1}>\frac{\beta_3}{\beta_1}{\rm\ and\ }
\frac{\gamma_3}{\gamma_2}<\frac{\beta_3}{\beta_2},
\end{equation}
and
\begin{equation}\label{appcond6}
\frac{\gamma_3}{\gamma_4}>\frac{\beta_3}{\beta_4}{\rm\ and\ }
\frac{\gamma_3}{\gamma_5}<\frac{\beta_3}{\beta_5},
\end{equation}
and
\begin{equation}\label{appcond7}
{\rm
min}\{\frac{\gamma_2}{\gamma_1},\frac{\gamma_5}{\gamma_4})>{\rm
max}\{\frac{\beta_2}{\beta_1},
\frac{\beta_5}{\beta_4},\frac{\beta_n}{\beta_6}\}.
\end{equation}

With a  routine calculation one can check  that
$$\gamma_1=\mu\gamma_3\frac{\beta_1}{\beta_3},$$
$$\gamma_2=(1+\eta)\gamma_3\frac{\beta_2}{\beta_3},$$
$$\gamma_4=h\gamma_3\frac{\beta_4}{\beta_3},$$
$$\gamma_5=(1+\lambda)\gamma_3\frac{\beta_5}{\beta_3}$$ is a
solution of the system of inequalities defined by Eqs.
(\ref{appcond5})--(\ref{appcond7}), where $\gamma_3>0$ is used to
satisfy the normalization condition:
$$\sum_{i=1}^5\gamma_i=1.$$  The parameters $\eta$, $\lambda$, $\mu$, and $h$ satisfy
$$0<\eta<\frac{\beta_1-\beta_2}{\beta_2},{\rm \ \ } 0<\lambda<\frac{\beta_4-\beta_5}{\beta_5}.$$
and
$$(1+\eta)\frac{\beta_2}{\beta_1}<\mu<{\rm
min}\{(1+\eta)\frac{\beta_2\beta_4}{\beta_1\beta_5},(1+\eta)\frac{\beta_2\beta_6}{\beta_1\beta_n},1\}
,$$
$$
(1+\lambda)\frac{\beta_5}{\beta_4}<h<{\rm
min}\{(1+\lambda)\frac{\beta_1\beta_5}{\beta_2\beta_4},(1+\eta)\frac{\beta_5\beta_6}{\beta_4\beta_n},1\}
.$$  Notice that $\beta_1>\beta_2$, $\beta_4>\beta_5$, and
$\beta_6>\beta_n$,  and such  a state $|\chi\rangle$ with
$\chi=(\gamma_1,\ldots, \gamma_5)$ always exists. So we have
actually constructed a class of states $|\chi\rangle$ with
dimension $5\times 5$ that can do partial entanglement recovery
for   the
transformation of $|\psi\rangle$ to $|\varphi\rangle.$ \hfill $\square$\\
\end{example}

\section{A polynomial time algorithm for partial entanglement recovery}
In this section we study partial entanglement recovery from the
algorithmic viewpoint. We present a polynomial algorithm  of time
complexity $O(n^2k^4)$ to decide whether $|\chi\rangle$ can be
used to recover some entanglement lost in the transformation of
$|\psi\rangle$ to $|\varphi\rangle$, where $n$ and $k$ are the
dimensions of $\varphi$ and $\chi$, respectively.

The key part of Problem 1 is to solve the majorization relation
$\psi\otimes \chi\prec \varphi\otimes \omega.$ As argued before,
the main difficulty here is how to deal with the order of the
tensor product $\varphi\otimes \omega$ when $\omega$ varies. We
will develop some techniques to overcome this difficulty. Notice
that the map from $\omega$ to $\varphi\otimes \omega$ is an affine
one. To make our discussions more general and more readable, in
what follows we consider affine maps instead of tensor products.

To be concise, some concepts are introduced first.
\begin{definition}\label{funcom}
Let $f$ and $g$ be real functions defined on $\mathcal{R}^m$, and
let $\mathcal{S}\subset \mathcal{R}^m$. $f$ and $g$ are said to be
comparable on $\mathcal{S}$ if

(i)  $\forall x\in \mathcal{S}$, $f(x)\geq g(x)$; or

(ii) $\forall x\in \mathcal{S}$, $f(x)\leq g(x)$.

\end{definition}

Let $F$ be a map from $\mathcal{R}^m$ to $\mathcal{R}^n$. We write
$F(x)=(f_1(x),\cdots, f_n(x)),$ where each $f_i$ is a real
function defined on $\mathcal{R}^m$.

\begin{definition} $F$ is said to have a fixed order on $\mathcal{S}$ if for any $1\leq i<j\leq
n$, $f_i$ and $f_j$ are comparable on $\mathcal{S}$.
\end{definition}

Suppose $F$ has a fixed order on $\mathcal{S}$, and assume that
whether $f_i$ and $f_j$ are comparable on $\mathcal{S}$ can be
determined in $O(1)$ time. Then there exists a common algorithm
which can sort the entries of $F(x)$ into non-increasing order for
any $x\in \mathcal{S}$ in $O(n\log_2 n)$ time. This fact is
extremely useful in the following discussions.

If some entries of $F$ are not comparable on $\mathcal{S}$, then
by the above definition $F$ does not have a fixed order. An
important question naturally arises: how many different orders can
$F$ have on $\mathcal{S}$?

\begin{definition}\label{differentorder}
$F$ is said to have at most $M$ different orders on $\mathcal{S}$
if there exists a decomposition of $\mathcal{S}$, say,
$\mathcal{S}_1,\cdots, \mathcal{S}_M$, such that

(i) $\mathcal{S}=\mathcal{S}_1\cup \cdots \cup\mathcal{S}_M$; and

(ii) $F$ has a fixed order on each $\mathcal{S}_i$, $i=1,\cdots,
M$.
\end{definition}

Now let $F$ be an affine map, $F(x)=Ax+b,$ where $A\in
\mathcal{R}^{n\times m}$ and $b\in \mathcal{R}^n$. At first
glance, $F$ may have $n!$ different orders on $\mathcal{R}^m$.
However, this is not always true. A somewhat surprising fact is
that the number of different orders of $F$ can be dramatically
reduced to $O(n^{2m})$ when $m$ is a constant.

\begin{lemma}\label{orderlemma}
$F$ has at most $O(n^{2m})$ different orders on $\mathcal{R}^m$.
\end{lemma}

{\it Proof.} For any $1\leq i<j\leq n$, the difference of $f_i(x)$
and $f_j(x)$ crosses  zero (from positive to negative, or negative
to positive) if and only if $x$ crosses the hyperplane determined
by the equation $f_i(x)-f_j(x)=0$, or more precisely,
$$\Gamma_{ij}=\{(x_1,\cdots,
x_m):\sum_{s=1}^m(a_{is}-a_{js})x_s+(b_i-b_j)=0\}.$$ It should be
noted that there are two cases where $\Gamma_{ij}$ does not define
a legal hyperplane. The first case is $\Gamma_{ij}=\emptyset$ and
the second one is $\Gamma_{ij}=\mathcal{R}^m$. We will exclude
these cases since in both of them $f_i$ and $f_j$ remain
comparable whatever $x$ varies. Denote
$$\Gamma=\{\Gamma_{ij}: \Gamma_{ij}\neq \emptyset {\rm\  and\ }\Gamma_{ij}\neq \mathcal{R}^m, 1\leq i<j\leq n\}.$$

The number of hyperplanes in $\Gamma$ is less than or equal to
$n(n-1)/2$. These hyperplanes divide $\mathcal{R}^m$ into at most
$O((n(n-1)/2)^m)=O(n^{2m})$ different parts. $F$ has a fixed order
on each part. With that we complete the proof.\hfill $\square$\\

It is obvious that the above lemma holds for any subset of
$\mathcal{R}^m$.

Lemma \ref{orderlemma} indicates that we can decompose
$\mathcal{R}^m$ into $O(n^{2m})$ parts,
$\mathcal{D}_1,\cdots,\mathcal{D}_M$, such that on each part, $F$
has a fixed order. In practice it is important to construct these
parts explicitly. To see how this procedure can be done
efficiently, let us first examine a special case where $m=1$.
\begin{example}\label{arr2}
Let $F(x)=(a_1x-b_1,\cdots, a_nx-b_n)$, where $x\geq 0$. For
simplicity  assume $a_i\neq a_j$, $b_i\neq b_j$, $a_i$, $b_i>0$
for any $1\leq i<j\leq n$.

By Lemma \ref{orderlemma}, $F$ has at most $O(n^2)$ different
orders when $x$ varies as a non-negative number. In what follows
we will show how to determine these orders explicitly.

Step 1. For each $1\leq i<j\leq n$, solve equation
$a_ix-b_i=a_jx-b_j$. The solution is given by
$\theta_{ij}=(b_i-b_j)/(a_i-a_j)$. Let $\Gamma=\{\theta_{ij}:1\leq
i<j\leq n\}\cup \{0\}$. The number of elements of $\Gamma$ is
denoted by $M$. It is easy to see that $M\leq n(n-1)/2+1$.

Step 2. Sort the elements in $\Gamma$ into non-decreasing order,
say $0=c_0<c_1<\cdots<c_{M-1}.$

Step 3. Construct a sequence of intervals:
$\mathcal{D}_1=[c_0,c_1]$, $\cdots$,
$\mathcal{D}_M=[c_{M-1},+\infty)$.

It is clear that $F$ has a fixed order on each interval. The above
procedure is completed in $O(n^2)+O(M\log_2 M)+O(M)=O(n^2\log_2
n)$ time.

It is notable that the leftmost interval $[c_0,c_1]$ can be
located in $O(n^2)$ time. This fact will be useful in the
following discussions.
\hfill $\square$\\
\end{example}

To deal with the general case, we need a lemma in computational
geometry. Let $\mathcal{H}$ be a set of $n$ hyperplanes in
$\mathcal{R}^d$ with $d>1$. Then $\mathcal{H}$ divides
$\mathcal{R}^d$ into $O(n^d)$ parts with pairwise disjoint
interiors. We call the set of these parts a $d$-arrangement of
$\mathcal{H}$. A celebrated result in computational geometry shows
that the $d$-arrangement of $\mathcal{H}$ can be enumerated
efficiently \cite{Edel86}.
\begin{lemma}\label{arrangement}
The $d$-arrangement of $n$ hyperplanes may be computed in time
$O(n^d)$.
\end{lemma}

Employing Lemma \ref{arrangement}, we can easily see that the
above decomposition $\mathcal{D}_1,\cdots,\mathcal{D}_M$ can be
computed in $O(n^{2m})$ time in the case that $m>1$.

With the aid of Lemma \ref{orderlemma}, we are able to solve a
majorization inequality of the form $c\prec Ax+b$ by using linear
programming methods.
\begin{lemma}\label{majaffine}
The majorization inequality  $c\prec F(x)$ can be solved in
$O(n^{2m+1}\log_2n)$ time, where $m$ is treated as a constant.
\end{lemma}

{\it Proof} By Lemma \ref{orderlemma}, $F$ has at most
$M=O(n^{2m})$ different orders on $\mathcal{R}^m$. Let us
decompose $\mathcal{R}^m$ into $M$ parts and enumerate them as
$\mathcal{D}_1,\cdots, \mathcal{D}_M$. On each part
$\mathcal{D}_i$, $F$ has a fixed order. This procedure needs time
$O(n^{2m})$. In what follows we will show on each part, the
majorization inequality $c\prec F(x)$ can be solved in $O(n\log_2
n)$ time by using standard methods of linear programming. Hence we
obtain an algorithm with time complexity
$O(n^{2m})+O(n^{2m}n\log_2 n)=O(n^{2m+1}\log_2n)$ to solve the
desired majorization inequality on $\mathcal{R}^m$.

Let us concentrate on a specific $\mathcal{D}_i$. An algorithm to
solve the majorization inequality on $\mathcal{D}_i$ is as
follows:

Step 1. Sort $c$ and $F(x)$ into non-increasing order,
respectively. Assume $c^{\downarrow}=(c^{(1)},\cdots,c^{(n)})$ and
$F^{\downarrow}(x)=(a^{(1)}(x),\cdots, a^{(n)}(x)).$

 Step 2. Transform the majorization inequality $c\prec F(x)$ into the following linear system of
 inequalities:
 \begin{equation}\label{affinemaj}
          \sum_{s=1}^l c^{(s)}\leq \sum_{s=1}^l a^{(s)}(x), \ \ 1\leq l\leq n,
 \end{equation}
with equality holding when $l=n$.

Step 3. Solve the system of inequalities in Eq. (\ref{affinemaj})
using standard techniques of linear programming.

Now let us  calculate the time complexity of each step. It is
obvious that $c$ can be sorted non-increasingly in $O(n\log_2 n)$
time. Since $F$ has a fixed order on $\mathcal{D}_i$, $F(x)$ can
also be sorted into non-increasing order in $O(n\log_2 n)$ time.
So Step 1 can be completed in $O(n\log_2n)$ time. To figure out
the time complexity of Step 2, we need the following simple fact:
the linear transform of $(y_1,\cdots, y_n)$ to
$(y_1,y_1+y_2,\cdots, y_1+y_2+\cdots+y_n)$ needs only $O(n)$ time.
So Step 2 needs $O(nm)=O(n)$ time. The time complexity of Step 3
needs a careful analysis. By applying the well-known Karmarkar's
algorithm in the theory of linear programming \cite{Karmarkar}
directly, Step 3 needs $O(n^{3.5})$ time. However, in
\cite{Megiddo}, it has been shown that linear programming can be
solved in linear time $O(n)$ when the dimension of variable $x$ is
fixed. Hence the total time to solve $c\prec F(x)$ on $D_i$ is
$$O(n\log_2n)+O(n)+O(n)=O(n\log_2n).$$

With that we complete the proof of Lemma \ref{majaffine}.\hfill $\square$\\

Now we are able to present our algorithms about partial
entanglement recovery. The first algorithm to solve Problem 1 is a
direct consequence of Lemma \ref{majaffine}.

\begin{theorem}\label{algorithm1}
Problem 1 is solvable in $O(n^{2k-1}\log_2n)$ time, where $k$ is
treated as a constant.
\end{theorem}

{\it Proof.} The key here is to solve the majorization inequality
$\psi\otimes \chi \prec \varphi\otimes \omega$. Notice that when
$\varphi$ is fixed, the map from $\omega$ to $\varphi\otimes$ is
an affine one. So Lemma \ref{majaffine} works. A subtle point here
is that $\omega$ is a $k$-dimensional probability vector and has
only $k-1$ independent parameters.  In addition, the relations
$\omega\prec \chi {\rm\ and\ }\chi^{\downarrow}\neq
\omega^{\downarrow} $ can easily be cast into linear constraints
of $\omega$. The total number of these constraints is at most
$O(k!)=O(1)$ when $k$ is a constant. Hence
the time complexity is in fact $O((nk)^{2(k-1)+1}\log_2n)+O(1)=O(n^{2k-1}\log_2n)$. \hfill $\square$\\

The main advantage of the above algorithm is that it can determine
all the resulting states $|\omega\rangle$ in the process of
partial entanglement recovery. However, this algorithm is
efficient only when $k$ is treated as a constant. If $k$  varies
freely, it will turn into exponential time complexity and cannot
be efficient anymore. To further reduce the time complexity, some
lemmas are necessary.

Let $\chi^{\downarrow}=(\gamma_1,\cdots, \gamma_k)$. For the sake
of convenience, we assume all $k$ entries of $\chi$ are distinct.
The general case can be considered similarly by using the compact
form of $\chi$. For each $1\leq i<j\leq k$ and $\epsilon>0$ we
introduce the following vector:
\begin{equation}\label{chiije}
\chi(i,j,\epsilon)=(\gamma_1,\cdots, \gamma_i-\epsilon, \cdots,
\gamma_j+\epsilon,\cdots, \gamma_k).
\end{equation}
To keep the order of $\chi(i,j,\epsilon)$ fixed when $\epsilon$
varies, the constraints
$$\gamma_i-\epsilon\geq \gamma_{i+1} {\rm\ and\ } \gamma_{j-1}\geq \gamma_{j}+\epsilon$$
should be satisfied. Let $\delta_{ij}$ be
$(\gamma_i-\gamma_{i+1})/2 $ if $j=i+1$, and be ${\rm min
}\{\gamma_i-\gamma_{i+1},\gamma_{j-1}-\gamma_j\}$ otherwise. Then
$\epsilon\in [0,\delta_{ij}]$.

The following two lemmas exhibit some interesting properties of
the solutions of Problem 1. Interestingly, the first lemma shows
that we only need to consider the solution $|\omega\rangle$ with
the Schmidt coefficient vector of a special
form given in Eq. (\ref{chiije}).\\

\begin{lemma}\label{reductionlemma}
Problem 1 has a solution if and only if there exist $1\leq i<j\leq
k$ and $\epsilon\in (0,\delta_{ij}]$ such that $\psi\otimes
\chi\prec \varphi\otimes\chi(i,j,\epsilon)$.
\end{lemma}

 {\it Proof.} Sufficiency: Suppose such $i$, $j$,
and $\epsilon$ do exist. It is easy to verify
$\chi(i,j,\epsilon)\prec \chi$ and $\chi^{\downarrow}\neq
\chi^{\downarrow}(i,j,\epsilon)$. These facts, together with the
hypothesis $\psi\otimes \chi\prec
\varphi\otimes\chi(i,j,\epsilon)$, indicate that
$\chi(i,j,\epsilon)$ is a solution of Problem 1.

Necessity: Assume Problem 1 has a solution $|\omega\rangle$. Then
we have $\psi\otimes\chi \prec \varphi\otimes \omega$,
$\omega\prec \chi$, and $\omega^{\downarrow}\neq
\chi^{\downarrow}$.  The existence of $i$, $j$, and $\epsilon$
such that $\psi\otimes \varphi\prec
\varphi\otimes\chi(i,j,\epsilon)$ follows directly from the
following two facts:

(a) If $\omega\prec \chi$ and $\chi^{\downarrow}\neq
\omega^{\downarrow}$ then there exist $1\leq i<j\leq k$ and
$\epsilon\in (0,\delta_{ij}]$ such that $\omega\prec
\chi(i,j,\epsilon)\prec \chi$. This is a direct consequence of
Lemma \ref{fact1} in Appendix B.

(b) Any state $|\chi'\rangle$ such that $\omega\prec\chi'\prec
\chi$ and $\chi^{\downarrow}\neq \chi'^{\downarrow}$ is also a
solution of Problem 1. This follows directly from our formulation
of Problem
1. \hfill $\square$\\

\begin{lemma}\label{keyfact}
If $|\chi(i,j,\epsilon_0)\rangle$ is a solution of Problem 1, then
for any $0<\epsilon<\epsilon_0$, $|\chi(i,j,\epsilon)\rangle$ is
also a solution.
\end{lemma}

{\it Proof.} Immediately from the formulation of Problem 1 and Eq.
(\ref{chiije}). \hfill $\square$\\

We are now in a position to state the main result of this section,
the promised algorithm of time complexity $O(n^2k^4)$.
\begin{theorem}\label{algorithm2}
Problem 1 is solvable in $O(n^2k^4)$ time.
\end{theorem}

{\it Proof.} By Lemma \ref{reductionlemma}, we only need to
consider the following problem: for each specific pair $(i,j)$
such that $1\leq i<j\leq k$, decide whether there exists
$\epsilon\in(0,\delta_{ij}]$ such that $\psi\otimes \chi\prec
\varphi\otimes \chi(i,j,\epsilon)$. In what follows we show that
this problem can be solved in $O(n^2k^2)$ time. Then by
enumerating all possible pairs of $(i,j)$, we get an $O(k(k-1)/2
n^2k^2)=O(n^2k^4)$ time algorithm to solve Problem 1.

Let us begin with two specific indices $i$ and $j$. By Lemma
\ref{orderlemma}, the number of the different orders of
$\varphi\otimes \chi(i,j,\epsilon)$  is at most $O((nk)^2)$ when
$\epsilon$ varies in $[0,\delta_{ij}]$. With Lemma \ref{keyfact}
in mind, it is enough to consider one special order among them.
More precisely, suppose the interval $[0,\delta_{ij}]$ is  divided
into $M$ parts (intervals), namely,
$$\mathcal{D}_1=[c_0,c_1], \mathcal{D}_2=[c_1,c_2],\cdots, \mathcal{D}_M=[c_{M-1},c_M],$$
where $0=c_0<c_1<\cdots <c_M=\delta_{ij}$, and $M=O((nk)^2)$. On
each interval $\varphi\otimes \chi(i,j,\epsilon)$ has a fixed
order.  By lemma \ref{keyfact}, if $|\chi(i,j,\epsilon_0)\rangle$
is a solution of Problem 1, then any  $|\chi(i,j,\epsilon)\rangle$
such that $0<\epsilon\leq {\rm min}\{\epsilon_0, c_1\}$  is also a
solution. So we need only to consider the leftmost interval
$\mathcal{D}_1$. Our algorithm goes as follows:

Step 1: Find $c_1$;

Step 2: Sort $\psi\otimes \chi$ and $\varphi\otimes
\chi(i,j,\epsilon)$ into non-increasing order, respectively, where
$\epsilon\in [0,c_1]$;

Step 3: Solve the system of inequalities induced by the
majorization relation $\psi\otimes \chi\prec \varphi\otimes
\chi(i,j,\epsilon)$.

Step 4: Output: if a solution of $\epsilon>0$ is obtained in Step
3, then Problem 1 has a solution $|\chi(i,j,\epsilon)\rangle$;
otherwise Problem 1 does not has a solution of the form
$|\chi(i,j,\epsilon)\rangle$ for fixed $i$ and $j$, and
$\epsilon\in(0,\delta_{ij}]$.

Step 1 requires that we search for the smallest positive elements
among $M$ items, which requires $O(M)=O((nk)^2)$ time (see also
Example \ref{arr2}). Step 2 needs $O(nk\log_2nk)$ time. Step 3
merely needs $O(nk)$ time since there is only a single parameter
$\epsilon$. Step 4 only needs $O(1)$ time. In sum, only
$$O((nk)^2)+O(nk\log_2 nk)+O(nk)+O(1)=O((nk)^2)$$ time is required. \hfill $\square$\\

In view of Theorem \ref{algorithm2}, we can say that Problem 1 is
efficiently solvable. It also suggests that we can study the
process of partial entanglement recovery using algorithmic
methods.

To conclude our discussions about Problem 1, we would like to
address an important issue for further study. In almost all the
results we obtained so far, we are only concerned with the
feasibility of partial entanglement recovery, while the efficiency
of this process has not been touched yet. These results are of
limited use in practice, when we hope to minimize entanglement
lost in LOCC transformations. In other words, we require the
resulting state $|\omega\rangle$ to be not only more entangled
than $|\chi\rangle$, but also  an ``optimal" one that we can
achieve in this process. Using entropy of entanglement as a
measure, we suggest the following optimization problem. We also
note that some aspects of the efficiency of partial entanglement
recovery have been discussed in \cite{SR02}.

{\it Open problem:\ } Given a triple of states
$(|\psi\rangle,|\varphi\rangle,|\chi\rangle)$ such that $\psi\prec
\varphi$, let $\Omega=\{\omega: \psi\otimes \chi\prec
\varphi\otimes \omega{\rm\ and\ }\omega\prec \chi \}.$ Maximize
$E(|\omega\rangle)$, subject to $\omega\in \Omega$.

In the above problem we remove the constraint
$\chi^{\downarrow}\neq \omega^{\downarrow}$. This makes $\Omega$
compact. Thus the continuous function $E(|\omega\rangle)$ can
attain its maximum on $\Omega$. Suppose $|\omega_0\rangle$ is one
of the states attaining the maximum. Noticing that the entropy of
entanglement decreases under majorization, we have the following
simple relation
$$E(|\chi\rangle)\leq E(|\omega_0\rangle)\leq
E(|\psi\rangle)-E(|\varphi\rangle)+E(|\chi\rangle),$$ where the
first inequality is from $\omega_0\prec \chi$, and the second
inequality is from $\psi\otimes \chi\prec \varphi\otimes \omega_0$
and the additivity of entropy of entanglement.  The first
inequality is an equality if and only if
$\omega^{\downarrow}=\chi^{\downarrow}$ for any $\omega\in
\Omega$, i.e., $|\chi\rangle$ cannot do partial entanglement
recovery for the transformation of $|\psi\rangle$ to
$|\varphi\rangle$. The second inequality is an equality if and
only if $(\psi\otimes \chi)^{\downarrow}=(\varphi\otimes
\omega_0)^{\downarrow}$. Theorem \ref{algorithm2} in fact provides
a polynomial time algorithm to determine whether the first
inequality holds strictly. How to design efficient algorithms to
find the optimal state $|\omega_0\rangle$ seems to be a
challenging and worthwhile problem.
\section{Some applications}
In this section, we establish  some interesting connections of
partial entanglement recovery to the generation of maximally
entangled states, quantum catalysis, mutual catalysis, and
multiple-copy entanglement transformation.

\subsection{How to obtain maximally entangled states by using partial entanglement recovery}

Maximally entangled  states play  a crucial role in many striking
applications of quantum entanglement such as quantum superdense
coding \cite{BS92} and quantum teleportation \cite{BBC+93}. It is
very important to generate such states in practical information
processing. Under the constraint of LOCC, a natural way to obtain
a maximally entangled state is to concentrate a large number of
partially entangled states \cite{BBPS96}. However, such a
concentrating protocol involves infinitely many copies of the
source state while in practice only finitely many copies can be
available. One can find various deterministic protocols based on
Nielsen's theorem \cite{NI99} and probabilistic protocols  based
on Vidal's theorem \cite{Vidal99} (see also \cite{LO97}). It has
been shown that two $2\times 2$ partially entangled states
sometimes can be concentrated into an EPR pair deterministically
\cite{FM00}. An extensive generalization of such a deterministic
concentration protocol was presented in \cite{FM01}, where the
maximal number of Bell states that can be concentrated from a
finite number of partially entangled states was derived. In what
follows, we consider deterministic transformations only.

The following theorem  shows that almost all deterministic
entanglement transformations can concentrate a partially entangled
pure state into a maximally entangled state with the same
dimension providing that they are close enough to each other.

\begin{theorem}\label{belllemma}\upshape
Let $|\psi\rangle$ be a state in $S^{o}(|\varphi\rangle)$ and let
$|\Phi^{+}\rangle=\sum_{i=1}^k \frac{1}{\sqrt{k}}|i\rangle
|i\rangle$ be a $k\times k$ maximally entangled state. Then there
exists a positive number $\delta$ such that for any $k\times k$
state $|\chi\rangle$ satisfying
$\||\chi\rangle-|\Phi^{+}\rangle\|<\delta$, the transformation of
$|\psi\rangle\otimes |\chi\rangle$ to $|\varphi\rangle\otimes
|\Phi^{+}\rangle$ can be realized with certainty by LOCC, i.e.,
$|\psi\rangle\otimes |\chi\rangle\rightarrow
|\varphi\rangle\otimes |\Phi^{+}\rangle.$
\end{theorem}

\textit{Proof.} This is a simple application of Lemma
\ref{lginterior}. Since $S^{o}(|\varphi\rangle)$ is not empty and
$l_u(|\Phi^{+}\rangle)=1>g_u(|\varphi\rangle)$, it follows from
Lemma \ref{lginterior} that
\begin{equation}\label{bellpair}
\psi\otimes \Phi^{+}\lhd \varphi\otimes \Phi^{+}.
\end{equation}
An arbitrary but small enough perturbation on $|\Phi^{+}\rangle$
in the left hand side of Eq. (\ref{bellpair}) can still keep the
relation `$\lhd$'. Hence
the existence of $\delta$ is proven. \hfill $\square$\\

The above theorem tells us that for any given $|\psi\rangle\in
S^o(|\varphi\rangle)$ and $k>1$, we can find a partially entangled
pure state $|\chi\rangle$ satisfying $ |\psi\rangle\otimes
|\chi\rangle\rightarrow |\varphi\rangle\otimes |\Phi^{+}\rangle$.
It is obvious that $|\chi\rangle$ depends  not only on
$|\varphi\rangle$ and $k$, but also $|\psi\rangle$. At first
glance, this seems to be contradicting our result about partial
entanglement recovery, which states the auxiliary state
$|\chi\rangle$ for partial entanglement recovery only depends on
the target state $|\varphi\rangle$ and the presence of equalities
in the majorization $\psi\prec \varphi$. The key point is  when we
consider whether $|\chi\rangle$ can be used to do partial
entanglement recovery for a transformation with the target state
$|\varphi\rangle$, the resulting state $|\omega\rangle$ is not
specified; while the resulting state here is given and is
maximally entangled. By Nielsen's theorem, $|\chi\rangle$ should
be determined by the relation $\psi\otimes \chi\prec
\varphi\otimes \Phi^+$, which obviously depends on the source
state, the target state and $k$.

Theorem \ref{belllemma} confirms the existence of the partially
entangled state $|\chi\rangle$. But it cannot yield a complete
characterization of $|\chi\rangle$. To obtain such a
characterization, we need to apply Nielsen's theorem and solve the
corresponding majorization relation directly. To illustrate this
procedure better, let us examine a simple case where
$|\psi\rangle$ and $|\varphi\rangle$ are both $2\times
2$-dimensional. In particular, the following example deals with
the case of $k=2$.

\begin{example}\label{example8}\upshape
Let $|\psi\rangle$, $|\varphi\rangle$, and $|\Phi^+\rangle$  be
three $2\times 2$ states with $\psi=(a,1-a)$,  $\varphi=(b,1-b)$,
and $\Phi^{+}=(\frac{1}{2},\frac{1}{2})$, where
$\frac{1}{2}<a<b\leq 1$. We are going to find a $2\times 2$
partially entangled  state $|\chi\rangle$ such that the
transformation of $|\psi\rangle$ to $|\varphi\rangle$ can
concentrate $|\chi\rangle$ into the maximally entangled state
$|\Phi^+\rangle$.

Suppose that $|\chi\rangle$  is of  the form $\chi=(p,1-p)$, where
$\frac{1}{2}<p<1$. By Nielsen's theorem, we only need
$|\chi\rangle$ to satisfy
$$\psi\otimes \chi\prec \varphi\otimes\Phi^{+}.$$ Notice that
$\varphi\otimes \Phi^+$ has only two distinct components
$\frac{1}{2}b$ and $\frac{1}{2}(1-b)$. By the definition of
majorization, the above equation holds if and only if
$$ap\leq \frac{1}{2}b$$ and
$$(1-a)(1-p)\geq \frac{1}{2}(1-b).$$
Hence $$\frac{1}{2}<p<{\rm min}\{\frac{b}{2a},
\frac{1-2a+b}{2(1-a)}\}.$$ Note that $\frac{1}{2}<a<b<1$, so the
above equation can be simplified into
\begin{equation}\label{2bell}
\frac{1}{2}<p\leq \frac{b}{2a},
\end{equation}
which is  exactly  the result obtained in \cite{FM00}.\hfill $\square$\\
\end{example}

More generally, suppose that the $k\times k$ auxiliary state
$|\chi\rangle$ is of the form $\chi=(\gamma_1,\ldots, \gamma_k)$.
Then to obtain a $k\times k$ maximally entangled state
$|\Phi^+\rangle$ from the above transformation of $|\psi\rangle$
to $|\varphi\rangle$, it suffices to have $\psi\otimes \chi\prec
\varphi\otimes \Phi^+$, which is equivalent to
$$\gamma_1 a\leq \frac{b}{k}$$ and
$$\gamma_k(1-a)\geq \frac{1-b}{k}.$$ Thus
\begin{equation}\label{kbell}
\frac{1-b}{k(1-a)}\leq \gamma_k<\gamma_1\leq \frac{b}{ka}.
\end{equation}
If $k=2$,  we can show that Eq. (\ref{kbell}) can be reduced to
Eq. (\ref{2bell}).

One can similarly consider the general case where both $k$ and $n$
are arbitrary positive integers.

\subsection{Partial entanglement recovery and quantum catalysis}

In the above discussions, we always assume that the source state
$|\psi\rangle$  is comparable to the target state
$|\varphi\rangle$, i.e., the transformation of $|\psi\rangle$ to
$|\varphi\rangle$ can be realized with certainty under LOCC. How
about the case where $|\psi\rangle$ and $|\varphi\rangle$ are not
comparable? The  general answer  to this question remains unknown.

In \cite{SR02} a special case where the transformation of
$|\psi\rangle$ to $|\varphi\rangle$ has  a catalyst state
$|c\rangle$ such that $|\psi\rangle\otimes |c\rangle\rightarrow
|\varphi\rangle\otimes |c\rangle$ \cite{JP99}, i.e., the
transformation of $|\psi\rangle$ to $|\varphi\rangle$ can be
realized under ELOCC,  was examined carefully. It was shown that
the problem of doing partial entanglement recovery for the
transformation of $|\psi\rangle$ to $|\varphi\rangle$ with
$\psi\nprec \varphi$ may be reduced to the problem of finding a
catalyst state $|c\rangle$ and then seeking a suitable auxiliary
state $|\chi\rangle$ to do partial entanglement recovery  for the
new transformation of $|\psi\rangle\otimes |c\rangle$ to
$|\varphi\rangle\otimes |c\rangle$ such that $\psi\otimes c\prec
\varphi\otimes c$. For this purpose, in \cite{SR02} an algorithm
of time complexity $O((nk)!)$ was proposed to find a $k\times k$
catalyst $|c\rangle$ for a transformation of $|\psi\rangle$ to
$|\varphi\rangle$ in which the source state and the target state
are both $n\times n$-dimensional.

However, the above algorithm is intractable since it is of
exponential time complexity. In \cite{SD03} a polynomial time
algorithm of $n$ for fixed $k$ was given. With the aid of this
efficient  algorithm, one can quickly determine whether an
$n\times n$ incomparable pair  has  a $k\times k$ catalyst. Then
by the results obtained in the present paper, such as Theorems
\ref{shannoncode}, \ref{generaltheorem1}, or
\ref{generaltheorem2}, a state $|\chi\rangle$ that can do partial
entanglement recovery  for the transformation of $|\psi\rangle$ to
$|\varphi\rangle$ can be explicitly constructed.

Therefore, if the transformation of $|\psi\rangle$ to
$|\varphi\rangle$ can be realized with certainty under ELOCC, then
we can find an auxiliary state $|\chi\rangle$ to do partial
entanglement recovery for this transformation.

\subsection{Partial entanglement recovery and mutual catalysis}

In \cite{XZZ02}, an interesting phenomenon named \textit{mutual
catalysis} was demonstrated. If $|\psi\rangle\nrightarrow
|\varphi\rangle$ and $|\alpha\rangle\nrightarrow |\beta\rangle$
but $|\psi\rangle\otimes |\alpha\rangle\rightarrow
|\varphi\rangle\otimes |\beta\rangle$, we say that $|\psi\rangle$
and $|\alpha\rangle$ can be mutually catalyzed by each other. The
trivial case such that  $|\psi\rangle\rightarrow |\beta\rangle$
and $|\alpha\rangle\rightarrow |\varphi\rangle$ is not necessary
to consider. With the help of the results obtained in previous
sections, one can easily construct many non-trivial instances with
the mutual catalysis effect. First, let us reexamine an example
from \cite{XZZ02}.

\begin{example}\label{example9}\upshape
Let $|\psi\rangle$, $|\varphi\rangle$, $|\alpha\rangle$, and
$|\beta\rangle$ be four states with $\psi=(0.33,0.32,0.3,0.05)$,
$\varphi=(0.6,0.2,0.14,0.06)$, $\alpha=(0.6,0.3,0.1,0)$, and
$\beta=(0.46,0.46,0.08,0)$. It is easy to see that both the
transformations of  $|\psi\rangle$ to $|\varphi\rangle$ and of
$|\alpha\rangle$ to $|\beta\rangle$ cannot happen with certainty
even under ELOCC. But we do have $|\psi\rangle\otimes
|\alpha\rangle\rightarrow |\varphi\rangle\otimes |\beta\rangle$ in a
non-trivial way. This is just the effect of mutual catalysis.

From another point of view, this example can be treated as  a
special instance of partial entanglement recovery. To see this,
let us relabel the above four states as follows:
$\chi=(0.6,0.3,0.1,0)$, $\omega=(0.6,0.2,0.14,0.06)$,
$\psi=(0.33,0.32,0.3,0.05)$,  and $\varphi=(0.46,0.46,0.08,0)$. It
is obvious that $\psi\lhd \varphi$. Noticing that
$l_u(|\omega\rangle)>g_u(|\varphi\rangle)=0$,  we have that
$\psi\otimes \omega\lhd \varphi\otimes \omega$ by Lemma
\ref{lginterior}. A small perturbation on $\omega$ will generate
$\chi=\omega+(0, 0.1,-0.04,-0.06).$ Note that
$E(|\chi\rangle)=1.2955<E(|\omega\rangle)=1.5472$. So the entropy
of entanglement of $|\chi\rangle$ is enhanced.
\hfill $\square$\\
\end{example}

The above example suggests a connection between partial
entanglement recovery and mutual catalysis.  More generally, any
pairs $\{|\psi\rangle, |\chi\rangle\}$ and $\{|\varphi\rangle,
|\omega\rangle\}$ such that $|\psi\rangle\otimes
|\chi\rangle\rightarrow |\varphi\rangle\otimes |\omega\rangle$,
$|\chi\rangle\nrightarrow |\omega\rangle$,
$|\chi\rangle\nrightarrow |\varphi\rangle$, and
$|\psi\rangle\nrightarrow |\varphi\rangle\otimes |\omega\rangle$
can be treated as  nontrivial instances of mutual catalysis. These
pairs can be easily obtained with the aid of lemma
\ref{lginterior}. Furthermore, one can choose the  state
$|\omega\rangle$  satisfying $\omega\prec \chi$ but $\chi\nprec
\omega$. We omit the construction details.

\subsection{Multiple-copy is essential for partial entanglement recovery}

Multiple-copy entanglement transformation is another interesting
topic in quantum entanglement theory. Let us review this concept
briefly. In \cite{SRS02}, it was demonstrated that sometimes
multiple copies of a source state may be transformed into the same
number of copies of a target state although the transformation
cannot happen for a single copy. That is, for some states
$|\psi\rangle$ and $|\varphi\rangle$, although the transformation
of $|\psi\rangle$ to $|\varphi\rangle$ cannot be realized with
certainty by LOCC, there may exist $m>1$ such that the
transformation of $|\psi\rangle^{\otimes m}$ to
$|\varphi\rangle^{\otimes m}$ can be achieved deterministically.
This kind of transformation that uses multiple copies of a source
state and then transforms all of them together into the same
number of copies of a  target state is intuitively called
`multiple-copy entanglement transformation', or MLOCC for short.
See \cite{SRS02}, \cite{DF03}, \cite{DFLY04}, and \cite{DF04} for
more about MLOCC.

It may be of interest to study the relations between partial
entanglement recovery and multiple-copy entanglement
transformation. To  our surprise, entanglement lost in a
multiple-copy entanglement transformation can be recovered more
easily than that in  a single-copy transformation when the
auxiliary state is specified. To demonstrate this point, we need
the following theorem as a useful tool.
\begin{theorem}\label{multiplecopy}\upshape
Let $|\varphi\rangle$ and $|\chi\rangle$ be two partially
entangled states.  If $|\chi\rangle$ has at least two distinct
nonzero Schmidt coefficients, then there exists a positive integer
$k_0$  such that for any $k\geq k_0$  and $|\psi\rangle\in
S^o(|\varphi\rangle)$, entanglement lost in the transformation of
$|\psi\rangle^{\otimes k}$ to $|\varphi\rangle^{\otimes k}$ can be
partially recovered by $|\chi\rangle$.
\end{theorem}
The most interesting part of the above theorem is that the choice
of $k_0$ only depends on $|\chi\rangle$ and $|\varphi\rangle$.

\textit{Proof.} First, applying Lemma 1 in \cite{DFLY04} yields
that $\psi\lhd \varphi$ implies $\psi^{\otimes k}\lhd
\varphi^{\otimes k}$ for any $k\geq 1$.  Second, notice that
$g_u(|\varphi\rangle^{\otimes k})=g_u^k(|\varphi\rangle)$ and
$g_u(|\varphi\rangle)<1$. By the assumption on $|\chi\rangle$, we
have $0<L_u(|\chi\rangle)<1$. Thus there exists $k_0\geq 1$ such
that $L_u(|\chi\rangle)>g_u^k(|\varphi\rangle)$ for any $k\geq
k_0$. Therefore, by Theorem \ref{shannoncode}, we deduce that
$|\chi\rangle$ can be used  to do partial  entanglement recovery
for the transformation of $|\psi\rangle^{\otimes k}$ to
$|\varphi\rangle^{\otimes k}$ such that $k\geq k_0$ and $|\psi\rangle\in S^o(|\varphi\rangle)$.  \hfill $\square$\\

Let us take now  two states $|\varphi\rangle$ and $|\chi\rangle$
such that $0<L_u(|\chi\rangle)<g_u(|\varphi\rangle)<1$. By Theorem
\ref{shannoncode}, $|\chi\rangle$ cannot do partial entanglement
recovery for any transformation with the target $|\varphi\rangle$
since $L_u(|\chi\rangle)<g_u(|\varphi\rangle)$. On the other hand,
it is easy to see that $|\varphi\rangle$ and $|\chi\rangle$
satisfy the assumptions of Theorem \ref{multiplecopy}. Hence,
there exists $k_0$ such that for any $k\geq k_0$ and
$|\psi\rangle\in S^o(|\varphi\rangle)$, entanglement lost in the
$k$-copy transformation, i.e., the transformation of
$|\psi\rangle^{\otimes k}$ to $|\varphi\rangle^{\otimes k}$, can
be partially recovered by $|\chi\rangle$.

We give an intuitive explanation for the above theorem. If the
auxiliary state $|\chi\rangle$ cannot do partial entanglement
recovery  for the transformations with the target state
$|\varphi\rangle$, then the target state is too uniform, and it is
too entangled relative to $|\chi\rangle$. So for any state
$|\psi\rangle$ that can be transformed into $|\varphi\rangle$
under LOCC, the extra entanglement left (except the necessary part
to finish the transformation of $|\psi\rangle$ to
$|\varphi\rangle$) is  not enough to be transferred into
$|\chi\rangle$. But if multiple copies of the source state are
provided, the extra entanglement will accumulate. Such extra
entanglement can be transferred into the state $|\chi\rangle$ when
it exceeds a threshold.

It is also interesting to investigate the partial entanglement
recovery power when multiple copies of the auxiliary state
$|\chi\rangle$ are available. We restrict ourselves to the special
case that $|\chi\rangle$ is $2\times 2$-dimensional. A surprising
result appears as the following:

\begin{theorem}\upshape
Let $|\chi\rangle$ be a $2\times 2$-dimensional partially
entangled state and $|\varphi\rangle$ be any partially entangled
state. If $L_u(|\chi\rangle)<g_u(|\varphi\rangle)$, then for any
$k\geq 1$, $|\chi\rangle^{\otimes k}$ cannot do partial
entanglement recovery for any transformation with target state
$|\varphi\rangle$.
\end{theorem}
Intuitively, if the auxiliary state $|\chi\rangle$ is  $2\times
2$-dimensional, then more copies of $|\chi\rangle$ do not provide
any extra power of  partial entanglement recovery if
$L_u(|\chi\rangle)<g_u(|\varphi\rangle)$. This is very reasonable
since the key point of such a recovery is the difference between
the entanglement resource of the source state and that of the
target state, which keeps invariant during the process of
increasing the number of copies of the auxiliary state.
\vspace{1em}

\textit{Proof.} By (3) of Lemma \ref{tensor} we have that
$L_u(|\chi\rangle^{\otimes k})=L_u(|\chi\rangle)$ since
$|\chi\rangle$ is a $2\times 2$ state. So by the assumption
$L_u(|\psi\rangle)<g_u(|\varphi\rangle)$, it follows from Theorem
\ref{shannoncode} that $|\chi\rangle^{\otimes k}$ cannot  be used
to do partial entanglement recovery for any transformation
with the target state $|\varphi\rangle$. \hfill $\square$\\

In the case when  $L_u(|\chi\rangle)=g_u(|\varphi\rangle)$,
however, the partial entanglement recovery capability of
$|\chi\rangle^{\otimes k}$ may be strictly more powerful than that
of $|\chi\rangle$  for suitably large  $k$. That is,
$|\chi\rangle^{\otimes k}$ can do partial entanglement recovery
for some transformation with the target state $|\varphi\rangle$
while $|\chi\rangle$ cannot. See the following example.

\begin{example}\label{example10}\upshape
Let $|\chi\rangle$ be a state with $\chi=(p,1-p)$, where
$\frac{1}{2}<p<1$. Then by Theorem \ref{shannoncode} we know that
$|\chi\rangle$ cannot  do partial entanglement recovery for any
transformation with target state $|\varphi\rangle$ such that
$\varphi=(p,p,p,p,1-p,1-p)/(2+2p)$.

However, by Theorem \ref{shannoncode} again,
$|\chi\rangle^{\otimes 2}$ with ${\chi^{\otimes
2}}=(p^2,p(1-p),p(1-p),(1-p)^2)$ can do partial entanglement
recovery for any transformation of $|\psi\rangle$ to
$|\varphi\rangle$ such that $|\psi\rangle$ is in
$S^{o}(|\varphi\rangle)$. \hfill $\square$\\
\end{example}

A more general result in this special case is: if
$|\varphi\rangle$ has only two distinct  nonzero Schmidt
coefficients,  then for a sufficiently large $k$,
$|\chi\rangle^{\otimes k}$ can always  do partial entanglement
recovery for any transformation of $|\psi\rangle$ to
$|\varphi\rangle$ with $|\psi\rangle\in S^{o}(|\varphi\rangle)$;
otherwise such a recovery is impossible for arbitrarily large $k$.

\section{Conclusion}
To summarize, we obtain a complete characterization of an
auxiliary bipartite entangled state $|\chi\rangle$ that can do
partial entanglement recovery for the transformation of
$|\psi\rangle$ to $|\varphi\rangle$ where $\psi$ is strictly
majorized by $\varphi$. It is interesting that the choice of the
auxiliary state can only depend on the target state
$|\varphi\rangle$ and the presence of the equalities in  the
majorization relation $\psi\prec \varphi$.  We further propose two
sufficient conditions for $|\chi\rangle$ that can be used to do
partial entanglement recovery for a class of transformations of
$|\psi\rangle$ to $|\varphi\rangle$ with $\psi\prec \varphi$.  We
also study the feasibility of partial entanglement recovery from
the algorithmic viewpoint. A polynomial algorithm of time
complexity $O(n^2k^4)$ is presented for deciding the possibility
of partial entanglement recovery.  As applications, we establish
some interesting connections of partial entanglement recovery to
the generation of maximally entangled states, quantum catalysis,
mutual catalysis, and multiple-copy entanglement transformation.
We hope the results presented here may help us to manipulate
quantum entanglement more economically.

\smallskip

\section*{Appendix A: Proof of  Lemma ~\ref{lginterior}}

Take $|\psi\rangle\in S^o(|\varphi\rangle)$. Assume
$\psi^{\downarrow}=(\alpha_1,\alpha_2,\ldots, \alpha_n)$,
$\varphi^{\downarrow}=(\beta_1,\beta_2,\ldots, \beta_n)$, and
$\chi^{\downarrow}=(\gamma_1,\gamma_2,\ldots, \gamma_k)$. If $k=1$
then the result follows trivially. In what follows we assume
$k>1$.

First we prove that if $l_u(|\chi\rangle)>g_u(|\varphi\rangle)$
then $|\psi\rangle\otimes |\chi\rangle$ is in
$S^{o}(|\varphi\rangle\otimes |\chi\rangle)$. In other words, we
shall prove
\begin{equation}\label{catamaj}
e_l(\psi\otimes \chi )<e_l(\varphi\otimes \chi)
\end{equation}
for any $1\leq l<nk$.

We  rewrite
\begin{equation}\label{maxdef}
e_l(\psi\otimes \chi )=\sum_{i=1}^{k}e_{l_i}(\gamma_i\psi),
\end{equation}
where $0\leq l_i\leq n$ and $\sum_{i=1}^{k}l_i=l$. Easily see that
$\gamma_i\psi\lhd \gamma_i\varphi$ for all $1\leq i\leq k$. So we
merely need to consider the following two cases:

Case 1: There exists $1\leq s\leq k$ such that $0<l_s<n$. In this
case, $e_{l_s}(\gamma_s\psi)<e_{l_s}(\gamma_s\varphi)$ holds. Then
Eq. (\ref{catamaj}) follows from
$$
\begin{array}{rl}
e_l(\psi\otimes
  \chi)&=\sum_{i=1}^{k}e_{l_i}(\gamma_i\psi)\\
  &<\sum_{i=1}^{k}e_{l_i}(\gamma_i\varphi)\\
  &\leq e_l(\varphi\otimes
\chi),
\end{array}
$$
where the equality is by Eq. (\ref{maxdef}), and the second
inequality is by the definition of $e_l(\varphi\otimes\chi)$.

Case 2: For any $1\leq i\leq k$, $l_i\in\{0,n\}$. Let $h$ be the
maximal index satisfying $l_h=n$. Then $1\leq
 h<k$; otherwise $h=k$ implies $l=nk$, which contradicts the
assumption  $l<nk$. Noticing
$l_u(|\chi\rangle)>g_u(|\varphi\rangle)$, we have
$\gamma_{h+1}/\gamma_h>\beta_n/\beta_1,$ or
\begin{equation}\label{lgcond}
\gamma_h\beta_n<\gamma_{h+1}\beta_1.
\end{equation}
By the definition of $e_{nh}(\psi\otimes \chi)$ and the assumption
on $h$,  we further have
$$
\begin{array}{rl}
e_{nh}(\psi\otimes\chi)&=\sum_{i=1}^{h}e_{n}(\gamma_i\psi)\\
&=\sum_{i=1}^{h-1}e_{n}(\gamma_i\varphi)+\sum_{i=1}^{n-1}\gamma_h\beta_i+
  \gamma_h\beta_n,
\end{array}
$$
where the second equality is due to $e_n(\psi)=e_n(\varphi)$.

Substituting Eq. (\ref{lgcond}) into the above equation, we have
$$
\begin{array}{rl}
e_{nh}(\psi\otimes
\chi)&<\sum_{i=1}^{h-1}e_{n}(\gamma_i\varphi)+\sum_{i=1}^{n-1}\gamma_h\beta_i+\gamma_{h+1}\beta_1\\
&\leq e_{nh}(\varphi\otimes \chi),
\end{array}
$$
where the second inequality is  by the definition of
$e_{nh}(\varphi\otimes
  \chi)$.

Therefore Eq. (\ref{catamaj}) holds for any $1\leq l<nk$. By the
definition, we have $|\psi\rangle\otimes |\chi\rangle$ is in
$S^{o}(|\varphi\rangle\otimes |\chi\rangle)$. (Note that
$e_{nk}(\psi\otimes \chi)=e_{nk}(\varphi\otimes \chi)$ holds for
any $\chi$ whenever $\psi\lhd \varphi$).

Conversely, suppose $|\psi\rangle\otimes |\chi\rangle $ is in
$S^{o}(|\varphi\rangle\otimes |\chi\rangle)$, while there exists
some $1\leq h<k$, such that
$$\frac{\gamma_{h+1}}{\gamma_{h}}\leq \frac{\beta_n}{\beta_1},$$
or equivalently,
$$
\gamma_h\beta_n\geq \gamma_{h+1}\beta_1.
$$

Then we have
$$
\begin{array}{rl}
e_{nh}(\varphi\otimes \chi)&=\sum_{i=1}^{h}e_n(\gamma_i\varphi)\\
&=\sum_{i=1}^{h}e_n(\gamma_i\psi)\\
 &\leq e_{nh}(\psi\otimes
\chi),
\end{array}
$$
which contradicts the assumption that $e_l(\psi\otimes
\chi)<e_l(\varphi\otimes \chi)$ for any $1\leq l<nk$.

With that we  complete the proof of Lemma \ref{lginterior}.

\section*{Appendix B: Proof of Theorem \ref{shannoncode}}

To prove Theorem \ref{shannoncode}, we need the following three
auxiliary facts about majorization.

\begin{lemma}\label{fact1}\upshape
Let $y\in \mathcal{R}^n$ with compact form
$y^{\downarrow}=(y_1'^{\oplus k_1},\ldots,y_s'^{\oplus k_s})$, and
let $x\in \mathcal{R}^n$ satisfying $x\prec y$ but $y\nprec x$.
Then there exists $z\in \mathcal{R}^n$ such that $x\prec z\prec y$
and
$$
\begin{array}{rl}
z^{\downarrow}=&[y_1'^{\oplus k_1},\ldots,y_i'^{\oplus
k_i-1},y'_i-\epsilon,\\
&\ldots,y_j'+\epsilon,y_j'^{\oplus k_j-1},\ldots, y_s'^{\oplus
k_s}],
\end{array}
$$
for some $1\leq i<j\leq s$ and $\epsilon>0$.
\end{lemma}
\textit{Proof.} This is a direct consequence of B.1. Lemma in \cite{MO79} (page 21).\hfill $\square$\\
\begin{lemma}\label{fact2}\upshape
If $x'\prec y'$ and $x''\prec y''$, then $(x',x'')\prec (y',y'')$.
\end{lemma}
\textit{Proof.} See part (i) of A.7. Lemma in \cite{MO79} (page
121).\hfill $\square$\\
\begin{lemma}\label{fact3}\upshape
Let $x^{\downarrow}=(x'^{\downarrow},x''^{\downarrow})$ and
$y^{\downarrow}=(y'^{\downarrow}$, $y''^{\downarrow})$. If $x\prec
y$ and $x'\prec y'$ (or $x''\prec y''$), then $x''\prec y''$
(resp. $x'\prec y'$).
\end{lemma}
\textit{Proof.} Suppose $x',y'\in \mathcal{R}^m$ and $x'', y''\in
\mathcal{R}^n$. By the assumption, we have
\begin{equation}\label{1} e_l(x'')=e_{m+l}(x)-e_m(x')
\end{equation}
and \begin{equation}\label{2}
e_l(y'')=e_{m+l}(y)-e_m(y')
\end{equation}
for any $1\leq l\leq n$.

Noticing $x'\prec y'$ and $x\prec y$, we also have
\begin{equation}\label{3}
e_m(x')=e_m(y')
\end{equation}
and
\begin{equation}\label{4}
e_l(x)\leq e_l(y)
\end{equation}
for any $1\leq l\leq m+n$ and the inequality is an equality if
$l=m+n$. Thus Eqs. (\ref{1})-- (\ref{4}) give
$$
e_l(x'')\leq e_l(y'')
$$
for any $1\leq l\leq n$,  with the equality holding when $l=n$.
That means  $x''\prec y''$. If $x\prec y$ and $x''\prec
y''$, we can prove $x'\prec y'$ similarly.\hfill $\square$\\

Now we proceed to the proof of Theorem \ref{shannoncode}. We first
deal with the case where all the nonzero Schmidt coefficients of
$|\chi\rangle$ are identical, i.e., $L_u(|\chi\rangle)=0$ or
$L_u(|\chi\rangle)=1$.

Assume $\chi^{\downarrow}=((\frac{1}{a})^{\oplus a}, 0^{\oplus
k-a})$. If $a=k$, then $|\chi\rangle$ is a $k\times k$ maximally
entangled state. There cannot be another $k\times k$ state
$|\omega\rangle$ that is more entangled than $|\chi\rangle$. So
partial entanglement recovery is not possible in this case.

Now suppose $1\leq a<k$. If $|\chi\rangle$ can do partial
entanglement recovery for some transformation of $|\psi\rangle$ to
$|\varphi\rangle$ such that $|\psi\rangle$ is in
$S(|\varphi\rangle)$, then the state $|\omega\rangle$ such that
$\psi\otimes \chi\prec \varphi\otimes \omega$ and $\omega\prec
\chi$, but $\chi^{\downarrow}\neq \omega^{\downarrow}$ should have
at least $a+1$ nonzero Schmidt coefficients. By the property of
majorization, the number of nonzero Schmidt coefficients of
$\psi\otimes \chi$ is not less than that of $\varphi\otimes
\omega$, i.e.,
$$
n''a\geq n'(a+1),
$$
where $n''$ denotes the number of nonzero Schmidt coefficients of
$|\psi\rangle$. Obviously, $n''\leq n$, thus we have
\begin{equation}\label{schmidtcondition}
na\geq n'(a+1).
\end{equation}

Conversely, if Eq. (\ref{schmidtcondition}) holds, we will show
that $|\chi\rangle$ can be used to do partial entanglement
recovery for any transformation of $|\psi\rangle$ to
$|\varphi\rangle$ with $|\psi\rangle\in S^o(|\varphi\rangle)$. Let
us take
$$
\chi(\epsilon)=((\frac{1}{a})^{\oplus a-1}, \frac{1}{a}-\epsilon,
\epsilon, 0^{\oplus k-a-1}),
$$
where $\epsilon$ is a suitably small positive number. We have that
$$
\chi(\epsilon)\prec \chi {\rm\ and\ } \chi^{\downarrow}\neq
\chi^{\downarrow}(\epsilon)
$$
for any $0<\epsilon<\frac{1}{2a}$. Noticing $|\psi\rangle\in
S^o(|\varphi\rangle)$, by Lemma \ref{lginterior} we have
$$ (\psi\otimes (\frac{1}{a})^{\oplus a})\lhd (\varphi\otimes
(\frac{1}{a})^{\oplus a}).
$$
Combining the above equation with Eq. (\ref{schmidtcondition})
gives
$$
(\psi\otimes (\frac{1}{a})^{\oplus a})\lhd (\varphi\otimes
((\frac{1}{a})^{\oplus a},0))^1_{na},
$$
where the notation $(x)_j^i$ denotes the segment
$(x^{\downarrow}_i,\ldots, x^{\downarrow}_j)$ of $x^{\downarrow}$.
Since a sufficiently small perturbation on the right-hand side of
the above equation cannot change the relation $\lhd$, we have that
\begin{equation}\label{appendzero}
(\psi\otimes (\frac{1}{a})^{\oplus a})\lhd (\varphi\otimes
((\frac{1}{a})^{\oplus a-1},\frac{1}{a}-\epsilon,\epsilon))^1_{na}
\end{equation}
for small enough positive number $\epsilon$.

Appending suitable number of zeros on the both sides of Eq.
(\ref{appendzero}) gives
$$
\psi\otimes \chi\prec \varphi\otimes \chi(\epsilon),
$$
which proves that $|\chi\rangle$ can do partial entanglement
recovery for the transformation of $|\psi\rangle$ to
$|\varphi\rangle$.

Now we turn to  the  general case that $|\chi\rangle$  has at
least two nonzero distinct Schmidt coefficients, i.e.,
$0<L_u(|\chi\rangle)<1$. We shall consider the following three
cases: (i) $L_u(|\chi\rangle)>g_u(|\varphi\rangle)$; (ii)
$L_u(|\chi\rangle)<g_u(|\varphi\rangle)$; and (iii)
$L_u(|\chi\rangle)=g_u(|\varphi\rangle)$. \vspace{1em}

First, we deal with case (i).  We shall prove that if
$L_u(|\chi\rangle)>g_u(|\varphi\rangle)$, then $|\chi\rangle$ can
do partial entanglement recovery for the transformation of
$|\psi\rangle$ to $|\varphi\rangle$ such that $|\psi\rangle$ is in
$S^{o}(|\varphi\rangle)$.

Suppose $\chi^{\downarrow}=(\gamma_1^{\oplus k_1},\ldots,
\gamma_m^{\oplus k_m})$. Then there exists $1\leq i<m$ such that
\begin{equation}\label{keycond}
g_u(|\varphi\rangle)<L_u(|\chi\rangle)=\frac{\gamma_{i+1}}{\gamma_i}<1.
\end{equation}

Let us denote $(\gamma_1^{\oplus k_1},\ldots, \gamma_{i}^{\oplus
k_i-1})$, $(\gamma_i,\gamma_{i+1})$, and $(\gamma_{i+1}^{\oplus
k_{i+1}-1}, \ldots, \gamma_{m}^{\oplus k_m})$ by $\gamma'$,
$\gamma''$, and $\gamma'''$, respectively. If $k_i=1$ or
$k_{i+1}=1$, we simply omit the meaningless terms $\gamma'$ or
$\gamma'''$, respectively. For any $|\psi\rangle\in
S^{o}(|\varphi\rangle)$, we have
\begin{equation}\label{blockmaj1}
\psi\otimes \gamma'\prec \varphi\otimes \gamma',
\end{equation}
\begin{equation}\label{blockmaj2}
\psi\otimes \gamma''\lhd \varphi\otimes \gamma'',
\end{equation}
and
\begin{equation}\label{blockmaj3}
\psi\otimes \gamma'''\prec \varphi\otimes \gamma''',
\end{equation}
where  Eq. (\ref{blockmaj2}) comes from Eq. (\ref{keycond}) and
Lemma \ref{lginterior}. So for a sufficiently small positive
number $\epsilon$, we have
\begin{equation}\label{blockmaj4}
\psi\otimes \gamma''\lhd \varphi\otimes \gamma''(\epsilon),
\end{equation}
where $\gamma''(\epsilon)=(\gamma_{i}-\epsilon,
\gamma_{i+1}+\epsilon)$. By Eqs. (\ref{blockmaj1}),
(\ref{blockmaj3}) and (\ref{blockmaj4}), applying Lemma
\ref{fact2} gives
$$
\psi\otimes \chi\prec \varphi\otimes \omega,
$$
where $\chi^{\downarrow}=(\gamma',\gamma'',\gamma'')$ and
$\omega^{\downarrow}=(\gamma',\gamma''(\epsilon),\gamma''')$. It
is obvious that $\omega\prec \chi$ but $\chi^{\downarrow}\neq
\omega^{\downarrow}$.

Second we deal with  case (ii). Suppose that
$L_u(|\chi\rangle)<g_u(|\varphi\rangle)$. We shall prove that
$|\chi\rangle$ cannot do partial entanglement recovery  for any
transformation $|\psi\rangle\rightarrow|\varphi\rangle$ such that
$|\psi\rangle$ is in $S(|\varphi\rangle)$.

By  contradiction, suppose that there exists a state
$|\omega\rangle$ such that $\psi\otimes \chi\prec\varphi\otimes
\omega$, $\omega\prec\chi$ and
$\chi^{\downarrow}\neq\omega^{\downarrow}$. For any state
$|\chi'\rangle$ such that $\omega\prec \chi'\prec \chi$ we have
\begin{equation}\label{6}
\psi\otimes \chi\prec \varphi\otimes \chi',
\end{equation}
where we have used the assumption $\psi\otimes \chi\prec
\varphi\otimes \omega$.

By Lemma \ref{fact1}, $\chi'$ can be chosen as
\begin{equation}\label{chiprime}
\chi'^{\downarrow}=(\gamma',\gamma''(\epsilon),\gamma'''),
\end{equation}
where $\gamma'=(\gamma_1^{\oplus k_1},\ldots,\gamma_i^{\oplus
 k_i-1})$, $\gamma''(\epsilon)=(\gamma_i-\epsilon,\ldots,\gamma_j+\epsilon)$,
$\gamma'''=(\gamma_j^{\oplus k_j-1},\ldots, \gamma_m^{\oplus
k_m})$, $1\leq i<j\leq m$, and $\epsilon$ is an arbitrarily
positive but small enough real number. In particular,
\begin{equation}\label{chi}
\chi^{\downarrow}=(\gamma',\gamma'',\gamma'''),
\end{equation}
where $\gamma''=\gamma''(0)=(\gamma_i,\ldots, \gamma_j)$. However,
we shall prove that such two indices $i$ and $j$ cannot exist, and
thus complete the proof of this case.

For simplicity, let $n={\rm dim}(\varphi)$. By  the assumption
$L_u(|\chi\rangle)<g_u(|\varphi\rangle)$, it follows that
\begin{equation}\label{Lgutarget}
\gamma_{s}\beta_n>\gamma_{s+1}\beta_1, {\rm\ for\ any\ }1\leq
s\leq m-1,
\end{equation}
where $\beta_1$ and $\beta_n$ are the greatest and the least
components of $\varphi$, respectively. Notice that $\psi\prec
\varphi$. Applying part (6) of Lemma \ref{property1} gives
$g_u(|\psi\rangle)\geq g_u(|\varphi\rangle)$, thus
$L_u(|\chi\rangle)<g_u(|\psi\rangle)$, or more explicitly,
\begin{equation}\label{Lgusource}
\gamma_{s}\alpha_n>\gamma_{s+1}\alpha_1, {\rm\ for\ any\ }1\leq
s\leq m-1,
\end{equation}
where $\alpha_1$ and $\alpha_n$ are the greatest and the least
components of $\psi$, respectively.

Eqs. (\ref{chi}) and (\ref{Lgusource}) imply
\begin{equation}\label{sourcedecom}
(\psi\otimes \chi)^{\downarrow}=((\psi\otimes
\gamma')^{\downarrow},(\psi\otimes
\gamma'')^{\downarrow},(\psi\otimes \gamma''')^{\downarrow}).
\end{equation}

Eqs. (\ref{chiprime}), (\ref{Lgutarget}), and $\epsilon>0$ imply
\begin{equation}\label{targetdecom}
(\varphi\otimes \chi')^{\downarrow}=((\varphi\otimes
\gamma')^{\downarrow},(\varphi\otimes
\gamma''(\epsilon))^{\downarrow},(\varphi\otimes
\gamma''')^{\downarrow}).
\end{equation}

Applying Lemma \ref{fact3} to Eqs. (\ref{sourcedecom}) and
(\ref{targetdecom}) yields
\begin{equation}\label{simplemaj}
\psi\otimes \gamma''\prec \varphi\otimes \gamma''(\epsilon),
\end{equation}
where we have used the assumption $\psi\prec\varphi$ and Eq.
(\ref{6}).

According to Eq. (\ref{Lgutarget}),  we can take a sufficiently
small positive number $\epsilon$ such that
$$
(\gamma_{i}-\epsilon) \beta_n>(\gamma_{i+1}+\epsilon)\beta_1.
$$
Then
$$
\begin{array}{rl}
e_n(\varphi\otimes
\gamma''(\epsilon))&=(\gamma_i-\epsilon)e_n(\varphi)\\
&<\gamma_ie_n(\psi)\\
&\leq e_n(\psi\otimes \gamma''),
\end{array}
$$
which contradicts Eq. (\ref{simplemaj}).

Finally,  we deal with case (iii), i.e.,
$L_u(|\chi\rangle)=g_u(|\varphi\rangle)$. This case is much more
complicated than the previous two cases. It is in fact the most
non-trivial part of  Theorem \ref{shannoncode}. Since this case is
of considerable interest, we will present a detailed proof for it.
To keep the proof as readable as possible, the lengthy proof is
divided into two easier lemmas. It is worth noting that both
lemmas are interesting in their own right.

The first lemma shows that an auxiliary state can do partial
entanglement recovery for a specific transformation if and only if
some of its segments can do partial entanglement recovery for the
same transformation.
\begin{lemma}\label{proposition1}\upshape
Let $|\psi\rangle$ and $|\varphi\rangle$ be two states such that
$\psi\prec \varphi$, and let $|\chi\rangle$ be a partially
entangled state with compact form
$\chi^{\downarrow}=(\gamma_1^{\oplus k_1},\ldots, \gamma_m^{\oplus
 k_m})$ for some $m>1$. If $L_u(|\chi\rangle)=g_u(|\varphi\rangle)$, then the following
 two statements are equivalent:

(i) $|\chi\rangle$ can do partial entanglement recovery for the
transformation of $|\psi\rangle$ to $|\varphi\rangle$;

(ii) There exists an index $i$ such that the unnormalized state
$|\chi'\rangle$ with $\chi'^{\downarrow}=(\gamma_{i}^{\oplus k_i},
\gamma_{i+1}^{\oplus k_{i+1}})$ can do partial entanglement
recovery for the transformation of $|\psi\rangle$ to
$|\varphi\rangle$, where $1\leq i<m$ and
$\frac{\gamma_{i+1}}{\gamma_i}=g_u(|\varphi\rangle)$.
 \end{lemma}

\textit{Proof.} The essential  part of the lemma is (i)
$\Rightarrow$ (ii). Suppose that $|\chi\rangle$ can do partial
entanglement recovery for the transformation of $|\psi\rangle$ to
$|\varphi\rangle$. That is,  there exists  a state
$|\omega\rangle$ satisfying $\psi\otimes \chi\prec \varphi\otimes
\omega$, $\omega\prec \chi$, and $\chi^{\downarrow}\neq
\omega^{\downarrow}$. Moreover, by Lemma \ref{fact2}, we can
assume that $|\omega\rangle$ is of the following form:
$$
\omega^{\downarrow}=(\gamma',\gamma''(\epsilon),\gamma'''),
$$
where $\gamma'=(\gamma_1^{\oplus k_1},\ldots, \gamma_{i-1}^{\oplus
k_{i-1}})$, $\gamma''(\epsilon)=(\gamma_i^{\oplus
k_i-1},\gamma_i-\epsilon,\ldots,
\gamma_j+\epsilon,\gamma_j^{\oplus k_j-1})$,
$\gamma'''=(\gamma_{j+1}^{\oplus k_{j+1}},\ldots, \gamma_m^{\oplus
k_m})$, $1\leq i<j\leq m$, and $\epsilon>0$. To make
$(\gamma''(\epsilon))^{\downarrow}=\gamma''(\epsilon)$ hold, we
have assumed that $\epsilon$ satisfies
$\gamma_p-\epsilon>\gamma_{p+1}+\epsilon$ for any $1\leq p\leq
m-1$. We also have
$$
\chi^{\downarrow}=(\gamma',\gamma''(0),\gamma'''),
$$
where $\gamma''(0)=(\gamma_i^{\oplus k_i},\ldots, \gamma_j^{\oplus
k_j})$.

By the assumptions $L_u(|\chi\rangle)=g_u(|\varphi\rangle)$ and
$\psi\prec \varphi$, we have
\begin{equation}\label{1sourcedecom}
(\psi\otimes \chi)^{\downarrow}=((\psi\otimes
\gamma')^{\downarrow},(\psi\otimes
\gamma''(0))^{\downarrow},(\psi\otimes \gamma''')^{\downarrow}).
\end{equation}
and
\begin{equation}\label{1targetdecom}
(\varphi\otimes \omega)^{\downarrow}=((\varphi\otimes
\gamma')^{\downarrow},(\varphi\otimes
\gamma''(\epsilon))^{\downarrow},(\varphi\otimes
\gamma''')^{\downarrow}).
\end{equation}

Applying Lemma \ref{fact3} to Eqs. (\ref{1sourcedecom}) and
(\ref{1targetdecom}) yields
\begin{equation}\label{1simplemaj}
\psi\otimes \gamma''(0)\prec \varphi\otimes \gamma''(\epsilon),
\end{equation}
where we have used the assumptions that $\psi\otimes \chi\prec
\varphi\otimes \omega$ and $\psi\prec\varphi$.

Therefore, for the simplicity of notations and without any loss of
generality, we can assume that $i=1$ and $j=m$. More directly, we
can write $\gamma''(0)$ and $\gamma''(\epsilon)$ as $\chi$ and
$\omega$, respectively. This, of course, will not cause any
confusion.

We shall prove $m=2$ and
$\frac{\gamma_2}{\gamma_1}=g_u(|\varphi\rangle)$ to complete the
proof of the lemma.

For simplicity, we assume that $n={\rm dim}(\psi)$ in the rest of
proof.

First, we prove $1<m\leq 3$. By contradiction, suppose that $m>3$.
Let us decompose
$$
\chi=(\chi'(0),\chi''(0))
$$
and
$$
\omega=(\chi'(\epsilon),\chi''(\epsilon)),
$$
where $$\chi'(\epsilon)=(\gamma_1^{\oplus
k_1-1},\gamma_1-\epsilon,\gamma_2^{\oplus k_2})$$ and
$$\chi''(\epsilon)=(\gamma_3^{\oplus k_3},\ldots,
\gamma_m+\epsilon, \gamma_m^{\oplus k_m-1}).$$

Again, $L_u(|\chi\rangle)=g_u(|\varphi\rangle)$ and $\psi\prec
\varphi$ give
$$
\gamma_3/\gamma_2\leq  g_u(|\varphi\rangle) {\rm\ and\ }
\gamma_3/\gamma_2\leq g_u(|\psi\rangle).
$$
That immediately yields
$$
(\psi\otimes \chi)^{\downarrow}=((\psi\otimes
\chi'(0))^{\downarrow},(\psi\otimes \chi''(0))^{\downarrow})
$$
and
$$
(\varphi\otimes \omega)^{\downarrow}=((\varphi\otimes
\chi'(\epsilon))^{\downarrow},(\varphi\otimes
\chi''(\epsilon))^{\downarrow}).
$$
So,
$$
\begin{array}{rl}
e_{(k_1+k_2)n}(\psi\otimes \chi)&=e_{(k_1+k_2)n}(\psi\otimes
\chi'(0))\\
&=k_1\gamma_1+k_2\gamma_2
\end{array}
$$
and
$$
\begin{array}{rl}
e_{(k_1+k_2)n}(\varphi\otimes
\omega)&=e_{(k_1+k_2)n}(\varphi\otimes
\chi'(\epsilon))\\
&=k_1\gamma_1+k_2\gamma_2-\epsilon,
\end{array}
$$
thus
$$
e_{(k_1+k_2)n}(\psi\otimes \chi)>e_{(k_1+k_2)n}(\varphi\otimes
\omega)
$$
for any small $\epsilon>0$.  This contradicts the assumption
$\psi\otimes \chi\prec \varphi\otimes \omega$. Hence $1< m\leq 3$

Second, we prove that for any  $1\leq i\leq m-1$, it holds that
${\gamma_{i+1}}/{\gamma_i}={\beta_n}/{\beta_1}$, where $\beta_1$
and $\beta_n$ are the greatest and the least components of
$\varphi$, respectively. By contradiction, we need to consider two
cases: (1) $m=2$ and (2) $m=3$.

(1) $m=2$. Suppose that $\gamma_2/\gamma_1<\beta_n/\beta_1$. Let
us choose a suitably small positive number $\epsilon$ such that

$$
(\gamma_2+\epsilon)\beta_1<(\gamma_1-\epsilon)\beta_n.
$$
A routine calculation shows  that
$$
\begin{array}{rl}
e_{k_1n}(\psi\otimes \chi)&=e_{k_1n}(\psi\otimes \gamma_1^{\oplus
k_1})\\
&=k_1\gamma_1
\end{array}
$$
and
$$
\begin{array}{rl}
e_{k_1n}(\varphi\otimes \omega)&=e_{(k_1-1)n}(\varphi\otimes
\gamma_1^{\oplus
k_1-1})+e_{n}((\gamma_1-\epsilon)\varphi)\\
&=k_1\gamma_1-\epsilon,
\end{array}
$$
which yields
$$
e_{k_1n}(\psi\otimes \chi)>e_{k_1n}(\varphi\otimes \omega)
$$
for any small $\epsilon>0$. That again contradicts  $\psi\otimes
\chi\prec \varphi\otimes \omega$.

(2) $m=3$. Suppose that $\gamma_2/\gamma_1<\beta_n/\beta_1$ or
$\gamma_3/\gamma_2<\beta_n/\beta_1$. We only consider the  case
where $\gamma_3/\gamma_2<\beta_n/\beta_1$,  and the left case is
similar to case (i).  Choose a suitably small positive number
$\epsilon$ such that
$$
(\gamma_3+\epsilon)\beta_1<\gamma_2\beta_n.
$$

Then a simple analysis shows
$$
\begin{array}{rl}
e_{(k_1+k_2)n}(\psi\otimes \chi)& = e_{k_1n}(\psi\otimes
\gamma_1^{\oplus k_1})+e_{k_2n}(\psi\otimes \gamma_2^{\oplus
k_2})\\
&=k_1\gamma_1+k_2\gamma_2
\end{array}
$$

and
$$
\begin{array}{rl}
e_{(k_1+k_2)n}(\varphi\otimes \omega) &
=e_{(k_1-1)n}(\varphi\otimes \gamma_1^{\oplus k_1-1})\\
& +e_{n}((\gamma_1-\epsilon)\varphi)+e_{k_2n}( \varphi\otimes
\gamma_2^{\oplus k_2})\\
&=k_1\gamma_1+k_2\gamma_2-\epsilon,
\end{array}
$$

which yields
$$
e_{(k_1+k_2)n}(\psi\otimes \chi)>e_{(k_1+k_2)n}(\varphi\otimes
\omega).
$$
That is a contradiction with $\psi\otimes \chi\prec \varphi\otimes
\omega$.

Third, we prove that $m=2$. By contradiction, we shall show that
if $m=3$ then $$\psi\otimes \chi\prec \varphi\otimes \omega$$
cannot hold for any  small enough positive number $\epsilon$,
where $\chi^{\downarrow}=(\gamma_1^{\oplus k_1},\gamma_2^{\oplus
k_2},\gamma_3^{\oplus k_3})$,
$\omega^{\downarrow}=(\gamma_1^{\oplus k_1-1},\gamma_1-\epsilon,
\gamma_2^{\oplus k_2},\gamma_3+\epsilon, \gamma_3^{\oplus
k_3-1})$,
$\gamma_2/\gamma_1=\gamma_3/\gamma_2=g_u(|\varphi\rangle)$.

To be specific,  let $\varphi^{\downarrow}=(\beta_1^{\oplus
m_1},\ldots, \beta_h^{\oplus m_h})$ for some $h\geq 2$. Obviously,
$n={\rm dim}(\psi)=\sum_{t=1}^hm_t$. Choose $\epsilon$ such that
$$
(\gamma_1-\epsilon)\beta_i>\gamma_1\beta_{i+1}
$$
and
$$
(\gamma_3+\epsilon)\beta_{i+1}<\gamma_3\beta_i
$$
for any $1\leq i\leq h-1$.

In addition to the above constraints, we also need $\epsilon$
satisfying
$$
(\gamma_1-\epsilon)\gamma_h>(\gamma_3+\epsilon)\beta_1.
$$
A direct calculation gives
$$
(\psi\otimes \chi)^{\downarrow}=(\varphi'(0),\varphi''(0))
$$
and
$$
(\varphi\otimes
\omega)^{\downarrow}=(\varphi'(\epsilon),\varphi''(\epsilon)),
$$
where
$$
\begin{array}{rl}
\varphi'(\epsilon)=&[\gamma_1\beta_1^{\oplus (k_1-1)m_1},
(\gamma_1-\epsilon)\beta_1^{\oplus m_1},\\
&\ldots, \gamma_1\beta_{h-1}^{\oplus (k_1-1)m_{h-1}},
(\gamma_1-\epsilon)\beta_{h-1}^{\oplus m_{h-1}}],
\end{array}
$$
and
$$
\begin{array}{rl}
\varphi''(\epsilon)=&[\gamma_1\beta_h^{\oplus
(k_1-1)m_h+k_2m_1},(\gamma_1-\epsilon)\beta_h^{\oplus m_h},\\
&\ldots,(\gamma_3+\epsilon)\beta_h^{\oplus m_h},
\gamma_3\beta_h^{\oplus (k_3-1)m_h}],
\end{array}
$$
where we have used $\gamma_1\beta_h=\gamma_2\beta_1$.

It can be readily verified that
$$
\begin{array}{rl}
e_{k_1n}(\psi\otimes
\chi)&=e_{k_1(n-m_h)}(\varphi'(0))+e_{k_1m_h}(\varphi''(0))\\
&=k_1\gamma_1
\end{array}
$$
and
$$
\begin{array}{rl}
e_{k_1n}(\varphi\otimes
\omega)&=e_{k_1(n-m_h)}(\varphi'(\epsilon))+e_{k_1m_h}(\varphi''(\epsilon))\\
&=(k_1\gamma_1-\epsilon)(1-m_h\beta_h)+e_{k_1m_h}(\varphi''(\epsilon)).
\end{array}
$$

To calculate $e_{k_1m_h}(\varphi''(\epsilon))$, we need to
consider the following two cases:

(a) $k_2m_1\geq m_h$. Then
$$
\begin{array}{rl}
 e_{k_1m_h}(\varphi''(\epsilon))&=e_{k_1m_h}(\gamma_1\beta_h^{\oplus
 k_1m_h})\\
 &=k_1m_h\gamma_1\beta_h,
\end{array}
$$
thus
$$
\begin{array}{rl}
e_{k_1n}(\varphi\otimes
\omega)&=k_1\gamma_1-\epsilon(1-m_h\beta_h)\\
&<e_{k_1n}(\psi\otimes \chi)
\end{array}
$$
providing $\epsilon>0$.

(b) $k_2m_1< m_h$. Then
$$
\begin{array}{rl}
e_{k_1m_h}(\varphi''(\epsilon))&=e_{l_1}(\gamma_1\beta_h^{\oplus
l_1})+e_{l_2}((\gamma_1-\epsilon)\beta_h^{\oplus l_2})\\
&=k_1m_h\gamma_1\beta_h-\epsilon(m_h-k_2m_1)\beta_h,
\end{array}
$$
where $l_1=(k_1-1)m_h+k_2m_1$ and $l_2=m_h-k_2m_1$. Thus
$$
\begin{array}{rl}
e_{k_1n}(\varphi\otimes
\omega)&=k_1\gamma_1-\epsilon(1-k_2m_1\beta_h)\\
&<k_1\gamma_1-\epsilon(1-m_h\beta_h)\\
&<e_{k_1n}(\psi\otimes \chi)
\end{array}
$$
providing $\epsilon>0$. In the above two cases we have used
$\gamma_1\beta_h=\gamma_2\beta_1$ to simplify the calculations.

Both the above two cases contradict $\psi\otimes \chi\prec
\varphi\otimes \omega$. Thus $m=3$ is impossible.

With that we complete the proof of Lemma \ref{proposition1}. \hfill $\square$\\

By Lemma \ref{proposition1}, under the condition
$L_u(|\chi\rangle)=g_u(|\varphi\rangle)$, we only need to consider
a special form of $|\chi\rangle$. More precisely, $\chi$  has only
two distinct components. The following lemma will just handle such
a special form of $|\chi\rangle$.

\begin{lemma}\label{proposition2}\upshape
Let $|\chi\rangle$ be a partially entangled  state with compact
form $\chi^{\downarrow}=(\beta_1^{\oplus k_1},\beta_2^{\oplus
k_2})$ for some $\beta_1>\beta_2>0$, and let $|\varphi\rangle$ be
another state satisfying $g_u(|\varphi\rangle)=L_u(|\chi\rangle)$.
Then $|\chi\rangle$ can do partial entanglement recovery for the
transformation of $|\psi\rangle$ to $|\varphi\rangle$ such that
$|\psi\rangle$ is in $S^o(|\varphi\rangle)$ if and only if
\begin{equation}\label{specialform}
\varphi^{\downarrow}=(\frac{\chi'^{\oplus m}}{C})^{\downarrow},
\end{equation}
where $\chi'$ is a segment of $\chi^{\downarrow}$ with two
distinct components, $C$ is a normalization factor, and $m\geq 1$.

Moreover, if $|\chi\rangle$ and $|\varphi\rangle$ don't satisfy
Eq. (\ref{specialform}), then $|\chi\rangle$ cannot do partial
entanglement recovery for any transformation of $|\psi\rangle$ to
$|\varphi\rangle$ such that $|\psi\rangle$ is in
$S(|\varphi\rangle)$.
\end{lemma}

\textit{Proof.} We first prove that if $|\chi\rangle$ can do
partial entanglement recovery for some transformation of
$|\psi\rangle$ to $|\varphi\rangle$ with $\psi\prec \varphi$,
i.e., there exists a state $|\omega\rangle$ satisfying
$\psi\otimes \chi\prec \varphi\otimes \omega$, $\omega\prec \chi
$, and $\chi^{\downarrow}\neq \omega^{\downarrow}$, then
$|\chi\rangle$ and $|\varphi\rangle$ should satisfy Eq.
(\ref{specialform}).

Suppose that $|\varphi\rangle$ and $|\omega\rangle$ have compact
forms
$$
\varphi^{\downarrow}=(\beta_1^{\oplus m_1},\ldots, \beta_h^{\oplus
m_h})
$$
and
$$
\omega^{\downarrow}=(\gamma_1^{\oplus
k_1-1},\gamma_1-\epsilon,\gamma_2+\epsilon,  \gamma_{2}^{\oplus
k_{2}-1}).
$$
We shall prove that if $\psi\otimes \chi\prec \varphi\otimes
\omega$ and $\omega\prec \chi $ for any sufficiently small
positive number $\epsilon$, then
\begin{equation}\label{condition4}
h=2 {\rm\ and\ } \frac{1}{k_2}\leq \frac{m_1}{m_2}\leq k_1.
\end{equation}
Or more compactly, $\varphi$ has the form as in Eq.
(\ref{specialform}).

The condition $L_u(|\chi\rangle)=g_u(|\varphi\rangle)$ is
equivalent to
\begin{equation}\label{equal}
\gamma_2\beta_1=\gamma_1\beta_h.
\end{equation}
For any $1\leq i\leq h-1$,  choose $\epsilon$ such that
\begin{equation}\label{epsilon1}
(\gamma_1-\epsilon)\beta_i>\gamma_1\beta_{i+1},
\end{equation}
and
\begin{equation}\label{epsilon2}
(\gamma_2+\epsilon)\beta_{i+1}<\gamma_2\beta_{i}.
\end{equation}

In addition to the above conditions, we also choose $\epsilon$
satisfying
\begin{equation}\label{epsilon3}
(\gamma_1-\epsilon)\beta_{h-1}>(\gamma_2+\epsilon)\beta_1,
\end{equation}
and
\begin{equation}\label{epsilon4}
(\gamma_1-\epsilon)\beta_h>(\gamma_2+\epsilon)\beta_2.
\end{equation}

By the condition $L_u(|\chi\rangle)=g_u(|\varphi\rangle)$ and
$\psi\prec \varphi$, it is easy to verify that
\begin{equation}\label{sourcetensor1}
(\psi\otimes \chi)^{\downarrow}=((\psi\otimes \gamma_1^{\oplus
k_1})^{\downarrow},(\psi\otimes \gamma_2^{\oplus
k_2})^{\downarrow}).
\end{equation}

Take
\begin{equation}\label{indexl}
n={\dim}(\varphi)=\sum_{i=1}^h{m_i}.
\end{equation}
Then by Eq. (\ref{sourcetensor1}),
\begin{equation}\label{condition5}
e_{k_1n}(\psi\otimes \chi)=k_1\gamma_1.
\end{equation}
By Eqs.  (\ref{equal})--(\ref{epsilon4}),  a careful analysis
gives
$$
(\varphi\otimes \omega)^{\downarrow}=(\varphi',\varphi''),
$$
where
$$
\begin{array}{rl}
\varphi'=&[\gamma_1\beta_1^{\oplus
m_1(k_1-1)},(\gamma_1-\epsilon)\beta_1^{\oplus
m_1},\\
&\ldots,\gamma_1\beta_{h-1}^{\oplus
m_{h-1}(k_1-1)},(\gamma_1-\epsilon)\beta_{h-1}^{\oplus m_{h-1}}]
\end{array}
$$
and
$$
\begin{array}{rl}
\varphi''=&[(\gamma_2+\epsilon)\beta_1^{\oplus
m_1},\gamma_1\beta_h^{\oplus m_h(k_1-1)+m_1(k_2-1)},\\
&(\gamma_1-\epsilon)\beta_h^{\oplus m_h},
\ldots,\gamma_2\beta_h^{\oplus m_h(k_2-1)}].
\end{array}
$$
So
\begin{equation}\label{llargest}
\begin{array}{rl}
e_{k_1n}(\varphi\otimes
\omega)&=e_{k_1(n-m_h)}(\varphi')+e_{k_1m_h}(\varphi'')\\
&=(k_1\gamma_1-\epsilon)(1-m_h\beta_h)+e_{k_1m_h}(\varphi'').
\end{array}
\end{equation}

We need to consider the following four cases according to the
values of $e_{k_1m_h}(\varphi'')$:

Case (a): $m_1>k_1m_h$. Then it is obvious that
\begin{equation}\label{caseA}
\begin{array}{rl}
e_{k_1m_h}(\varphi'')&=e_{k_1m_h}((\gamma_2+\epsilon)\beta_1^{\oplus
k_1m_h})\\
&=k_1m_h(\gamma_2+\epsilon)\beta_1.
\end{array}
\end{equation}
Combining Eqs.  (\ref{llargest}) with (\ref{caseA}), we have
$$e_{k_1n}(\varphi\otimes \omega)=k_1\gamma_1+\epsilon(k_1m_h\beta_1+m_h\beta_h-1),$$
where we have used the relation $\gamma_1\beta_h=\gamma_2\beta_1$
to simplify the calculations.

Since $\psi\otimes \chi\prec \varphi\otimes \omega$, it follows
that $e_{k_1n}(\varphi\otimes \omega)\geq e_{k_1n}(\psi\otimes
\chi)$, i.e.,
$$
k_1\gamma_1+\epsilon(k_1m_h\beta_1+m_h\beta_h-1)\geq k_1\gamma_1.
$$
Or equivalently,
\begin{equation}\label{contradiction2}
k_1m_h\beta_1+m_h\beta_h\geq 1.
\end{equation}
However, by $m_1>k_1m_h$ and $\sum_{i=1}^h m_i\beta_i=1$, it
follows that
$$
k_1m_h\beta_1+m_h\beta_h<m_1\beta_1+m_h\beta_h\leq 1,
$$
which contradicts Eq. (\ref{contradiction2}).

Case (b): $m_h\leq m_1\leq k_1m_h$. It is easy to calculate that
\begin{equation}\label{caseB}
\begin{array}{rl}
e_{k_1m_h}(\varphi'')&=e_{m_1}((\gamma_2+\epsilon)\beta_1^{\oplus
m_1})\\
&{\rm\ \ }+e_{k_1m_h-m_1}(\gamma_1\beta_h^{\oplus
k_1m_h-m_1})\\
&=\epsilon m_1\beta_1+k_1m_h\gamma_1\beta_h.
\end{array}
\end{equation}
By Eqs.  (\ref{llargest}) and (\ref{caseB}), it follows that
$$
e_{k_1n}(\varphi\otimes
\omega)=k_1\gamma_1+\epsilon(m_1\beta_1+m_h\beta_h-1).
$$

Since $\psi\otimes \chi\prec \varphi\otimes \omega$, it follows
that $e_{k_1n}(\varphi\otimes \omega)\geq e_{k_1n}(\psi\otimes
\chi)$, i.e.,
$$
k_1\gamma_1+\epsilon(m_1\beta_1+m_h\beta_h-1)\geq k_1\gamma_1.
$$
Or equivalently,
\begin{equation}\label{contradiction3}
m_1\beta_1+m_h\beta_h\geq 1.
\end{equation}

It is easy to verify that Eq. (\ref{contradiction3}) holds if and
only if $h=2$.

Case (c): $m_1\leq m_h\leq k_2m_1$. Similar to Case (b),
$\psi\otimes \chi\prec \varphi\otimes \omega$ holds for any small
enough positive $\epsilon$  if and only if $h=2$.

Case (d): $m_h>k_2m_1$. Similar to Case (a), this also causes a
contradiction.

Summarizing the above four cases, we obtain that $|\varphi\rangle$
should satisfy  Eq. (\ref{condition4}), which is equivalent to Eq.
(\ref{specialform}).

Now we turn to prove that the condition in Eq. (\ref{specialform})
is also sufficient for partial entanglement recovery. Suppose that
$|\varphi\rangle$ and $|\omega\rangle$ are with compact forms
$$
\varphi^{\downarrow}=(\beta_1^{\oplus m_1},\beta_2^{\oplus m_2})
$$
and
$$
\omega^{\downarrow}=(\gamma_1^{\oplus
k_1-1},\gamma_1-\epsilon,\gamma_2+\epsilon,  \gamma_{2}^{\oplus
k_{2}-1}),
$$
where
\begin{equation}\label{assumption}
\frac{\beta_2}{\beta_1}=\frac{\gamma_2}{\gamma_1}{\rm\ and\ }
\frac{1}{k_2}\leq \frac{m_1}{m_2}\leq k_1.
\end{equation}
Take $|\psi\rangle\in  S^o(|\varphi\rangle)$. We shall prove that
for a sufficiently small positive number $\epsilon$, the
transformation of $|\psi\rangle\otimes |\chi\rangle$ to
$|\varphi\rangle\otimes |\omega\rangle$ can be realized with
certainty under LOCC.

By  the assumptions $L_u(|\chi\rangle)=g_u(|\varphi\rangle)$ and
$\psi\prec \varphi$, it is easy to verify that
\begin{equation}\label{sourcetensor3}
(\psi\otimes \chi)^{\downarrow}=(\psi',\psi''),
\end{equation}
where $\psi'=(\psi\otimes \gamma_1^{\oplus k_1})^{\downarrow}$ and
$\psi''=(\psi\otimes \gamma_2^{\oplus k_2})^{\downarrow}$.
Similarly,
\begin{equation}\label{targettensor3}
(\varphi\otimes\chi)^{\downarrow}=(\varphi',\varphi''),
\end{equation}
where $\varphi'=(\varphi\otimes \gamma_1^{\oplus
k_1})^{\downarrow}$ and $\varphi''=(\varphi\otimes
\gamma_2^{\oplus k_2})^{\downarrow}$.

By Eq. (\ref{assumption}), it holds that
$\gamma_1\beta_2=\gamma_2\beta_1$. Hence we also have
\begin{equation}\label{segpartion1}
\varphi'=(\gamma_1\beta_1^{\oplus k_1m_1},\gamma_2\beta_1^{\oplus
k_1m_2})
\end{equation}
and
\begin{equation}\label{segpartion2}
\varphi''=(\gamma_1\beta_2^{\oplus k_2m_1},\gamma_2\beta_2^{\oplus
k_2m_2}).
\end{equation}
Similarly,
\begin{equation}\label{targettensor4}
(\varphi\otimes
\omega)^{\downarrow}=(\varphi'(\epsilon),\varphi''(\epsilon)),
\end{equation}
where
\begin{equation}\label{segperturbation1}
\begin{array}{rl}
\varphi'(\epsilon)=&[\gamma_1\beta_1^{\oplus
(k_1-1)m_1},(\gamma_1-\epsilon)\beta_1^{\oplus m_1},\\
&(\gamma_2+\epsilon)\beta_1^{\oplus m_1},\gamma_2\beta_1^{\oplus
k_1m_2-m_1}]
\end{array}
\end{equation}
and
\begin{equation}\label{segperturbation2}
\begin{array}{rl}
\varphi''(\epsilon)=&[\gamma_1\beta_2^{\oplus
m_1k_2-m_2},(\gamma_1-\epsilon)\beta_2^{\oplus m_2},\\
&(\gamma_2+\epsilon)\beta_2^{\oplus m_2},\gamma_2\beta_2^{\oplus
(k_2-1)m_2}].
\end{array}
\end{equation}
Note that  Eqs.  (\ref{segperturbation1}) and
(\ref{segperturbation2}) are well-defined since we have Eq.
(\ref{assumption}). We have also assumed that $\epsilon$ in Eqs.
(\ref{segperturbation1}) and (\ref{segperturbation2}) satisfies
the following constraints:
\begin{equation}\label{epsilon7}
(\gamma_1-\epsilon)\beta_1>(\gamma_2+\epsilon)\beta_1 {\rm\ and\ }
(\gamma_2+\epsilon)\beta_2<(\gamma_1-\epsilon)\beta_2.
\end{equation}

Since $\psi\lhd \varphi$, by Eqs.  (\ref{sourcetensor3}) and
(\ref{targettensor3}), applying Lemma \ref{lginterior} gives
\begin{equation}\label{strict1}
\psi'\lhd \varphi' {\rm\  and\ } \psi''\lhd \varphi''.
\end{equation}
A careful observation caries out that  $\varphi'(\epsilon)$ and
$\varphi''(\epsilon)$ are obtained by adding perturbations on
$\varphi'$ and $\varphi''$, respectively. So we have
\begin{equation}\label{strict2}
\psi'\lhd \varphi'(\epsilon) {\rm\  and\ } \psi''\lhd
\varphi''(\epsilon)
\end{equation}
for a sufficiently small positive number $\epsilon$.

Thus by Eqs.  (\ref{sourcetensor3}), (\ref{targettensor4}), and
(\ref{strict2}), applying Lemma \ref{fact2} gives
\begin{equation}
\psi\otimes \chi\prec \varphi\otimes \omega.
\end{equation}
It is easy to see that  $\omega\prec \chi$ and
$\chi^{\downarrow}\neq\omega^{\downarrow}$ providing the positive
number $\epsilon$ small enough. In other words, $|\chi\rangle$ can
do partial entanglement recovery for any transformation of
$|\psi\rangle$ to $|\varphi\rangle$ such that $|\psi\rangle$ is in
$S^o(|\varphi\rangle)$.

With that we complete the proof of  Lemma \ref{proposition2}.
\hfill $\square$\\

\section*{Appendix C: Proof of Theorem \ref{generaltheorem1}}

For simplicity, we denote $I_{\psi,\varphi}$ and
$D_{\psi,\varphi}$ by $I$ and $D$, respectively. We only need to
show that for any $1\leq l<{\rm dim}(\psi){\rm dim}(\chi)$, one of
the following two cases holds:

Case 1:  $e_l(\psi\otimes \chi)<e_l(\varphi\otimes \chi)$; or

Case 2:  $e_l(\psi\otimes \chi)= e_l(\varphi\otimes \chi)$, but
both sides are not related to $\chi^i$ ($i\in D$) and both of them
remain unchanged by an arbitrary but small enough perturbations on
$\chi^i$($i\in D$). Here we should point out that $\sum \chi^i$ is
supposed to be kept as a constant for each $i\in D$ during the
perturbations to guarantee that $|\chi\rangle$ is a valid quantum
state.

For this purpose, we rewrite $e_l(\psi\otimes \chi)$ as follows:
$$e_l(\psi\otimes \chi)=\sum_{i=1}^m\sum_{j=1}^m e_{l_{i,j}}(\psi^i\otimes
\chi^j),$$ where $\sum_{i,j}l_{i,j}=l{\rm \ and\ }0\leq
l_{i,j}\leq {\rm dim}(\psi^i\otimes \chi^j).$

It is easy to see that
\begin{equation}\label{majreform}
\begin{array}{rl}
e_l(\psi\otimes \chi)&=\sum_{i,j}e_{l_{i,j}}(\psi^i\otimes
\chi^j)\\
&\leq \sum_{i,j}e_{l_{i,j}}(\varphi^i\otimes \chi^j)\\
&\leq e_l(\varphi\otimes \chi),
\end{array}
\end{equation}
where the first inequality  follows from $\psi^i\otimes
\chi^j\prec \varphi^i\otimes \chi^j$ and the second one follows
from the definition of $e_l(\varphi\otimes \chi)$. If one of these
inequalities is strict, then Case 1 holds, and the proof is
completed; otherwise we only need to prove that Case 2 holds.

More precisely, we only need to show that if $e_l(\psi\otimes
\chi)= e_l(\varphi\otimes \chi)$ then for any $1\leq i\leq m$ and
$j\in D$, $l_{i,j}$ can only take two values: $0$ or ${\rm
dim}(\psi^i\otimes \chi^j)$ . Notice that $I\cup D=\{1,\ldots,
m\}$ and $I\cap D=\emptyset$. It suffices to prove $l_{i,j}\in \{0
, {\rm dim}(\psi^i\otimes \chi^j)\}$ for two cases: (1) $i\in I$,
$j\in D$, and (2) $i\in D$, $j\in D$.

Let us consider the case when $i\in I$ and $j\in D$ first. By Eq.
(\ref{embdedcond}), we have
\begin{equation}\label{useembdded}
\varphi^i\otimes \chi^j\sqsubset \varphi^j\otimes \chi^i, {\rm\
for\ all\ } i\in I {\rm\ and\ } j\in D.
\end{equation}
That is, the values of the extreme components of $\varphi^i\otimes
\chi^j$ are strictly bounded by those of $\varphi^j\otimes
\chi^i$. Thus, we have
$$(\varphi^i\otimes \chi^j)^{\downarrow}_1<(\varphi^j\otimes \chi^i)^{\downarrow}_1$$
and
$$(\varphi^i\otimes \chi^j)^{\uparrow}_1>(\varphi^j\otimes \chi^i)^{\uparrow}_1.$$
Hence by the assumption that all inequalities in Eq.
(\ref{majreform}) hold with equalities and the definition of
$e_l(\varphi\otimes \chi)$, together with the above two equations,
we have
$$l_{j,i}=0\Rightarrow l_{i,j}=0$${\rm\ and\ }$$
l_{j,i}={\rm dim}(\psi^j\otimes \chi^i)\Rightarrow l_{i,j}={\rm
dim}(\psi^i\otimes \chi^j).$$ So, in order to prove $l_{i,j}\in
\{0,{\rm dim}(\psi^i\otimes \chi^j)\}$ in the case of $i\in I$ and
$j\in D$, we only need to show that $l_{j,i}\in \{0,{\rm
dim}(\psi^j\otimes \chi^i)\}$ for $i\in I$ and $j\in D$. Or
equivalently, to show $l_{i,j}\in \{0,{\rm dim}(\psi^i\otimes
\chi^j)\}$ for $i\in D$ and $j\in I$. (Here we interchange the
indices $i$ and $j$ for convenience.)

So combining this with the case of $i\in D$ and $j\in D$, the only
thing left to be proven is that
\begin{equation}\label{twovalue}
l_{i,j}\in \{0, {\rm dim}(\psi^i\otimes \chi^j)\}{\rm\ for\
all\rm\ }i\in D{\rm\ and\ }1\leq j\leq m.
\end{equation}

By Eq. (\ref{moreuniformcon}) and Lemma \ref{lginterior} we have
\begin{equation}\label{usestrictmaj}
\psi^i\otimes \chi^j\lhd \varphi^i\otimes \chi^j {\rm\ for\ all\ }
i\in D{\rm \ and\ }1\leq j\leq m.
\end{equation}
If there exist $s\in D$ and $1\leq t\leq m$ such that
$0<l_{s,t}<{\rm dim}(\psi^s\otimes \chi^t),$ then by Eq.
(\ref{usestrictmaj}) we have
\begin{equation}\label{strictmajcond}
e_{l_{s,t}}(\psi^s\otimes \chi^t)< e_{l_{s,t}}(\varphi^s\otimes
\chi^t).
\end{equation}
It follows that the first inequality in Eq. (\ref{majreform})
strictly holds,  which contradicts $e_l(\psi\otimes \chi)=
e_l(\varphi\otimes \chi)$. So Eq. (\ref{twovalue}) holds. With
that we complete the proof of Theorem \ref{generaltheorem1}.

\section*{Acknowledgement}

The authors are very grateful to the anonymous referees for their
detailed comments and instructive suggestions that helped to
improve the presentation of this paper greatly. In particular,
they pointed out a technical error in the proof of Lemma 2.2 in an
earlier version of this paper. The simple proof of statement (2)
of Lemma 2.2 in this version is due to one of them. The
algorithmic approach to partial entanglement recovery has been
motivated heavily by their suggestions. The authors are also
indebted to the colleagues in the Quantum Computation and Quantum
Information Research Group for many useful discussions. A special
acknowledgement is given to Yongzhi Cao for his careful reading of
this paper.


\begin{thebibliography}{99}

\bibitem{BB84}   C.~H.~Bennett and G.~Brassard, {\it Proceedings of IEEE International
Conference on Computers, Systems, and Signal Processing}, pp.
175--179, IEEE, New York, 1984.
\bibitem{BS92}   C.~H.~Bennett and S.~J.~Wiesner, \lq\lq Communication via one- and two-particle operators on Einstein-Podolsky-Rosen states," {\it Phys. Rev. Lett.}, vol. 69, no. 20, pp.
2881--2884, 1992.
\bibitem{BBC+93} C.~H.~Bennett, G.~Brassard, C.~Crepeau, R.~Jozsa, A.~Peres, and
W. K.~Wootters, \lq\lq Teleporting an unknown quantum state via
dual classical and Einstein-Podolsky-Rosen channels," {\it Phys.
Rev. Lett.}, vol. 70,  no. 13, pp. 1895--1899, 1993.
\bibitem{M00}    M.~A.~Nielsen and I.~L.~Chuang, {\it Quantum Computation and
Quantum Information,} Cambridge University Press, Cambridge, 2000.
\bibitem{BBPS96} C.~H.~Bennett, H.~J.~Bernstein, S.~Popescu, and B.~Schumacher, \lq\lq Concentrating partial entanglement by local
operations,"  {\it Phys. Rev. A}, vol. 53, no. 4, pp. 2046--2052,
1996.
\bibitem{NI99}   M.~A.~Nielsen, \lq\lq Conditions for a class of entanglement transformations," {\it Phys. Rev. Lett.}, vol. 83, no.2,  pp. 436--439, 1999.
\bibitem{MO79}   A.~W.~Marshall and I.~Olkin, {\it Inequalities: Theory of Majorization
and Its Applications,}  New York, American: Academic Press, 1979.
\bibitem{AU82}   P.~M.~Alberti and A.~Uhlmann, {\it Stochasticity and Partial Order:
Doubly Stochastic Maps and Unitary Mixing,} Dordrecht, Boston,
1982.
\bibitem{Shor02}   C.~H.~Bennett, P.~W.~Shor, J.~A.~Smolin, and A.~V.~Thapliyal, \lq\lq
Entanglement-assisted classical capacity of a quantum channel and
the reverse Shannon theorem," \textit{IEEE Trans. Inf. Theory},
vol. 48, no. 10, pp. 2637--2655, Oct. 2002.
\bibitem{FM00}   F.~Morikoshi, \lq\lq Recovery of entanglement lost in entanglement manipulation," {\it Phys. Rev. Lett.}, vol. 84, no. 14, pp. 3189--3192, 2000.
\bibitem{FM01}   F.~Morikoshi and M.~Koashi, \lq\lq Deterministic entanglement concentration," {\it Phys. Rev. A}, vol.
64, no. 2,  art. 022316, 2001.
\bibitem{SVF01}  S.~Bandyopadhyay, V.~Roychowdhury, and F.~Vatan, \lq\lq Partial recovery of entanglement in
bipartite-entanglement transformations," {\it Phys. Rev. A}, vol.
65, no. 4, art. 040303 (Rapid Communications), 2001.
\bibitem{SVF01b}  S.~Bandyopadhyay, V.~Roychowdhury, and F.~Vatan, \lq\lq Partial recovery of entanglement in
bipartite-entanglement transformations". Available online:
http://arxiv.org/abs/quant-ph/0105019.
\bibitem{SR02}  S.~Bandyopadhyay, and V.~Roychowdhury, \lq\lq Efficient entanglement-assisted transformation for bipartite pure states,"
{\it Phys. Rev. A}, vol. 65, no. 4, art. 042306, 2002.
\bibitem{SRS02}  S.~Bandyopadhyay, V.~Roychowdhury, and U.~Sen, \lq\lq Classification of nonasymptotic bipartite pure-state
entanglement transformations," {\it Phys. Rev. A}, vol. 65, no. 5,
art. 052315, 2002.
\bibitem{LO97} H.-K.~Lo and S.~Popescu, \lq\lq Concentrating entanglement by local actions: beyond mean
values," {\it Phys. Rev. A}, vol. 63, no. 2, art. 022301, 2001.
Available online: http://arxiv.org/abs/quant-ph/9707038.
\bibitem{Vidal99}G.~Vidal, \lq\lq Entanglement of pure states for a single copy," {\it Phys. Rev. Lett.},  vol. 83, no. 5, pp. 1046--1049, 1999.
\bibitem{JP99}   D.~Jonathan and M.~B.~Plenio, \lq\lq Entanglement-assisted local manipulation of pure quantum states," {\it Phys. Rev. Lett.}, vol. 83, no. 17, pp.
3566--3569, 1999.
\bibitem{DF03}   R.~Y.~Duan, Y.~Feng, X.~Li, and M.~S.~Ying, \lq\lq Trade-off between multiple-copy transformation and entanglement cataysis,"
{\it Phys. Rev. A}, vol. 71, no. 6, art. 062306, 2005. Available
online: http://www.arXiv.org/abs/quant-ph/0312010.
\bibitem{DFLY04} R.~Y.~Duan, Y.~Feng, X.~Li, and M.~S.~Ying, \lq\lq Multiple-copy entanglement transformation and entanglement cataysis,"  {\it Phys. Rev. A}, vol. 71, no. 4, art. 042319, 2005.
Available online: http://www.arXiv.org/abs/quant-ph/0404148.
\bibitem{DF04}   R.~Y. Duan, Y.~Feng, and M.~S.~Ying, \lq\lq An equivalence of entanglement-assisted transformation and multiple-copy transformation."  Available online:
http://www.arXiv.org/abs/quant-ph/0404046.
\bibitem{SD03}   X.~M.~Sun, R.~Y.~Duan, and M.~S.~Ying, \lq\lq The existence of quantum entanglement
catalysts," \textit{IEEE Trans. Inf. Theory}, vol. 51, no. 1, pp.
75--80, Jan. 2005.
\bibitem{XZZ02}  X.~L.~Feng, Z.~Y.~Wang, and Z.~Z.~Xu, \lq\lq Mutual catalysis of entanglement transformations for pure entangled states," {\it Phys. Rev. A}, vol. 65,
no. 2, art. 022307, 2002.
\bibitem{Edel86} H.~Edelsbrunner, J.~O'Rourke, and R.~Seidel,
\lq\lq Constructing arrangements of lines and hyperplanes with
applications," \textit{SIAM J. Comput.}, vol. 15, pp. 341--363,
1986.
\bibitem{Karmarkar} N.~Karmarkar, \lq\lq A new polynomial-time
algorithm for linear programming," \textit{Combinatorica}, vol. 4,
pp. 373--395, 1984.
\bibitem{Megiddo} N.~Megiddo, \lq\lq Linear programming in linear time  when the dimension is
fixed," \textit{J. ACM}, vol. 31, pp. 114--127, 1984.
\end{thebibliography}
\end{document}